\documentclass[aps,twocolumn,longbibliography,floatfix]{revtex4-2}
\usepackage{graphicx}
\usepackage{subcaption}
\usepackage{bm}
\usepackage{color}
\usepackage{amsmath}
\usepackage{float}
\usepackage{gensymb}
\usepackage{array}
\usepackage{tabularx}
\usepackage{multirow}
\usepackage[bookmarks, colorlinks=false]{hyperref}
\hypersetup{colorlinks=true, allcolors=blue}

\begin{document}
\title{Simulations of Nanocrystalline Iron Formation under High Shear Strain}
\author{Ivan Tolkachev}
\email{Ivan.Tolkachev@eng.ox.ac.uk}
\affiliation{Department of Engineering Science, University of Oxford, United Kingdom}
\author{Pui-Wai Ma}
\email{Leo.Ma@ukaea.uk}
\affiliation{United Kingdom Atomic Energy Authority, Culham Campus, Abingdon, Oxfordshire, OX14 3DB, United Kingdom}
\author{Daniel Mason}
\email{Daniel.Mason@ukaea.uk}
\affiliation{United Kingdom Atomic Energy Authority, Culham Campus, Abingdon, Oxfordshire, OX14 3DB, United Kingdom}
\author{Felix Hofmann}
\email{Felix.Hofmann@eng.ox.ac.uk}
\affiliation{Department of Engineering Science, University of Oxford, United Kingdom}

\begin{abstract}
High-shear methods have long been used in experiments to refine grain structures in metals, yet the underlying mechanisms remain elusive. We demonstrate a refinement process using molecular dynamic simulations of iron, wherein nanocrystalline structures are generated from initially perfect lattices under high-shear strain. The simulation cells undergo a highly disordered state, followed by an atomic reordering and grain coarsening, resulting in nanograins. We explore the dependence on parameters such as temperature, heat dissipation rate, shear strain rate, and carbon impurity concentration. Higher temperatures lead to the formation of larger and longer grains. The faster heat dissipation sample initially yields more small grains, but their number subsequently reduces, and is lower than the slower heat dissipation sample at approximately $\gamma = 1.5$. Slower strain rates do not promote nanograin formation. The presence of carbon impurities appears to have little effect on grain formation. This detailed analysis affords insight into the mechanisms that control the formation of nanograins under high-shear conditions.
\end{abstract}
\maketitle

\section{Introduction}
Grain refinement can be achieved through various methods, including the use of chemical refiners or spray-forming techniques \cite{GOVENDER2014109}. An attractive approach to accomplish grain refinement is through severe plastic deformation techniques, such as accumulative roll bonding and equal channel angular pressing \cite{AgarwalIOP}. High-shear methods, such as high-pressure torsion , can significantly alter the microstructure of metals. Numerous materials, including Al, Cu, and Mg alloys, have shown a reduction in grain size with increasing shear strain \cite{ZHILYAEV2005277,ITO200932,JAMALIAN2019142,BRYLA2018323,Yusuf, STRANGWARDPRYCE2023101468}. Yusuf \textit{et al.} \cite{Yusuf} observed a reduction in grain size with increasing shear strain for 316L stainless steel, which also correlates with an increase in Vickers microhardness. Despite numerous experimental studies attempting to elucidate the mechanisms of nanocrystal formation with increased shear strain, the results are inconclusive \cite{Borodachenkova17}.

The leading theory for grain refinement suggests that dislocations are generated in the material due to shear strain. These dislocations accumulate and form subgrain boundaries. As the shear strain increases, dislocations are annihilated at these boundaries, leading to a rise in the misorientation angle between grains. Some of the formed dislocations are not absorbed and persist, forming low-angle grain boundaries, which perpetuates the grain refinement process \cite{ITO200932, Borodachenkova17, XU20223506, XiuyanLi}.

Other factors can also influence grain refinement, such as the number and distribution of precipitates within the material \cite{STARINK20095796, LIU12, BORODACHENKOVA2015150}, as well as the presence of twinning deformation \cite{LUO2019157, Wenqian}. Isik \textit{et al.} \cite{MuratIsik2016M2016052} observed a grain subdivision and grain refinement through a deformation-induced martensitic phase transformation, from $\gamma \rightarrow \epsilon$ phase in Co-28Cr-6Mo alloy.

Numerous computational methods have been employed to explore the mechanism of grain refinement through plastic shearing. Finite element has been used to understand how the HPT process can modify the samples \cite{Verleysen,Pereira2019MF201906,KIM2001617,YOON200832}. Finite element analysis simulations greatly depend on the constitutive model used during the simulation \cite{ZHANG2017175}. As such, the physics of a given finite element system are defined by the constitutive model, which may not allow other physical mechanisms of grain refinement to exist within the simulation, which in turn affects the observable deformation modes. Nevertheless, numerous finite element studies have been conducted that draw conclusions on possible grain refinement mechanisms under shear strain.

Wei \textit{et al.} \cite{Wei12030351} investigated the mechanisms of grain refinement in single crystal Ni subjected to high-pressure torsion, employing crystal plasticity finite element simulations. The results suggest the existence of two rotation modes that directly contributed to the formation of grain boundaries with high misorientations. Additionally, FEA simulations of high angle pressing were conducted by Frydrych \textit{et al.} \cite{Frydrych_2018} to explore the mechanism of grain refinement in an already nanocrystalline face-centered cubic material. This study confirmed that initial grain orientation was a major indicator of a grain's susceptibility to further refinement. The study showed that grains which had larger Taylor factors were more prone to refinement. A larger Taylor factor corresponds to greater plastic work done, for a given slip range \cite{MECKING1996465}. The work of Frydrych \cite{Frydrych_2018} makes no further attempt to quantify the orientations, making it difficult to obtain exact lattice orientations that are more prone to refinement.

Molecular dynamics (MD) simulations have been employed to study microstructural evolution in metals subjected to severe plastic deformation \cite{Schiotz12, Merzhievsky12, KOTRECHKO200692, Wang_2017, agarwal_paun, Husain12, Nikonov2018, BulatovPaper, TRAMONTINA20149, LuuShock, WenShock, HEALY2014105, GuoFeComp}. The work of Nikonov \cite{Nikonov2018} used MD simulations to shear a perfect, single crystal simulation cell. However, the results were only used for shear stress vs. strain comparisons with a polycrystalline box, and no conclusions were reached about the single crystal shear. The work of Guan \textit{et al.} \cite{GUAN2022111105} attempted to replicate the high-pressure torsion process using an MD simulation of Aluminum. In \cite{GUAN2022111105}, the initial box was already in a nanocrystalline state before the shearing took place, and thus, there was no investigation into the actual formation of nanocrystals.

The work of Zepeda-Ruiz \textit{et al.} \cite{BulatovPaper} used MD simulations to apply compression to tantalum at high strain rates. It was found that under certain strain conditions, the mechanism of structural evolution changed. For example, the initially pristine simulation cell yielded by deformation twinning whilst a cell that contained 24 dislocation loops yielded through dislocation motion and multiplication. Similarly, Healy and Ackland \cite{HEALY2014105} applied compressive and tensile strain to iron nanopillars and also found the deformation mechanisms to be dislocation glide and twinning.

Guo \textit{et al.} \cite{GuoFeComp} used MD to apply strain to iron nanowires at high strain rates. This work showed the formation of nanocrystalline iron under high compressive strain and attributed this to a body-centred cubic (BCC) to hexagonal close packed structural transition, followed by a reverse process. However, this study was limited to and engineering shear strain of $\gamma = 0.4$. It is important to explore higher strain levels to examine the microstructural evolution of iron subjected to large strain. 

The precise underlying processes responsible for the development of a nanocrystalline microstructure via high shear methods remains elusive. There is a significant disparity between the shear rates, samples sizes, and timescales in our simulations when compared with experimental data. Nonetheless, molecular dynamics simulations on iron can offer insights into the mechanisms through which shear could induce the formation of nanocrystalline structures. The saturation grain size for pure metals subjected to severe plastic deformation is commonly on the order of several hundred nanometers \cite{Pippan}, however, large strain rates like those present in this study have been used experimentally to reduce the annihilation of dislocations \cite{XiuyanLi} allowing for the production of much smaller grains. Furthermore, \cite{BulatovPaper} found that the stress-strain and dislocation density response of simulated material was very similar when considering a 33 million atom cell compared to a 268 million atom cell, implying that smaller simulation cells are useful for capturing larger scale material behaviour.

Iron is chosen as the material of the current study because of its diverse uses in industrial applications. Especially, in advanced fission and future fusion reactors \cite{Baluc2004,ZINKLE2013735,ZINKLEBUSBY}, iron-based alloys, such as ferritic/martensitic steels, are  chosen as the structural material. Ferritic/martensitic steels have the same body-centred cubic crystal structure as iron. It was shown experimentally that ferritic/martensitic steels exhibited less neutron irradiation-induced swelling than austenitic steels with face-centred cubic structure \cite{Garner_JNM_2000}. Experimental data on Eurofer 97 \cite{STRANGWARDPRYCE2023101468}, an ferritic/martensitic steel, showed extensive grain refinement under severe plastic deformation.

In the following, we first discuss our simulation methods. Then, we present our simulation results and discuss potential mechanisms underlying the observed phenomena. Where possible, we draw comparisons with experimental studies of severe plastic deformation.

\section{Simulation setup}

All MD simulations were carried out using LAMMPS \cite{LAMMPS}. We have not accounted for atomic spin in our simulations as it is difficult to quantify both strain and magnetisation at large strain values \cite{DosSantonSpin}, and its inclusion may not necessarily reveal new microstructural evolution effects. Simulation cells were created using Atomsk \cite{ATOMSK}. We constructed perfect crystal simulation cells with $80\times 80\times80$ unit cells, where each unit cell contains 2 atoms in body-centred cubic structure with a lattice parameter of 2.8665 {\AA}, corresponding to Fe. Each simulation cell contains 1,024,000 atoms. The starting crystal orientations are x = [100], y = [010], and z = [001]. Periodic boundary conditions are applied in all three directions. It is common practice to start with an initially perfect simulation cell when subjecting metals to high-strain \cite{BulatovPaper}. This is known to overestimate the strength of the crystal giving an upper-bound estimate of the perfect crystal strength, due to the initially defect free volume. In addition, we investigate the effect of including symmetry-breaking defects into the unstrained cell. The results of this can be found in Appendix A1.

The work of Zepeda-Ruiz \textit{et al.} \cite{BulatovPaper} found similar stress-strain and dislocation density responses between different simulation sizes. At present, there is no definitive criterion to determine the ideal system size for molecular dynamics simulations and Streitz \textit{et al.} \cite{Streitz} suggested a simulation size of 8 million atoms is required to produce size-independent results. As such, we also constructed a 8,192,000 atom simulation cell and subjected this to shear strain, with the results presented in Appendix A2. This larger simulation cell produced very similar results to the smaller box size. Furthermore, Tramontina \textit{et al.} \cite{TRAMONTINA20149} conducted high strain rate MD simulations in tantalum and found that cells with a $50\times 50$ unit cell cross-section were large enough to produce accurate results. As such, a cell size of 1,024,000 atoms is used throughout this work. Based on our simulations, we cannot rule out the existence of size effects, but we do not further consider these here. 

We also examined the effect of carbon impurities. We created simulation cells by adding 102 carbon atoms into perfect simualtion cells at random positions. This corresponds to a carbon impurity content of 100 appm. 

We adopted the interatomic potential for iron  developed by Ackland \textit{et al.} \cite{GJAckland_2004}, which has been widely used to investigate the microstructure evolution of iron  \cite{Hao2022,Li2023,Terentyev20111063,Li2012259}. For carbon inclusive simulations, we used the Hepburn-Ackland Fe-C interatomic potential \cite{HepburnAckland}, which was developed based on the aforementioned Ackland iron potential. It has been used in several studies concerned with the effect of carbon on the microstructural evolution of iron \cite{LiXieWen,TERENTYEV2011272,GRANBERG201523,GUNKELMANN20124901}.

The trajectory of a system of interacting atoms is governed by the Langevin equation of motion:
\begin{eqnarray}
\frac{d\mathbf{r}_i}{dt}&=& \mathbf{v}_i\\
m\frac{d\mathbf{v}_i}{dt}&=& \mathbf{F}_i - b\mathbf{v}_i +\mathbf{f}_i
\end{eqnarray}
where $\mathbf{r}_i$, $\mathbf{v}_i$, and $\mathbf{F}_i=-\partial U/\partial \mathbf{r}_i$ are position, velocity, and force associated with atom $i$. $U$ is the interatomic potential energy. $m$ is the mass of an atom. The temperature is controlled by the Langevin thermostat, where the damping parameter, $b$, is related to the delta-correlated fluctuating force $\mathbf{f}_i$ according to the fluctuation-dissipation theorem \cite{Chandrasekhar_Rev_Mod_Phys_1943,Kubo_RPP_1966}, namely,
\begin{equation}
    \langle \mathbf{f}_i (t)\rangle = 0,
\end{equation}
\begin{equation}
    \langle f_{i\alpha}(t) f_{j\beta}(t') \rangle = \mu \delta_{ij}\delta_{\alpha\beta}\delta(t-t'),
\end{equation}
where $\alpha$ and $\beta$ are Cartesian axis directions, and
\begin{equation}
    \mu = 2 b k_B T.
\end{equation}
The fluctuation and dissipation of atoms can be considered as a result of electron-phonon interaction, so b describes the strength of electron-phonon coupling. We used the electron-phonon coupling parameter for iron according to Mason \textit{et al.} \cite{Mason_JPCM_2015}, such that $b=6.875$ eV fs {\AA}$^{-2}$. We denote this as $b_1$. The value of $b_1$ is obtained in Mason \textit{et al.} \cite{Mason_JPCM_2015} by utilising an experimentally derived electron-phonon $\lambda$ value \cite{AllenEph} for iron. To investigate the damping strength effect, we also used a damping parameter $b_2=10b_1$ in some simulations.

Before applying any shear, the simulation cells are thermalised to particular temperatures. The cell volumes are also relaxed isotropically, so they attain stress-free conditions. This is done under isobaric conditions with zero hydrostatic stress.

Shear was imposed by continually deforming the simulation cell. The shear strain is applied on the \textit{xy} plane with displacement in the x direction. This was done by imposing a cell tilt factor change, which is effectively an engineering strain. A total simulation time of 335 ps was used, with a maximum final shear strain of $\gamma = 10$. This means the shear rate $d\gamma/dt$ is approximately $2.985\times 10^{10}$ s$^{-1}$. We also performed the same simulation with shear rates 10, 100, and 1000 times lower to investigate the shear rate effect to the maximum extent available with our computational resources.

In addition, we performed a conjugate gradient shearing simulation on an initially undeformed simulation cell. In each step, the simulation cell was tilted to produce 0.01 shear strain, and the atom positions were remapped to the current box state. This was followed by a conjugate gradient energy minimisation. This simulation corresponds to the athermal limit.

To avoid boundary self-interactions, we remap the simulation cell when the tilting is more than 0.5 of the cell length. The cell vector and the positions of atoms are remapped along the shearing direction, with the simulation cell going from a 0.5 to a -0.5 tilt. It is important to note that, due to the periodic boundary conditions, the remapping makes no change to the local atomic environment.

We used OVITO \cite{Stukowski_2010} for analysis. Grains are identified using the grain segmentation modifier. The minimum grain size was chosen as 50 atoms. We discuss the effect of using a different number of atoms as the minimum grain size in Appendix A5. Since this algorithm requires the local crystal orientation of each atom, the polyhedral template matching (PTM) modifier \cite{Larsen_2016} is adopted beforehand. PTM can determine the local crystal orientation of each atom in terms of a quaternion, where each quaternion can be projected into Rodrigues space \cite{DAI2015144}. Then, the local crystal orientation of each atom in this space can be visualised by mapping orientation to an RGB colouring scheme \cite{ALBOU20103022}. Dislocation lines are detected using the dislocation analysis modifier \cite{Stukowski_2012}.

\section{Results}

We first present the results of our simulations exploring the influence of shear strain on initially perfect, single-crystalline iron. Next, we investigate the behaviour of samples under various conditions, including differing temperatures, damping parameters, strain rates, and the presence of carbon impurities.

\subsection{Shear-induced nanocrystalline structure}

\begin{figure*}[t]
\begin{subfigure}{0.3\textwidth}
\caption{}
\includegraphics[width=\linewidth]{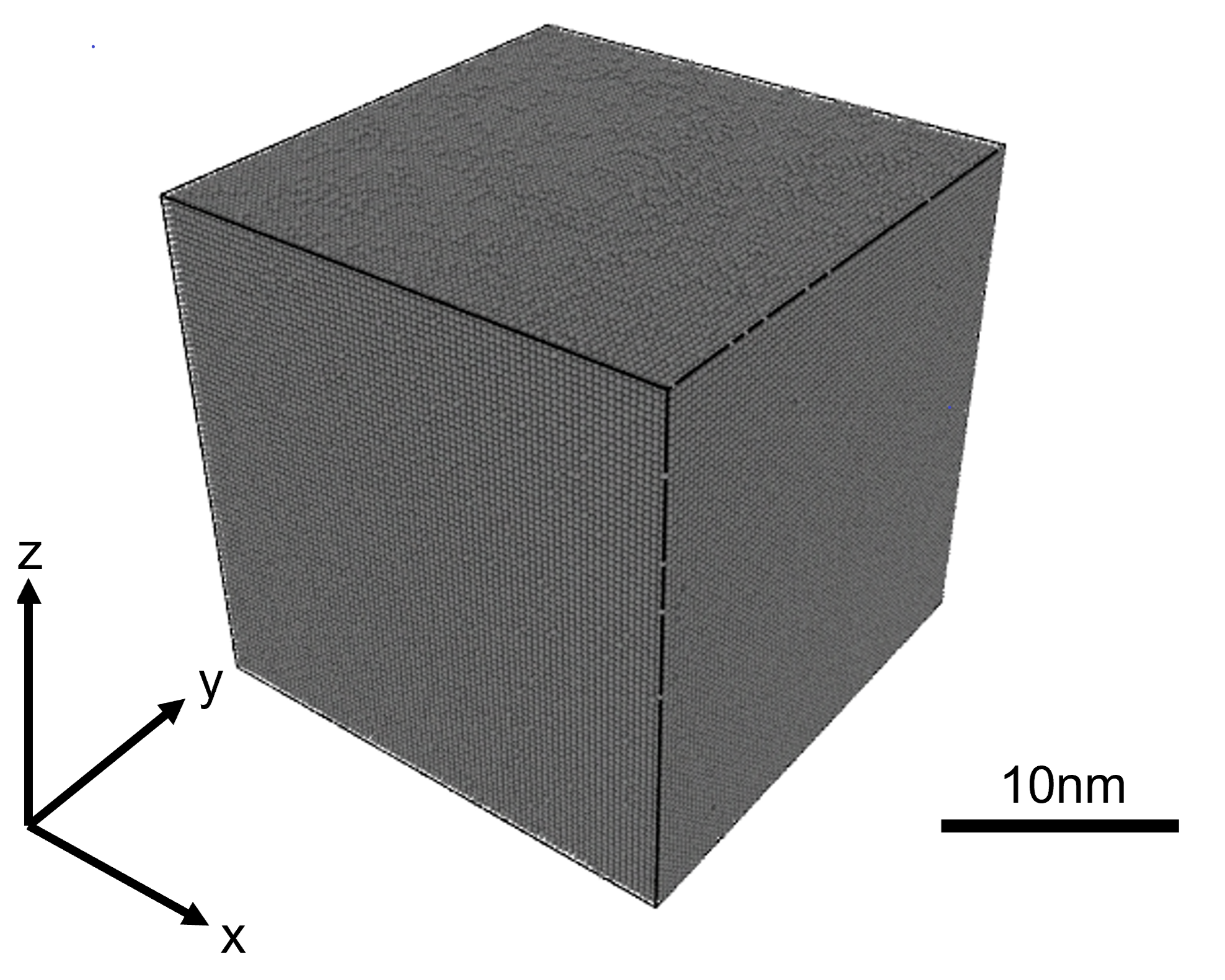}
\label{fig:Fig1a}
\end{subfigure}
\begin{subfigure}{0.3\textwidth}
\caption{}
\includegraphics[width=\linewidth]{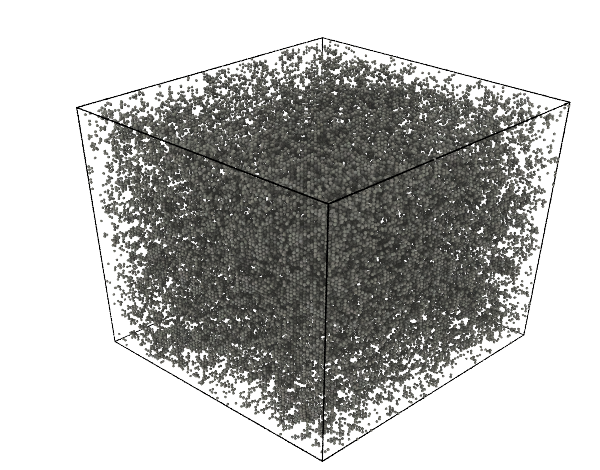}
\label{fig:Fig1b}
\end{subfigure}\
\begin{subfigure}{0.3\textwidth}
\caption{}
\includegraphics[width=\linewidth]{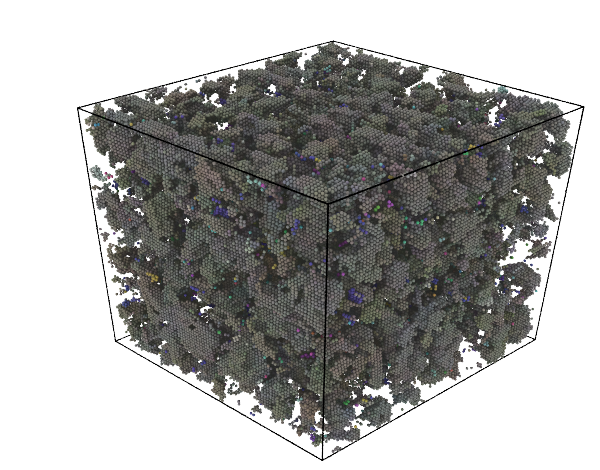}
\label{fig:Fig1c}
\end{subfigure}

\medskip
\begin{subfigure}{0.3\textwidth}
\caption{}
\includegraphics[width=\linewidth]{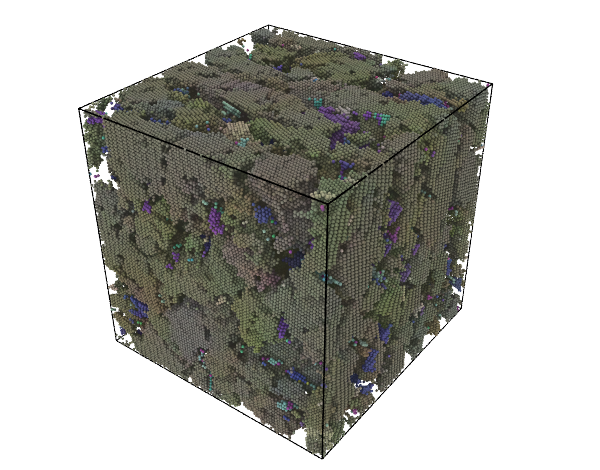}
\label{fig:Fig1d}
\end{subfigure}
\begin{subfigure}{0.3\textwidth}
\caption{}
\includegraphics[width=\linewidth]{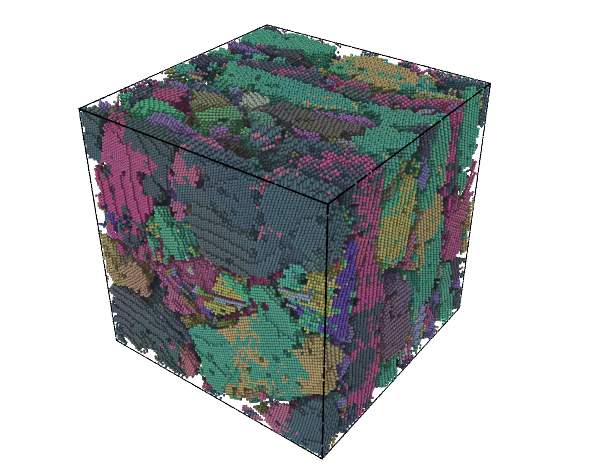}
\label{fig:Fig1e}
\end{subfigure}
\begin{subfigure}{0.3\textwidth}
\caption{}
\includegraphics[width=\linewidth]{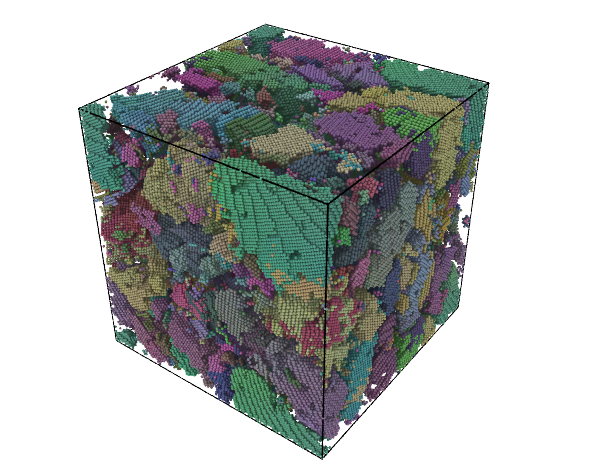}
\label{fig:Fig1f}
\end{subfigure}

\medskip
\begin{subfigure}{0.3\textwidth}
\caption{}
\includegraphics[width=\linewidth]{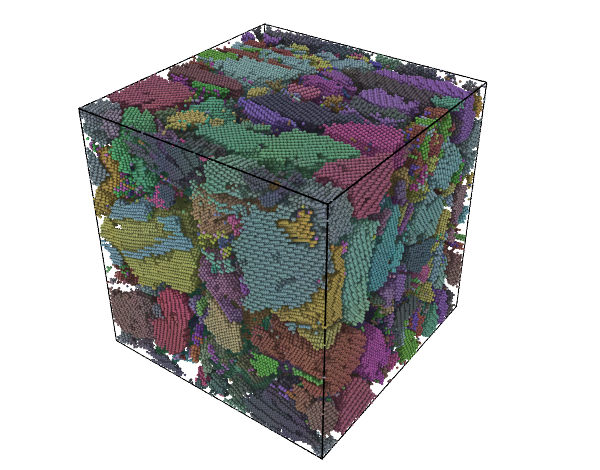}
\label{fig:Fig1g}
\end{subfigure}

\caption{Grain refinement process with atoms coloured by local crystallographic orientations. Atoms not identified as BCC by polyhedral template matching are not shown. Temperature $T=$300 K. Strain rate $d\gamma/dt= 1/33.5$ ps$^{-1}\approx 2.985\times 10^{10}$ s$^{-1}$ and damping parameter $b=6.875$ eV fs {\AA}$^2$. (a) 0 Strain (b) Highly Disordered - 0.27 Strain (c) Reordering - 0.32 Strain (d) 1 Strain (e) 4 Strain (f) 7 Strain (g) 10 Strain.} \label{fig:1}
\end{figure*}

\begin{figure*}[t]
\begin{subfigure}{0.45\textwidth}
\caption{}
\includegraphics[width=\linewidth]{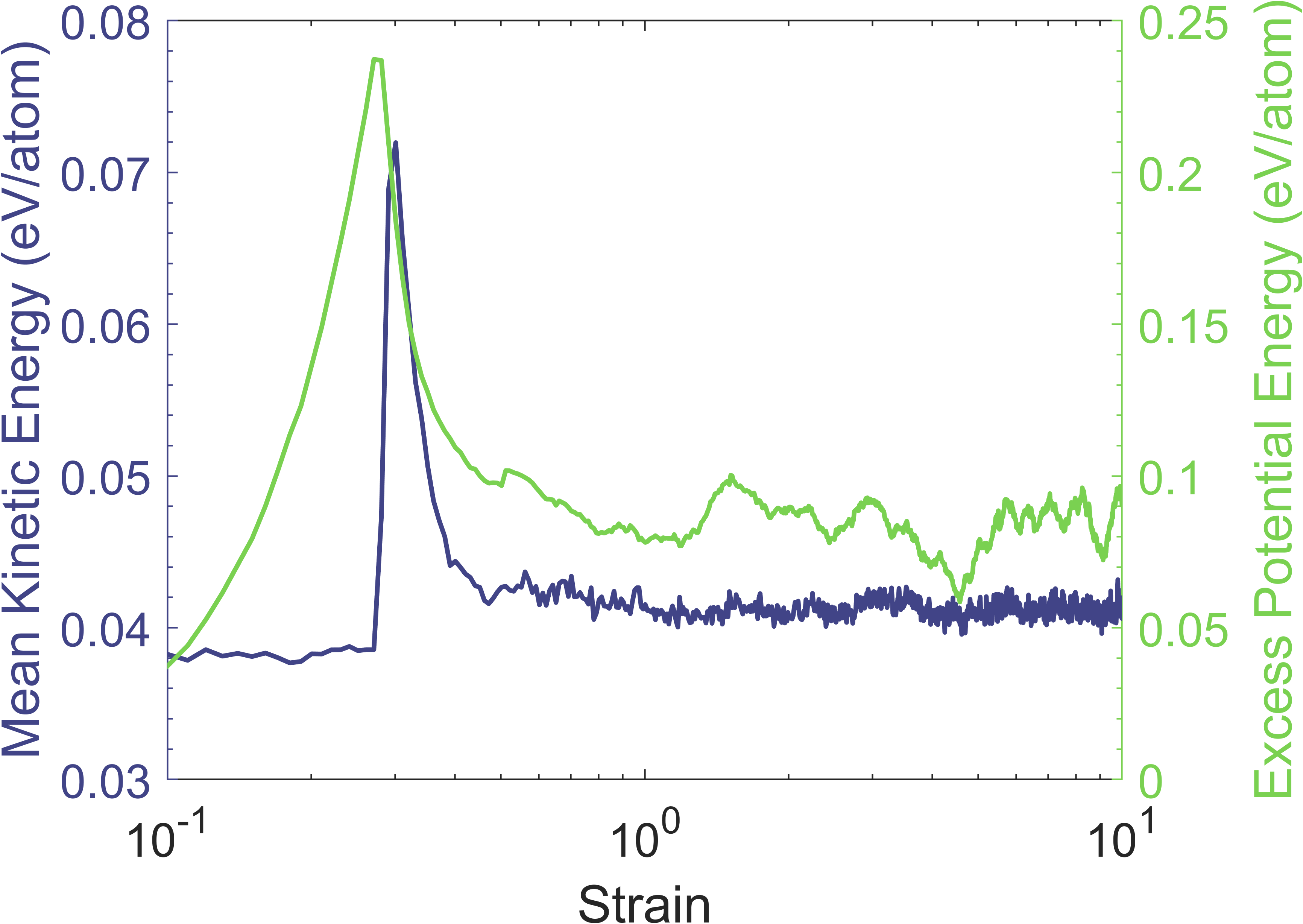}
\label{fig:Fig2a}
\end{subfigure}
\hspace{1em}
\begin{subfigure}{0.45\textwidth}
\caption{}
\includegraphics[width=\linewidth]{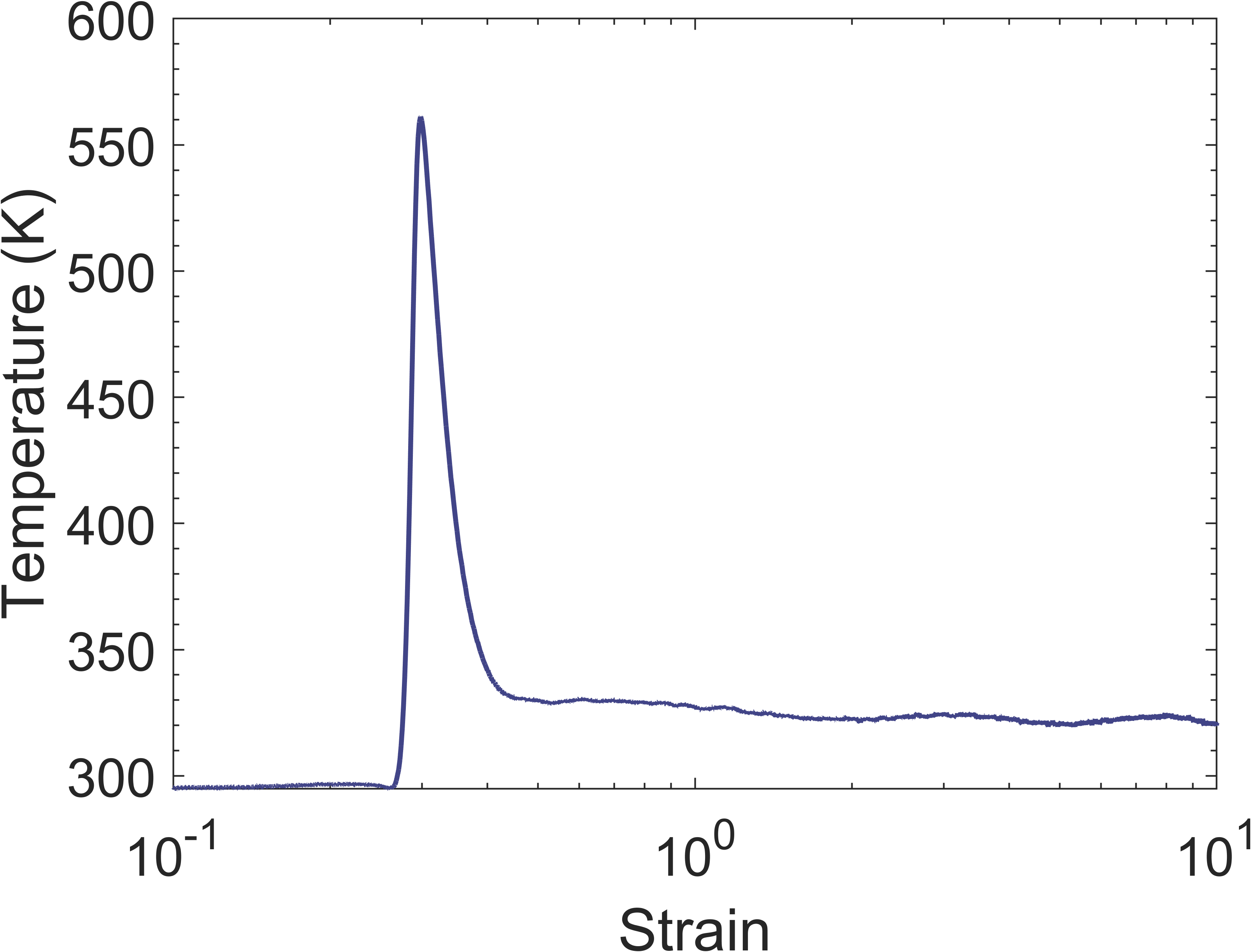}
\label{fig:Fig2b}
\end{subfigure}
\medskip

\begin{subfigure}{0.45\textwidth}
\caption{}
\includegraphics[width=\linewidth]{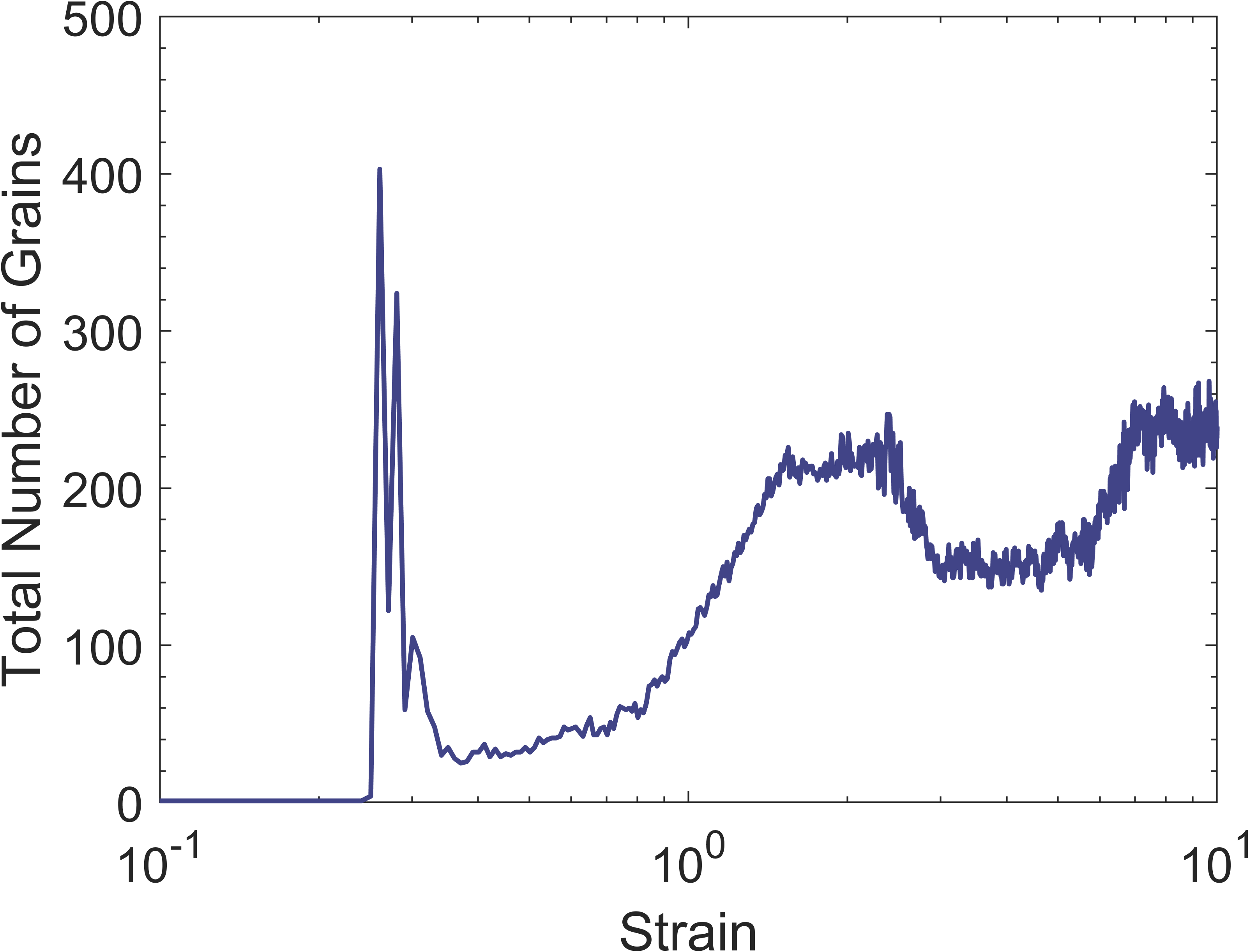}
\label{fig:Fig2c}
\end{subfigure}
\hspace{1em}
\begin{subfigure}{0.45\textwidth}
\caption{}
\includegraphics[width=\linewidth]{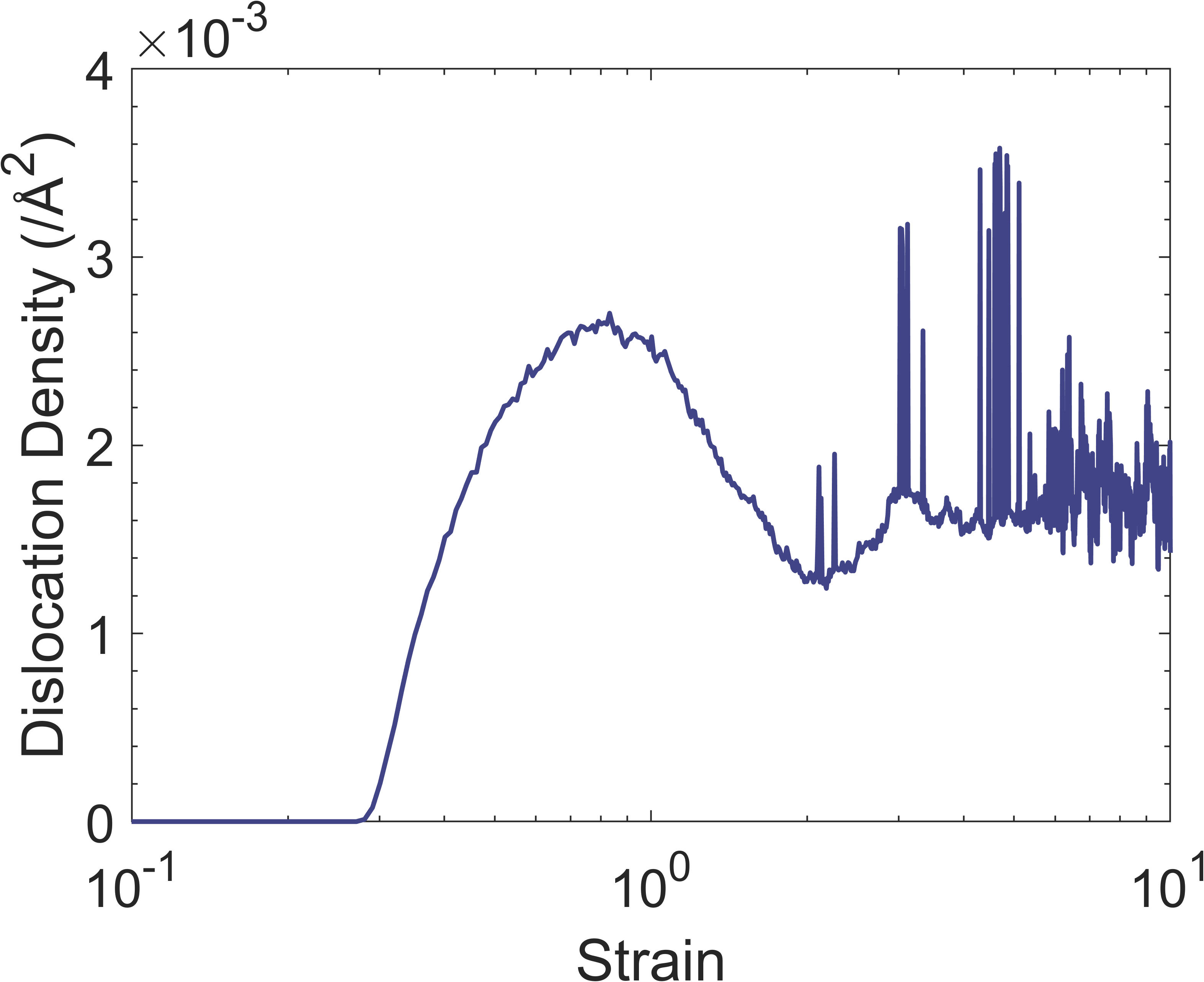}
\label{fig:Fig2d}
\end{subfigure}
\medskip

\begin{subfigure}{0.45\textwidth}
\caption{}
\includegraphics[width=\linewidth]{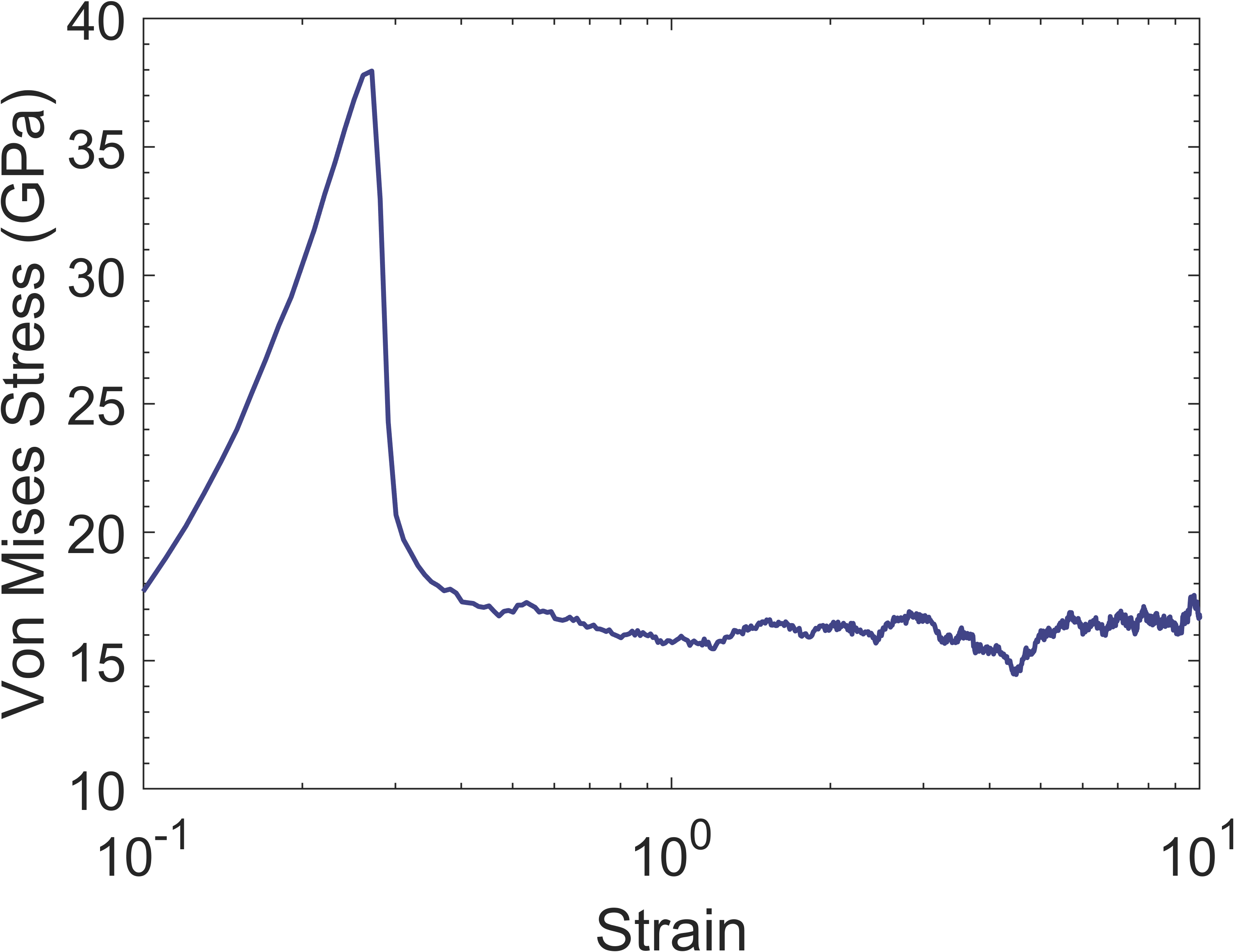}
\label{fig:Fig2e}
\end{subfigure}

\caption{Benchmark simulation data plotted as a function strain. Temperature = 300 K. Strain rate $d\gamma/dt= 1/33.5$ ps$^{-1}\approx 2.985\times 10^{10}$ s$^{-1}$ and damping parameter $b=6.875$ eV fs {\AA}$^2$. (a) Kinetic and Excess Potential Energy (b) Temperature (c) Number of Grains (d) Dislocation Density (e) Von Mises Stress.} \label{fig:2}
\end{figure*}

Figure \ref{fig:1} shows a key result of this work. Figures \ref{fig:Fig1a} to \ref{fig:Fig1g} show the simulation cell at different shear strain. The results reveal a process in which nanograins are formed due to shear strain in a real-space representation. This simulation was run for 335 ps, up to $\gamma=10$, at a constant strain rate, at a temperature of 300 K, and with a constant damping parameter $b_{1}$. We will refer to this as the \textit{benchmark} simulation. The colouring of each atom in Figure \ref{fig:1} is based on its local crystallogrpahic orientation. Only atoms detected as being part of body-centred cubic crystal structures are shown. Since a similar colour means a similar local crystal orientation, one can clearly observe the formation of nanograins in real space.

Figure \ref{fig:Fig1a} shows the initially perfect crystal simulation cell. It is subjected to a shear strain in the \textit{xy} direction, which causes the crystal lattice to deviate from its perfect body-centred cubic structure. At $\sim\gamma = 0.27$, there is a high level of atomic disorder, as shown in Figure \ref{fig:Fig1b}. Most atoms deviate from the perfect body-centred cubic structure. We can consider that atoms are now in a disordered state. Henceforth, when we are referring to a disordered state, we are referring to atoms which the PTM modifier designates as non-BCC.

Shortly after that, a reordering process begins, as shown in Figure \ref{fig:Fig1c}, at $\sim\gamma = 0.3$, where many small grains are visible. Between Figure \ref{fig:Fig1c} and \ref{fig:Fig1d}, grain formation occurs, with the local atomic crystal orientations still largely similar. By Figure \ref{fig:Fig1e}, at $\gamma = 4$, the local atomic crystal orientations are largely dissimilar, with grains exhibiting distinct orientations. It appears that the grain refinement process continues between 4 and 7 strain, see Figure \ref{fig:Fig1f}. Then, the grain number and sizes reach a dynamic quasi-steady state. The stability of the produced nanocrystalline structure, at $\gamma = 10$, is further discussed in Appendix A6.

Grain refinement has been observed for different metals in simulations \cite{GUAN2022111105, NI2011327, GuoFeComp} and experiments \cite{ZHANG201739}. These investigations typically focus on very low strain values. A fundamental difference is that the current simulations start with a perfect crystal structure and continue up to $\gamma = 10$. Our results show that when shear strain is applied, atoms become highly disordered, followed by reordering into a polycrystalline structure. However, this does not always occur under different simulation conditions, as will be discussed below.

For a more detailed analysis, we can consider Figure \ref{fig:2} alongside Figure \ref{fig:1}. Figure \ref{fig:Fig2a} shows the average excess atomic potential and kinetic energy as a function of shear strain. Here, and throughout the subsequent figures, the excess potential energy is given as the absolute value of energy exceeding the unstrained, thermalised state.

When the shear strain is first applied, there is a gradual increase in the excess potential energy up to $\sim0.24$ eV/atom. The initial excess potential energy increase is parabolic, suggesting a linear elastic response, as the elastic potential energy is proportional to $\gamma^2$. The point of maximum excess potential energy occurs at $\gamma = 0.27$, which is the point shown in Figure \ref{fig:Fig1a}. The disordered state is a state of high potential energy. 

Then, the excess potential energy in the system gradually reduces to $\sim 0.09$ eV/atom and remains roughly at this value for the remaining shearing process. At $\gamma = 0.27$, where the excess potential energy reaches its peak, we also observe a rapid increase in the kinetic energy of the system, as shown in Figure \ref{fig:Fig2a}, from a value of 0.039 eV/atom to 0.072 eV/atom, peaking at $\gamma = 0.3$. This increase is followed by a sharp decrease, reaching a value of $\sim 0.042$ eV/atom, at which the simulation plateaus during continued shearing. Since the estimated system temperature is directly proportional to the kinetic energy, we observe a substantial increase in temperature at $\gamma = 0.27$, as shown in Figure \ref{fig:Fig2b}. The maximum temperature reaches 560 K, far below 1,811 K, the melting temperature of iron.

Figure \ref{fig:Fig2c} illustrates the number of grains detected by OVITO. The number of grains undergoes a sharp increase at $\gamma = 0.25$ and peaks at $\gamma = 0.27$ with approximately 400 grains. This corresponds to the highly disordered state. After that, there is a rapid decrease in grain number, reaching a minimum of 25 grains at $\sim\gamma = 0.3$, coinciding with the decrease in excess potential energy. Following this, the number of grains once again rises to $\sim 230$ at $\gamma = 2$. Then, there is a decrease to about 160 at $\gamma = 3$. Another increase in grain count is observed after reaching $\gamma = 6$, plateauing at a value of $\sim 240$ until the end of the simulation. This corresponds to a mean grain volume of $\sim 51\ nm^3$ and a mean grain diameter of $4.6\ nm$ using a spherical grain approximation. 

In Figure \ref{fig:Fig2d}, we also observe an increase in dislocation density starting at $\gamma = 0.27$. The dislocation density is given as the total line length divided by the volume of the simulation cell. Dislocation density rises to $\sim2.6 \times 10^{-3}/$\AA$^2$ at $\gamma = 0.82$, followed by a decrease to $1.2 \times 10^{-3}/$\AA$^2$ at $\sim\gamma = 2$. The dislocation density experiences a slight increase to $1.6 \times 10^{-3}/$\AA$^2$ at $\gamma = 3$ and subsequently remains relatively constant during further shearing.

The average atomic von Mises stress, as shown in Figure \ref{fig:Fig2e}, and the potential energy per atom are well correlated. There is an initial increase in von Mises stress, followed by a rapid decrease when the highly disordered state is formed at $\gamma = 0.27$, the yield point. Nikonov \cite{Nikonov2018} also found an increase in stress during the single crystal shearing of BCC iron. The same rapid increase, followed by a sharp decrease, was observed and attributed to lattice reorientation. It is noteworthy that the excess potential energy never decreases to zero, as some excess energy,  associated with dislocations and grain boundaries generated during deformation, is stored within the material \cite{ALANEME201919}.

As shown in Figure \ref{fig:1}, the atoms become highly disordered, with peak disorder occurring at $\gamma = 0.27$, which corresponds to a maximum in the excess potential energy. This maximum is a threshold value after which the atoms within the lattice move and reorganise themselves, as made evident by the increase in kinetic energy in Figure \ref{fig:Fig2a}, and the reordering of atoms at $\gamma = 0.3$ in Figure \ref{fig:Fig1c}. This point of maximum excess potential energy is also the point of yielding, corresponding to maximum stress in Figure \ref{fig:Fig2e}.

The increasing shear strain causes a breaking of the symmetry in the crystal lattice that changes the potential energy landscape. As the bonds between the atoms are stretched, the excess potential energy of the system increases. As we start with an initially pristine sample, the bonds are allowed to stretch greatly before yielding, and the PTM modifier no longer assigns the structure as body-centred cubic, which is discussed later. Since atoms already have a high excess potential energy, and a high level of atomic disorder, thermal vibrations cause a local breaking of the symmetry, eventually triggering the onset of yielding.

Subsequently, there is a surge in system temperature, as shown in Figure \ref{fig:Fig2b}, which is due to the increase in atomic velocities. This is attributed to yielding and the production of dislocations as can be seen in Figures \ref{fig:Fig2d} and \ref{fig:Fig2e}. As our system is attached to a Langevin thermostat, the system temperature reduces back to the target temperature. Alongside the cool-down, the system undergoes an atomic reordering, causing grain nucleation. The grains are tiny at $\sim\gamma = 0.3$, see Figure \ref{fig:Fig1c}. The microstructure remains dominated by highly disordered regions. Then, grain coarsening starts and continues, as observed in Figure \ref{fig:Fig2c}. This is driven by the excess free energy of the disordered region manifesting as grain boundaries, which are known to possess excess free energy \cite{Rohrer2011GrainBE}. This excess free energy provides the driving force for atomic transport and subsequent grain growth \cite{MullinsGrainGrowth}.

Experimental studies have exhibited similar trends for nanocrystal formation. Studies on Fe-8\% Al \cite{castanFe8Al}, TiAl \cite{WAN201711}, AZ91 Mg \cite{EBRAHIMI20122066}, and NiTi \cite{ZHANG2013124}, all used some form of hot deformation to induce strain into the material, followed by quenching. These studies observed small, recrystallised grains compared to the original structure.

Our decision to start with an initially perfect simulation cell is in accordance with common practice \cite{BulatovPaper} in molecular dynamics simulations. The work in Appendix A1 introduces symmetry breaking defects into the unstrained iron. In agreement with \cite{BulatovPaper}, we find that introducing symmetry breaking defects reduces the yield strength of the material. However, the final steady state is similar to that reached when beginning with an initially perfect simulation cell, and nanograin formation is observed independent of the starting configuration. As such, we only consider initially pristine simulation cells in this section.

Whilst typical metals used for engineering contain pre-existing defects, there is also a body of literature which considers the microstructural evolution of single crystalline, initially defect free, microwhiskers and nanowires of face-centred cubic \cite{C5NR03902A, YueCopperNano, YueCopperNano2, Kiely, Corcoran, LiXiaodong} and body-centred cubic \cite{HUANG2009193, CHEN2014114, WangTungsten, MarichalTungsten, HuangMoly, KIM20102355, MinHan, XIE2013439, MarichalTungsten2} metals. These experimental studies found that, in the absence of grains and initial defects, the yield strength of the metal is far greater, as there are no initial defects, which is apparent for our simulations when considering Figure \ref{fig:Fig2e} and also agrees with the MD results of Zepeda-Ruiz \textit{et al.} \cite{BulatovPaper}.

The mechanical testing of these small-volume BCC materials found that plasticity was dislocation-mediated \cite{HUANG2009193, CHEN2014114, WangTungsten, MarichalTungsten, HuangMoly, KIM20102355, MinHan, XIE2013439, MarichalTungsten2}. At the onset of yielding in our benchmark simulation, at $\gamma = 0.27$, the dislocation density begins to increase and continues to rise until $\gamma = 0.82$. At the same time, the kinetic energy increases. These observed behaviours are consistent with the pristine microwhisker and nanopillar experimental observations where at yield, the production of dislocations abruptly increases and atom reordering occurs, increasing the kinetic energy as the atoms move. 

Shearing of the initially pristine sample causes the atoms to deviate from their lattice positions. Since no initial defects are seeded into the material, the yield strength of the iron is very high, in agreement with the microwhisker experiments. As such, the atomic bonds are allowed to stretch to lengths much greater than usually encountered, corresponding to the large increase in excess potential energy. This causes the template matching modifier to break down, whereby it no longer detects the lattice as having a body-centred cubic structure. This phenomenon describes the disordered state and is the reason for the large increase in grain number at $\gamma = 0.27$, where the small pockets of body-centred cubic structure that the modifier can identify are detected as grains. The analysis in Appendix A5 shows that, when a larger cutoff is used to detect the grains, the initial spike in grain number is absent. As such, the spike in grain number at $\gamma = 0.27$, shown in Figure \ref{fig:Fig2c}, is an artifact of utilising the PTM modifier beyond its limits. We continue to use the lower cutoff of 50 atoms throughout this work as it allows us to more accurately capture the existence of small grains. A large cutoff may not necessarily capture an accurate number of grains. Those grains which are smaller than the cutoff will see their atoms adopted into neighbouring grains as orphan atoms. This may lead to different neighbouring grains being considered as one crystal, reducing the accuracy of microstructure representation.

In pristine material, dislocations are initially formed abruptly from many sources \cite{XIE2013439, WeinbergerDisloc}. By calculating the Schmidt factors \cite{schmid1950plasticity} for our benchmark simulation in Appendix A3, we show that 20 out of the possible 48 slip systems are activated. As dislocations move around the re-orientated cell and begin to pile up, grain boundaries are formed, leading to a reduction in dislocation density and increase in grain number, as shown in \ref{fig:Fig2c} and \ref{fig:Fig2d}.

Figure \ref{fig:Fig14a} in Appendix A1 shows the number of grains in the sheared simulation cells that contain initially seeded defects. Materials with a 15 \AA{}
prismatic dislocation loop and 1000 vacancies do not exhibit the spike in grain number at $\gamma = 0.27$ seen in the benchmark simulation. These defects act as sites for dislocation nucleation and multiplication \cite{TRAMONTINA20149, DENG2010234}. As such, the presence of defects mitigates the creation of the disordered state. The seeded defects reduce the stress required to yield the material, which does not allow the atomic bonds to stretch beyond the strain at which a body-centred cubic structure can be reliably detected using the PTM modifier.

The main limitation of using molecular dynamics simulations to explore the microstructural evolution of iron is the timescale associated with these simulations. Here, we have reported nanocrystal formation and have discussed the possible mechanisms for their creation however, these are not all of the mechanisms, and there will be some further slow evolution of the dislocation and grain structure. Appendix A6 shows that further evolution does exist and that the changes in microstructure decrease with time, but we cannot reach the long timescale limit with these simulations. It is worth noting that the system does not immediately revert to the pristine, single crystalline state, which gives an indication of the stability of the nanocrystals.

In Fig \ref{fig:Fig2d}, the dislocation density in the simulation cell increases and peaks at $\sim\gamma = 0.82$. Dislocations also possess excess free energy \cite{Rohrer2011GrainBE}, which may contribute to grain growth. We note an important point regarding the dislocation analysis modifier: low-angle grain boundaries can be recognised as arrays of dislocations. Therefore, it is unclear if the detected dislocations are grain boundaries or dislocations within grains. There is a new approach that may resolve this situation: Ma \textit{et al.} \cite{Ma_JNM_2023} suggested an algorithm to calculate the shortest distance of any atom from the determined grain boundaries. By eliminating atoms close to the grain boundaries, one can estimate the dislocation line density inside grains. However, we have not adopted this method in our current work, because the definition of grains is somewhat ambiguous in such a highly disordered structure. Additional analysis is provided in Appendix A4.

\begin{figure*}[t]
\begin{subfigure}{1.0\textwidth}
\includegraphics[width=\linewidth]{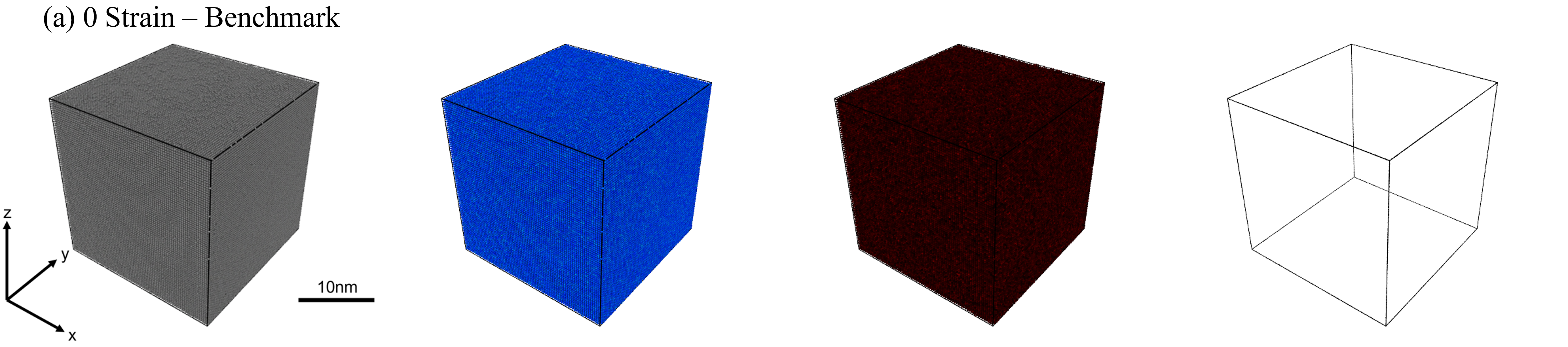}
{\phantomsubcaption\label{fig:Fig3a}}
\end{subfigure}\
\begin{subfigure}{1.0\textwidth}
\includegraphics[width=\linewidth]{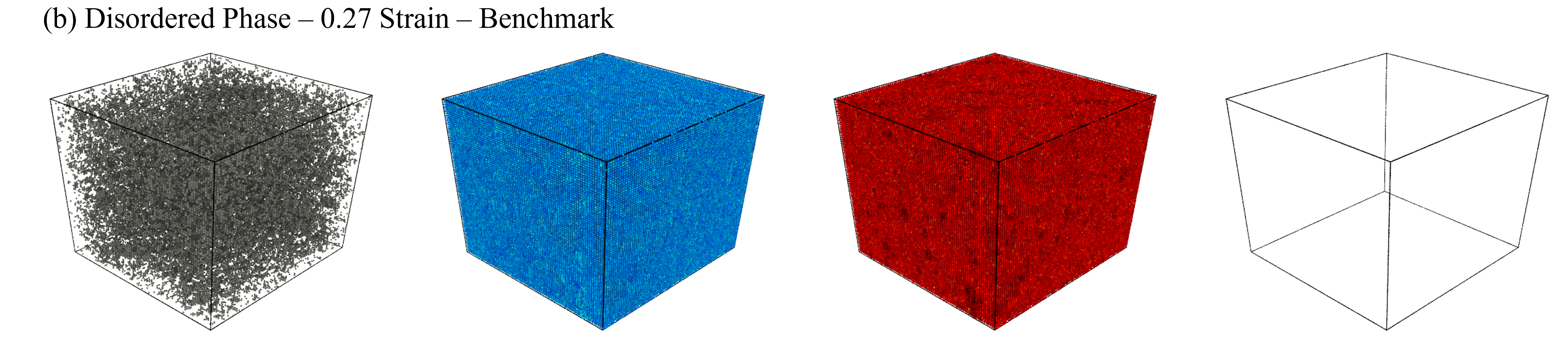}
{\phantomsubcaption\label{fig:Fig3b}}
\end{subfigure}\
\begin{subfigure}{1.0\textwidth}
\includegraphics[width=\linewidth]{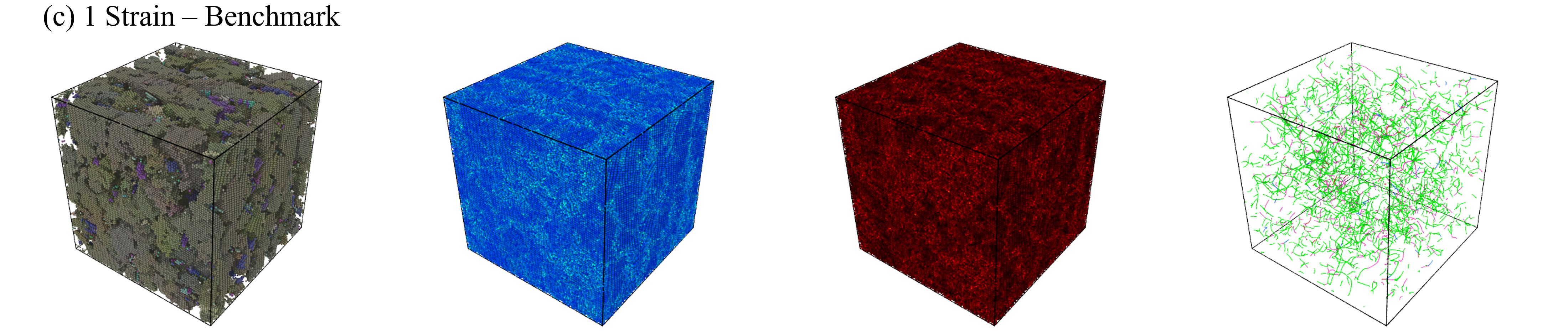}
{\phantomsubcaption\label{fig:Fig3c}}
\end{subfigure}
\begin{subfigure}{1.0\textwidth}
\includegraphics[width=\linewidth]{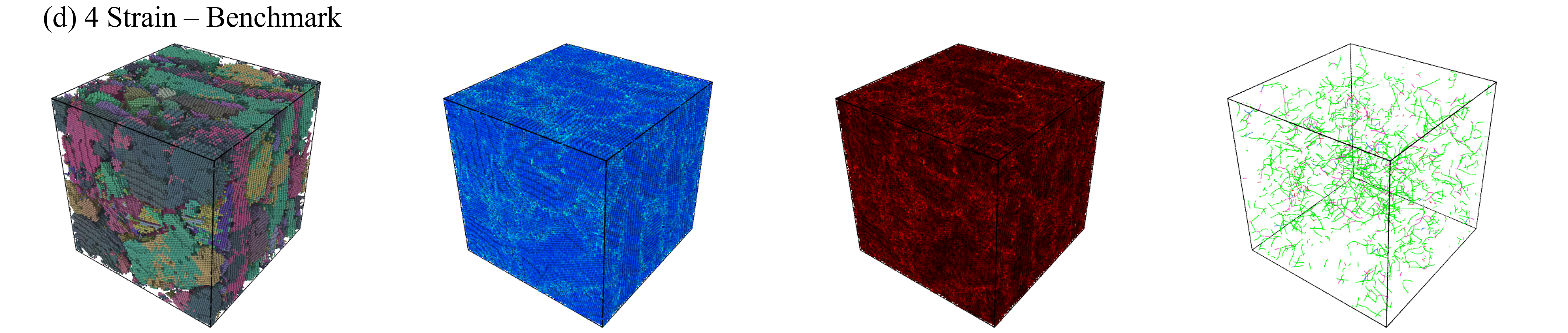}
{\phantomsubcaption\label{fig:Fig3d}}
\end{subfigure}\
\begin{subfigure}{1.0\textwidth}
\includegraphics[width=\linewidth]{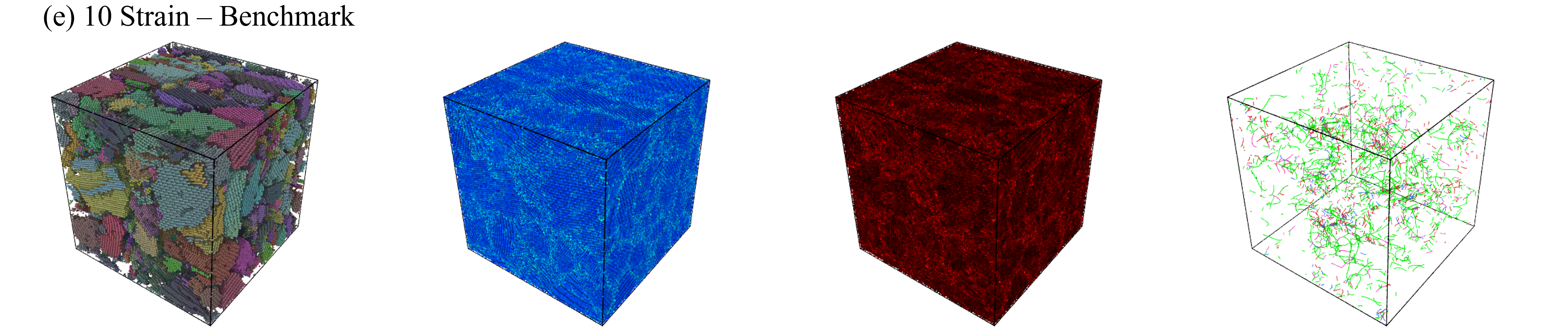}
{\phantomsubcaption\label{fig:Fig3e}}
\end{subfigure}\
\begin{subfigure}{0.8\textwidth}
\includegraphics[width=\linewidth]{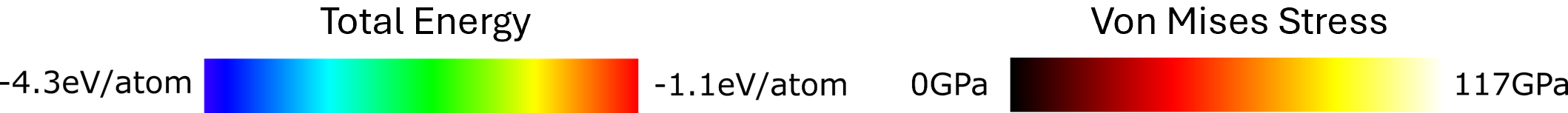}
{\phantomsubcaption\label{fig:Fig3f}}
\end{subfigure}\
\caption{Benchmark simulation visualisations. Strain rate $d\gamma/dt= 1/33.5$ ps$^{-1}\approx 2.985\times 10^{10}$ s$^{-1}$ and damping parameter $b=6.875$ eV fs {\AA}$^2$. Colour key found above. Left - Atoms coloured by local crystal orientations. Atoms not identified as BCC by polyhedral template matching are not shown; Centre-left - Total energy; Centre-right - Von Mises stress; Right - Dislocation analysis; for different levels of strain in benchmark simulation.} 
\label{fig:3}
\end{figure*}

\begin{figure*}[t]
\begin{subfigure}{0.45\textwidth}
\caption{} 
\includegraphics[width=\linewidth]{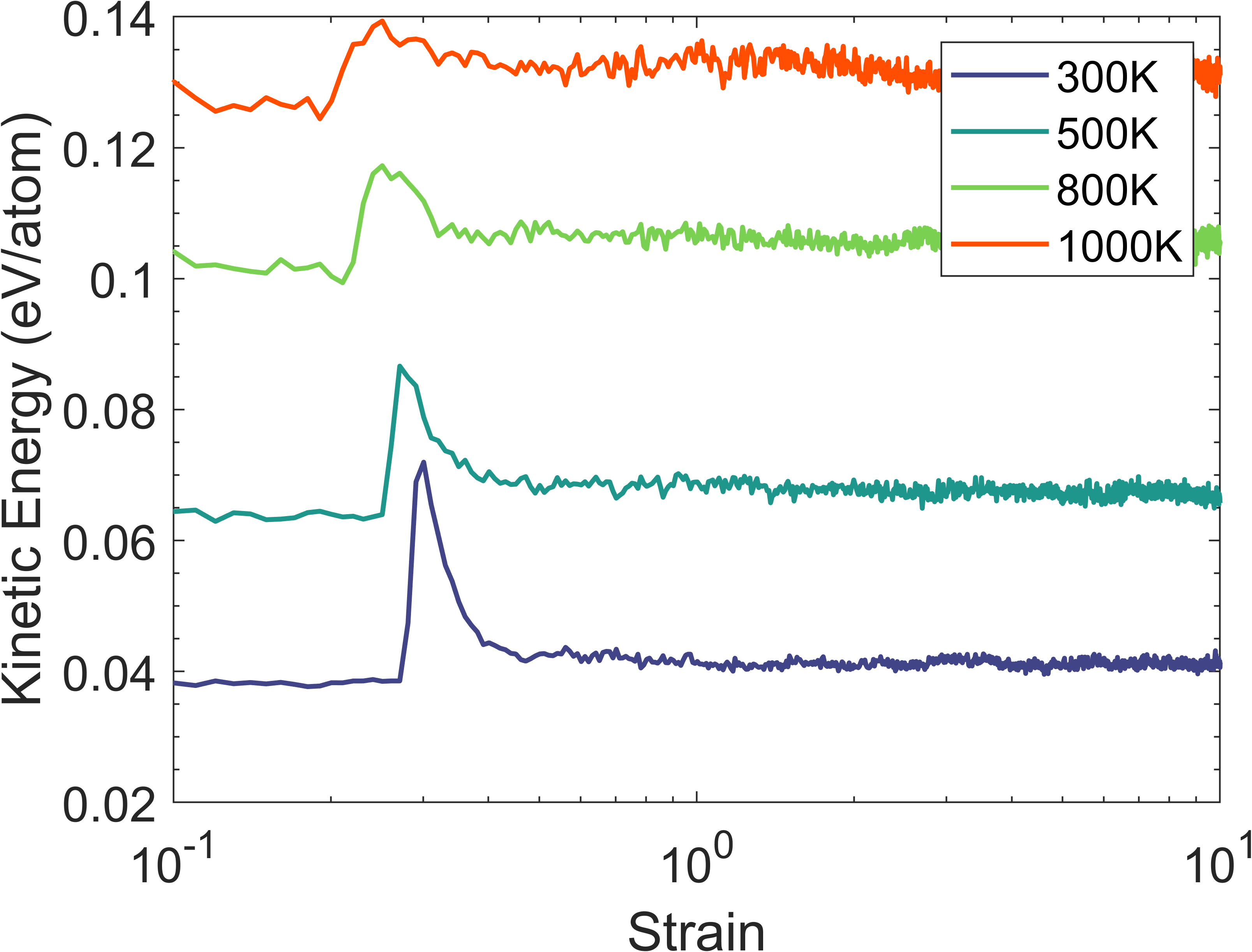}
\label{fig:Fig4a}
\end{subfigure}
\hspace{1em}
\begin{subfigure}{0.45\textwidth}
\caption{} 
\includegraphics[width=\linewidth]{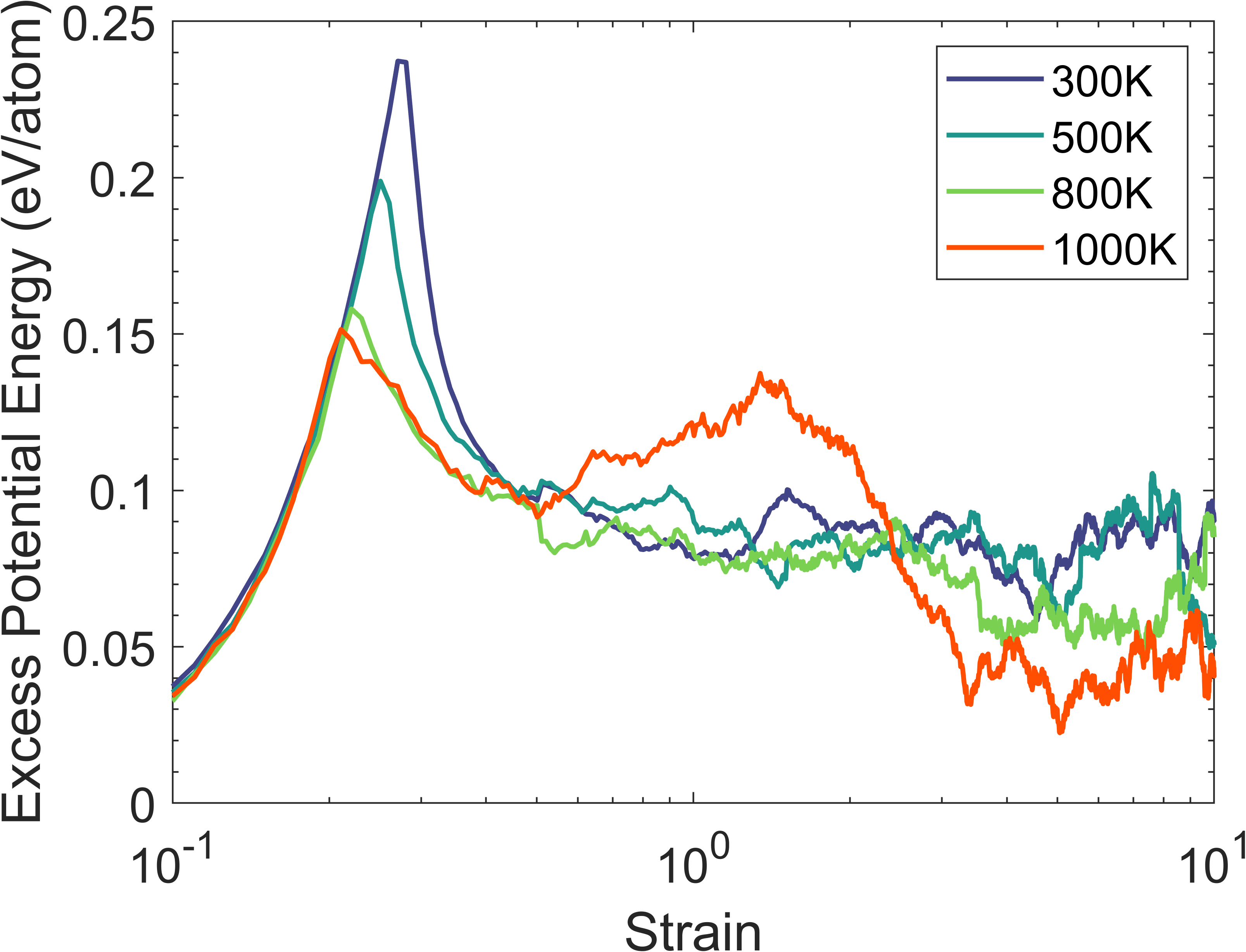}
\label{fig:Fig4b}
\end{subfigure}

\medskip
\begin{subfigure}{0.45\textwidth}
\caption{} 
\includegraphics[width=\linewidth]{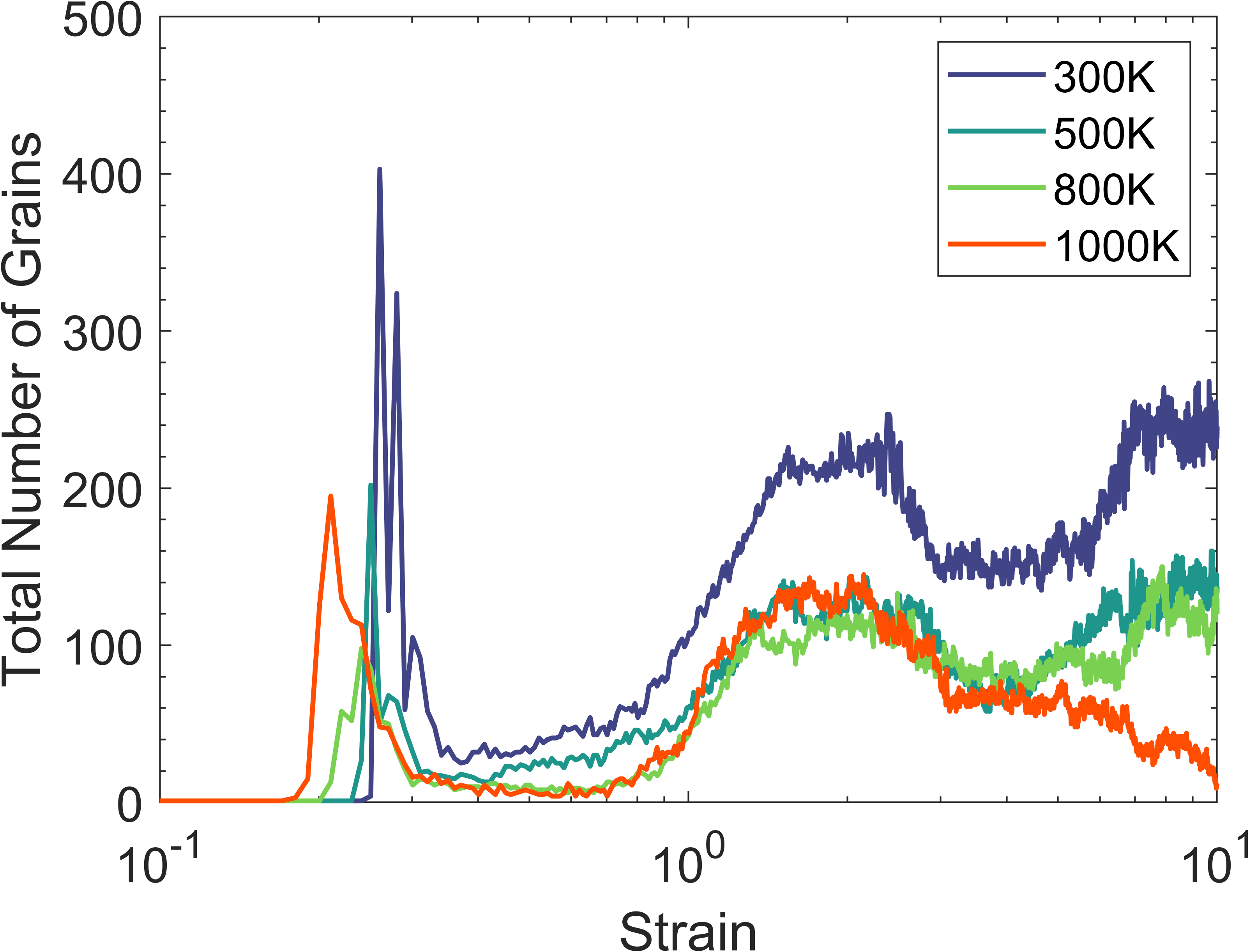}
\label{fig:Fig4c}
\end{subfigure}
\hspace{1em}%
\begin{subfigure}{0.45\textwidth}
\caption{} 
\includegraphics[width=\linewidth]{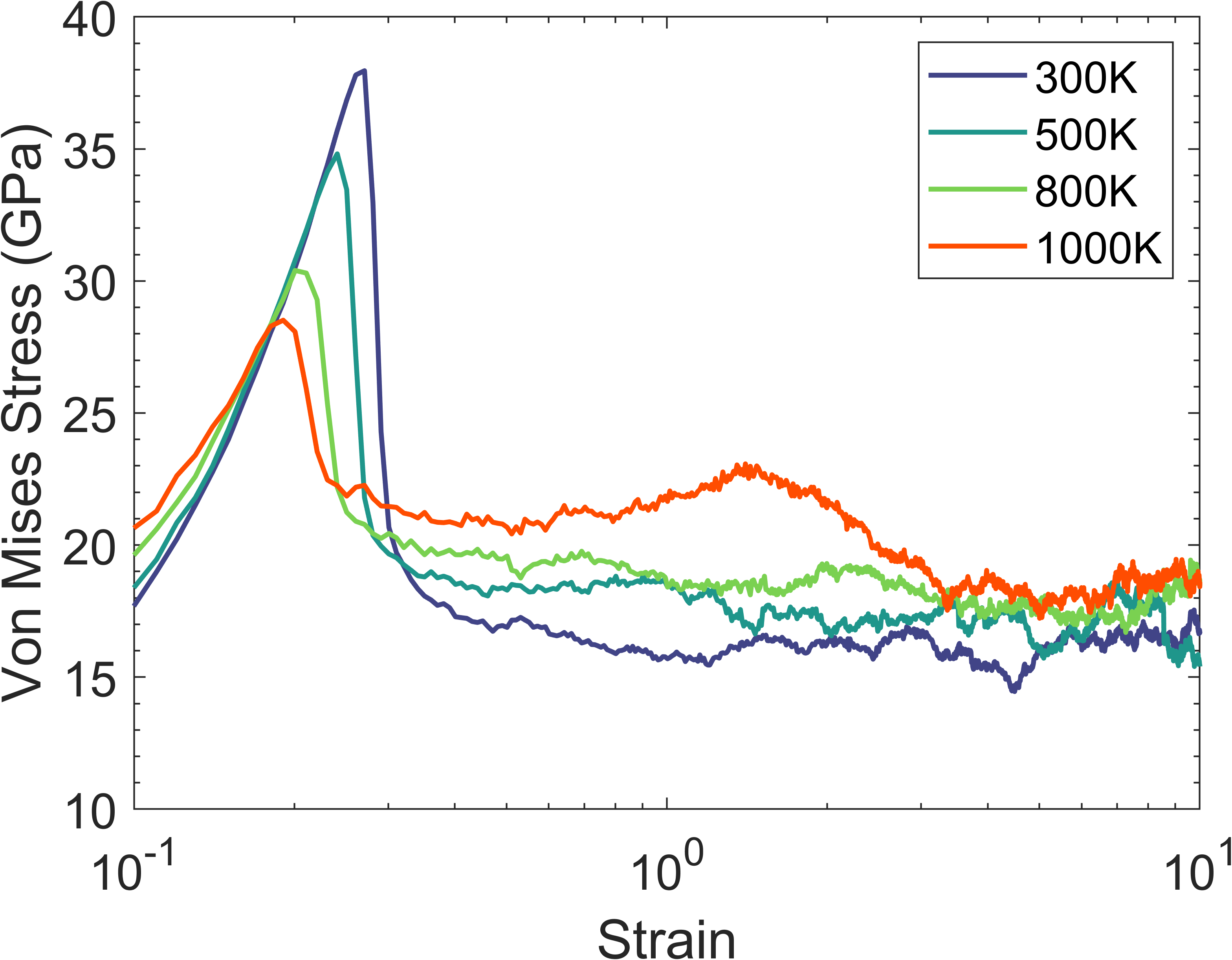}
\label{fig:Fig4d}
\end{subfigure}

\medskip
\begin{subfigure}{0.45\textwidth}
\caption{} 
\includegraphics[width=\linewidth]{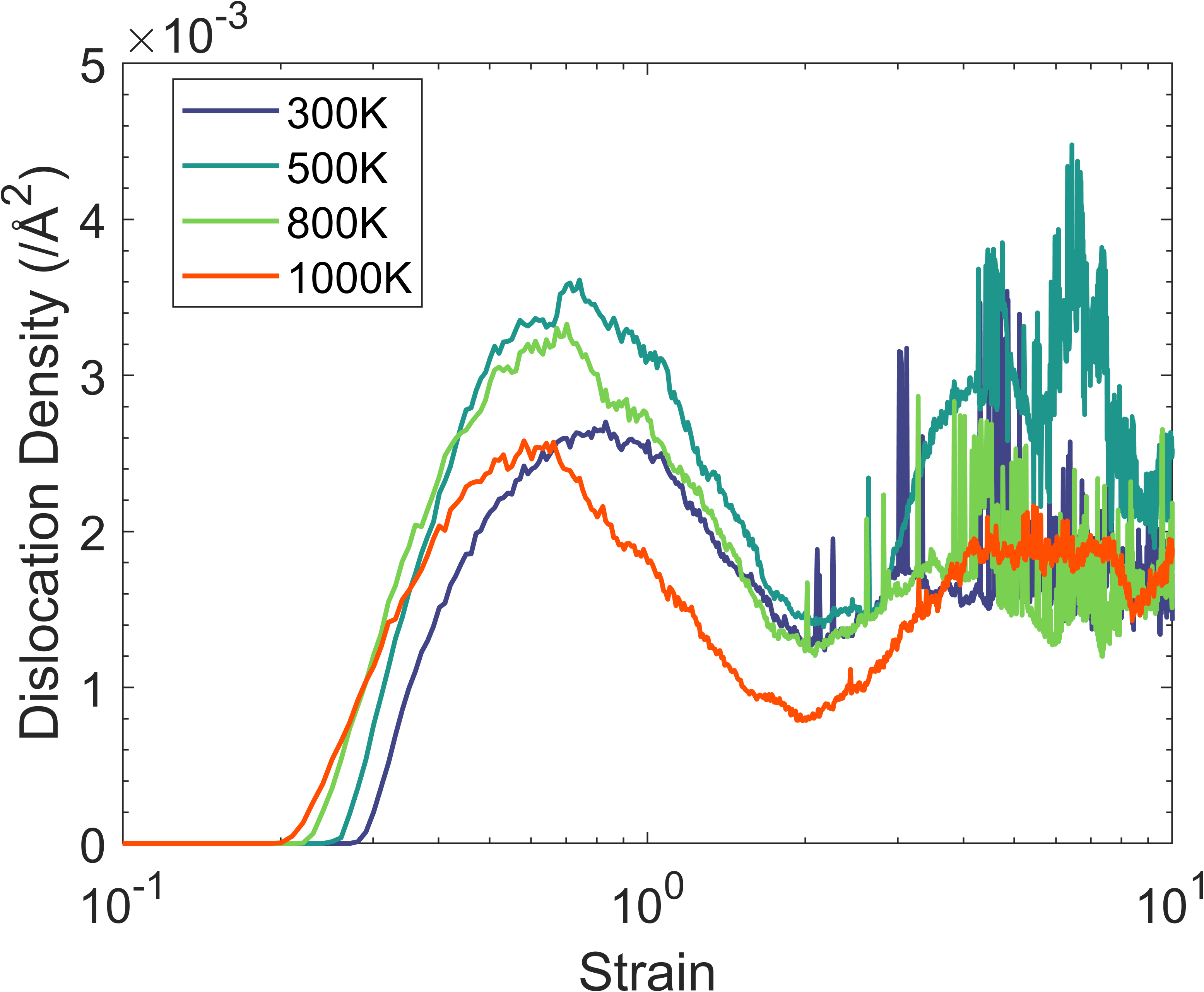}
\label{fig:Fig4e}
\end{subfigure}

\caption{Simulation data plotted as a function of strain at different thermostat temperatures. Strain rate $d\gamma/dt= 1/33.5$ ps$^{-1}\approx 2.985\times 10^{10}$ s$^{-1}$ and damping parameter $b=6.875$ eV fs {\AA}$^2$. Purple line - 300 K, Blue line - 500 K, Green line - 800 K, Orange line - 1000 K. (a) Kinetic Energy (b) Excess Potential Energy (c) Number of Grains (d) Von Mises Stress (e) Dislocation Density.} \label{fig:4}
\end{figure*}

Figure \ref{fig:3} shows the atomic crystal orientation, total energy, von Mises stress, and dislocation lines at different shear strain values. The total energy is the sum of atomic potential and kinetic energy. In the dislocation analysis, green lines represent dislocations with Burgers vector $\mathbf{b}=\frac{1}{2}\langle 111 \rangle$, and pink lines represent $\mathbf{b}=\langle100\rangle$.
Figure \ref{fig:3} shows that most dislocation lines have $\mathbf{b}=\frac{1}{2}\langle 111\rangle$. Only few dislocations have $\mathbf{b}=\langle100\rangle$. This agrees with the experimental findings in the literature on iron, iron-chromium alloys and ferritic/martensitic steels \cite{KLIMENKOV2011124,DETHLOFF2016471,SCHAUBLIN2017427}.

Next, we consider the total energy and von Mises stress, shown in Figure \ref{fig:3}. Initially, the atoms within the box have low energy and low stress. As the atoms become disordered, they experience a high energy and stress state, as shown by the colouring in Fig. \ref{fig:Fig3b}. As reordering occurs, it is evident that some atoms return to the low energy and stress state, whilst others remain in the high energy and stress state (Fig. \ref{fig:Fig3c}). As shear strain is continually applied, we can observe areas of both high and low energy and stress. The areas of low energy and low stress correspond to atoms within the grains and are BCC in nature, whilst the areas of high energy and high stress correspond to atoms which are not BCC in nature, and are not shown in the crystal orientation images. As such, Figure \ref{fig:3} shows that the atoms between the grains have high energy and high stress, which further confirms the presence of grain boundaries in the simulation box \cite{Rohrer2011GrainBE}.

\subsection{Temperature}
To probe the underlying mechanisms for nanocrystal formation from an initially perfect crystal under high-shear strain, certain simulation variables were altered one-by-one. We first changed the thermostat temperature, whilst keeping all other conditions unchanged. Three additional simulations were carried out with thermostat temperatures T = 500 K, 800 K, and 1000 K. Figure \ref{fig:4} shows the kinetic energy, the excess potential energy, the number of grains, the dislocation density and the von Mises stress of the different simulations.

Figure \ref{fig:Fig4a} shows the change in kinetic energy throughout the simulations. We observe a notable difference in the increase in kinetic energy associated with the formation of the disordered phase at different thermostat temperatures. For example, the kinetic energy increases from $\sim 0.039$ eV/atom to 0.072 eV/atom for the 300 K simulation, an increase of 0.033 eV/atom, whilst the 800 K simulation increases from 0.1 eV/atom to $\sim 0.117$ eV/atom, an increase of 0.017 eV/atom, much less than the increment at 300 K. We can also observe in Figure \ref{fig:Fig4a} that the initial increase in kinetic energy occurs earlier for the higher temperature simulations: at $\gamma = 0.27$ for the 300 K, $\gamma = 0.25$ for the 500 K, $\gamma = 0.2$ for the 800 K, and $\gamma = 0.19$ for the 1000 K simulations. After this increase and subsequent reduction, the kinetic energy remains relatively constant for all simulations.

The increase in kinetic energy coincides with a decrease in excess potential energy. Figure \ref{fig:Fig4b} shows how the excess potential energy reaches its peak earlier for the higher temperature simulations. As before, the 300 K simulation shows a peak at 0.24 eV/atom at $\gamma = 0.27$. The 500 K and 800 K simulations exhibit excess potential energy peaks of 0.20 eV/atom at $\gamma = 0.25$, and 0.16 eV/atom at $\gamma = 0.22$, respectively. Finally, the 1000 K simulation has a peak excess potential energy of 0.15 eV/atom at $\gamma = 0.21$. The higher temperature simulations generally show lower excess potential energy after the initial spike, with the 800 K simulation hovering at $\sim 0.07$ eV/atom after $\gamma = 2$, whilst the 300 K simulation plateaus at $\sim 0.08$ eV/atom. 


Inspection of Figures \ref{fig:Fig4c} provides interesting insight into the effect of the thermostat temperature change. All simulations experience an initial spike in grain number, with the spike occurring at a lower strain for higher temperature simulations. Small pockets of body-centred cubic structure are again detected as grains due to the breakdown of the PTM modifier. Interestingly, the magnitude of the increase does not appear to correlate with thermostat temperature. For example, whilst the 300 K simulation increases to $\sim 400$ grains, the 1000 K simulation only increases to $\sim 195$ grains, which is less than the 500 K simulation, but more than the 800 K simulation.

All simulations show a rapid decrease in grain number between $\gamma = 0.19$ and $\gamma = 0.25$, followed by an increase in grain number at $\sim$2 strain. The 300 K simulation notably has the most grains at the local maximum point of $\gamma = 2$. The simulations at other temperatures have fewer grains at this point, with the 500 K, 800 K, and 1000 K simulations having roughly the same number of grains up to $\gamma = 4$. The grain number in the 500 K and 800 K simulations increases and saturates after $\gamma = 7$. The 1000 K simulation experiences a reduction in grain number after $\gamma = 4$, and is as low as 11 grains at $\gamma = 10$. 

\begin{figure*}[b]
\begin{subfigure}{1.0\textwidth}
\includegraphics[width=\linewidth]{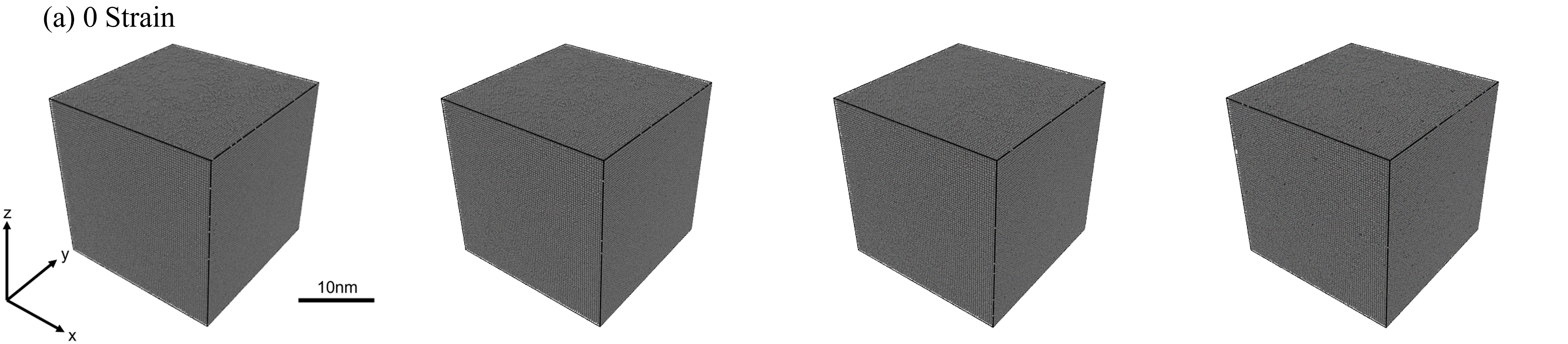}
{\phantomsubcaption\label{fig:Fig5a}}
\end{subfigure}
\begin{subfigure}{1.0\textwidth}
\includegraphics[width=\linewidth]{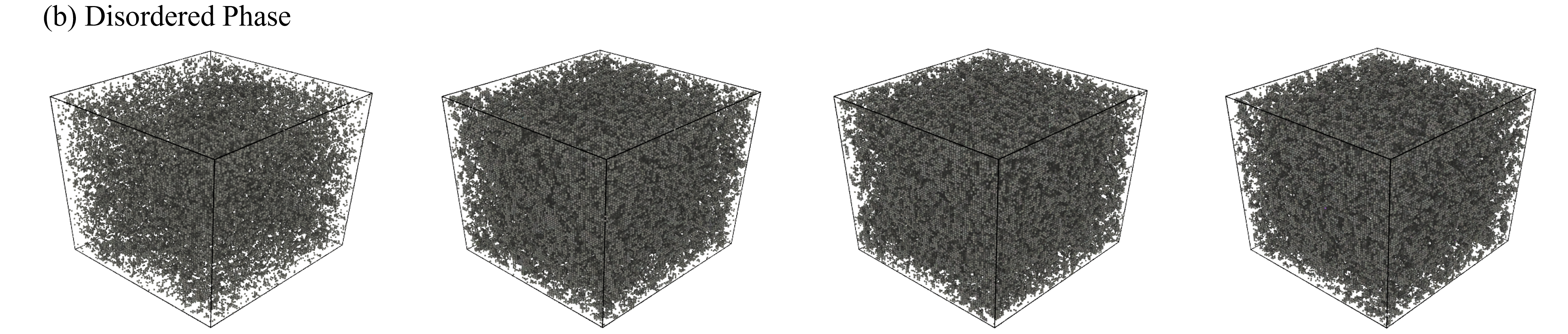}
{\phantomsubcaption\label{fig:Fig5b}}
\end{subfigure}\
\begin{subfigure}{1.0\textwidth}
\includegraphics[width=\linewidth]{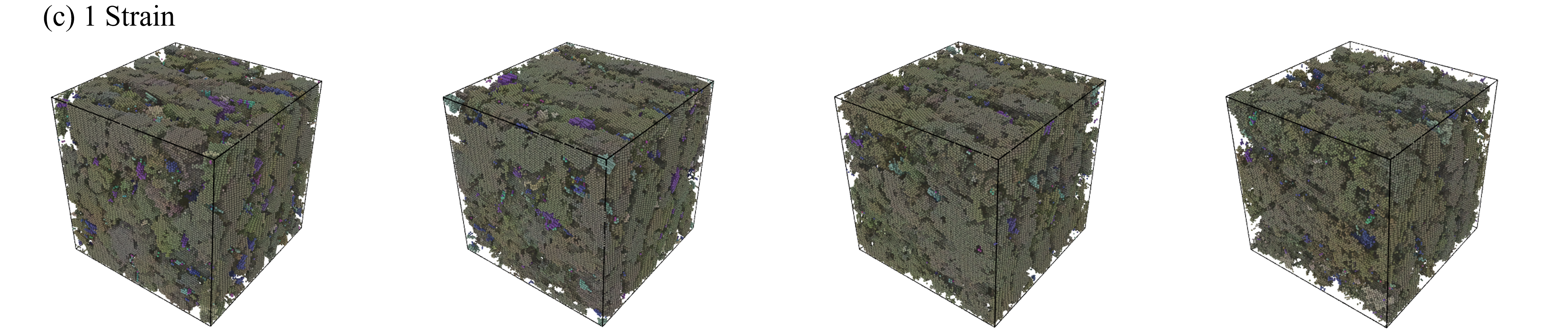}
{\phantomsubcaption\label{fig:Fig5c}}
\end{subfigure}\
\begin{subfigure}{1.0\textwidth}
\includegraphics[width=\linewidth]{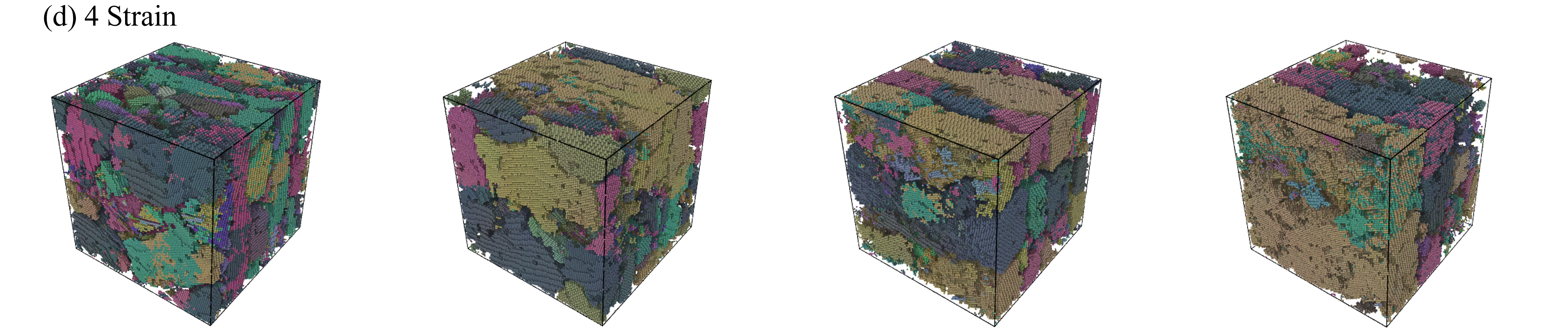}
{\phantomsubcaption\label{fig:Fig5d}}
\end{subfigure}\
\begin{subfigure}{1.0\textwidth}
\includegraphics[width=\linewidth]{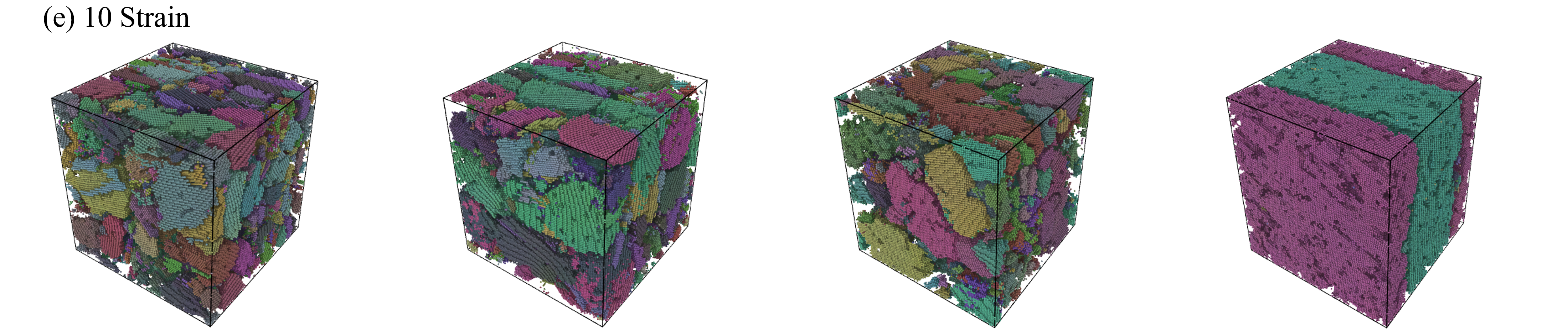}
{\phantomsubcaption\label{fig:Fig5e}}
\end{subfigure}\

\caption{Atoms coloured by local crystal orientation for different temperature simulations. Atoms not identified as BCC by polyhedral template matching are not shown. Strain rate $d\gamma/dt= 1/33.5$ ps$^{-1}\approx 2.985\times 10^{10}$ s$^{-1}$ and damping parameter $b=6.875$ eV fs {\AA}$^2$. Left - T = 300 K simulation; Centre-left - T = 500 K simulation; Centre-right - T = 800 K simulation; Right - T = 1000 K simulation.} \label{fig:5}
\end{figure*}

Figure \ref{fig:Fig4d} shows the von Mises stress. The stress directly correlates with the excess potential energy. This is evident in the figure as the decrease in stress after the initial increase occurs at exactly the same strain value as the excess potential energy decrease (see Figure \ref{fig:Fig4b}). As made evident by Figure \ref{fig:Fig4d}, the higher temperature simulations show a lower maximum von Mises stress and, as the simulation temperature decreases, the maximum von Mises stress increases. The subsequent decrease in von Mises stress is also larger in magnitude for the lower temperature simulations, and the value at which the stress plateaus with shearing is lower for the lower temperature simulations. 

We observe in Figure \ref{fig:Fig4e} that the dislocation densities for the higher temperature simulations follow the same pattern as the benchmark simulation. For each simulation, there is a marked increase in dislocation density between $\gamma = 0.19$ and $\gamma = 0.25$, which corresponds to the point when kinetic energy begins to increase, and the potential energy decreases, showing the onset of yielding. This is followed by a maximum point in dislocation density, at $\gamma = 0.82$ for the 300 K simulation, $\gamma = 0.74$ for the 500 K simulation, $\gamma = 0.7$ for the 800 K simulation, and $\gamma = 0.6$ for the 1000 K simulation. We note that the maximum value of dislocation density does not correlate with temperature. Both the 500 K and 800 K simulations have higher dislocation densities, whilst the 300 K and 1000 K simulations have lower maximum densities. As with the benchmark simulation, the dislocation densities decrease up to $\gamma = 2$, after which there is another increase and saturation in dislocation density for all simulations.

The higher temperature simulations reach the yield point sooner and show a lower peak in excess potential energy in Figure \ref{fig:Fig4b}. Higher temperature simulations inherently have larger potential energy due to the principle of equipartition of energy. At higher temperatures, a system under strain should reach the point of instability, in terms of potential energy, sooner. This, in turn, causes the kinetic energy to increase at a lower strain value, meaning that the higher temperature simulations experience the initiation of yielding at a lower strain, which is expected in metals \cite{Smallman}. This also explains why the spike in grain number occurs earlier for the higher temperature simulations.

Interestingly, in Figures \ref{fig:Fig4c} and \ref{fig:Fig4e}, we observe that the peak in grain number after reorientation (at $\gamma = 0.2$) directly corresponds to a minimum value in dislocation density after the formation of the disordered phase. This finding aligns with previous MD simulations of nanocrystalline Cu \cite{Schiotz12}. The study found that at high dislocation density, a system is highly unstable when nanocrystalline grains are present. With applied strain, the dislocations were able to glide towards and annihilate at grain boundaries.

\begin{figure*}[t]
\begin{subfigure}{0.45\textwidth}
\caption{} 
\includegraphics[width=\linewidth]{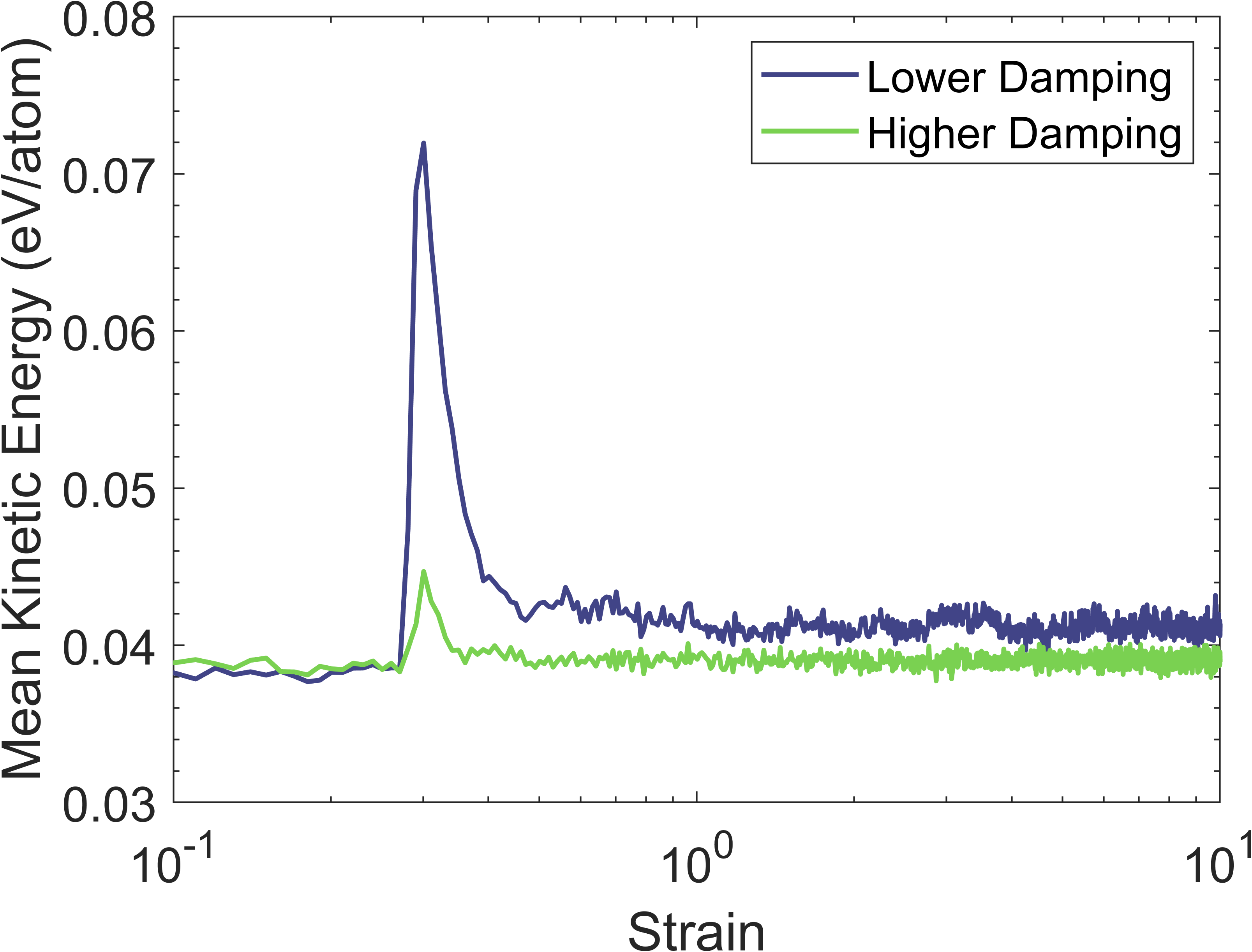}
\label{fig:Fig6a}
\end{subfigure}
\hspace{1em}
\begin{subfigure}{0.45\textwidth}
\caption{} 
\includegraphics[width=\linewidth]{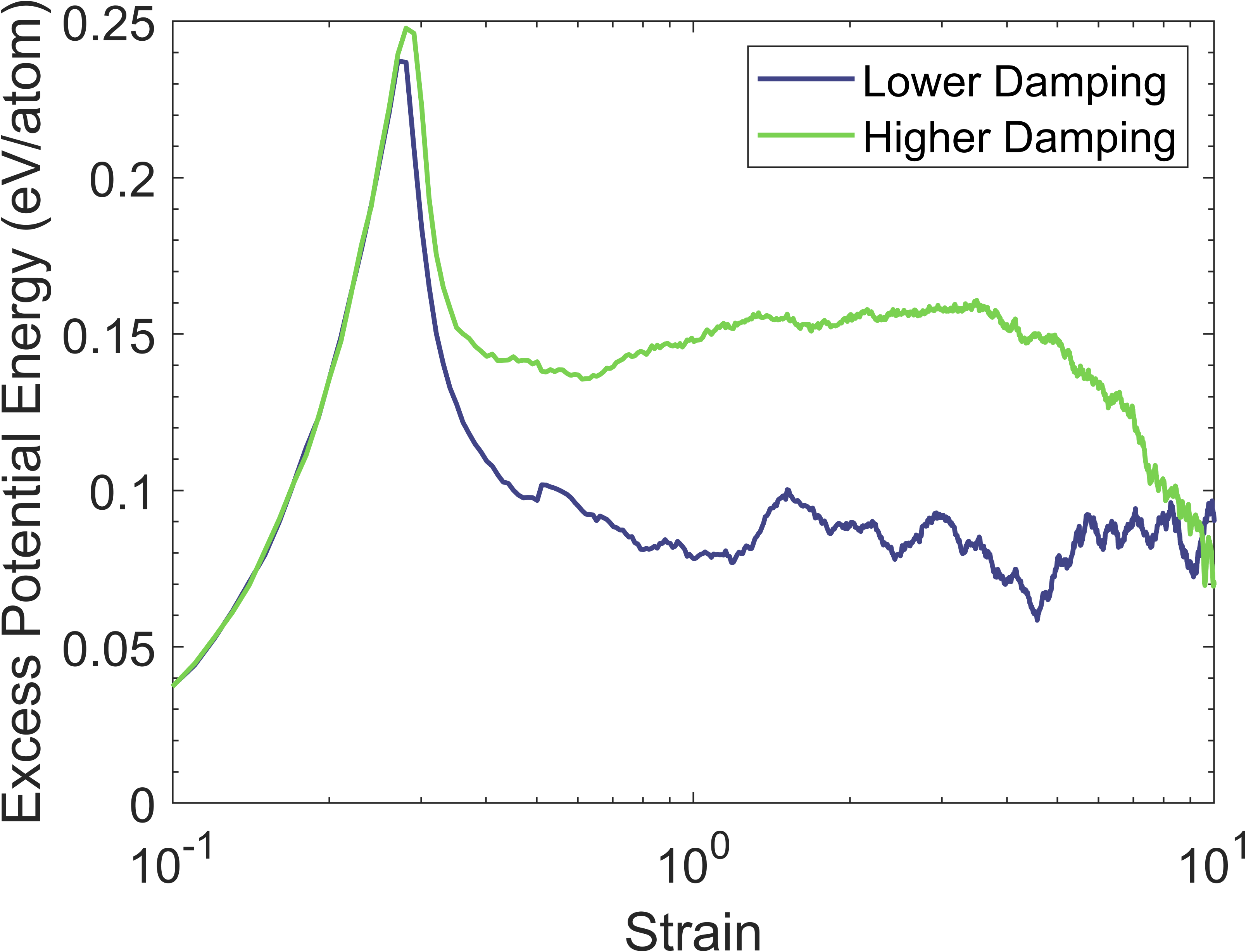}
\label{fig:Fig6b}
\end{subfigure}

\medskip
\begin{subfigure}{0.45\textwidth}
\caption{} 
\includegraphics[width=\linewidth]{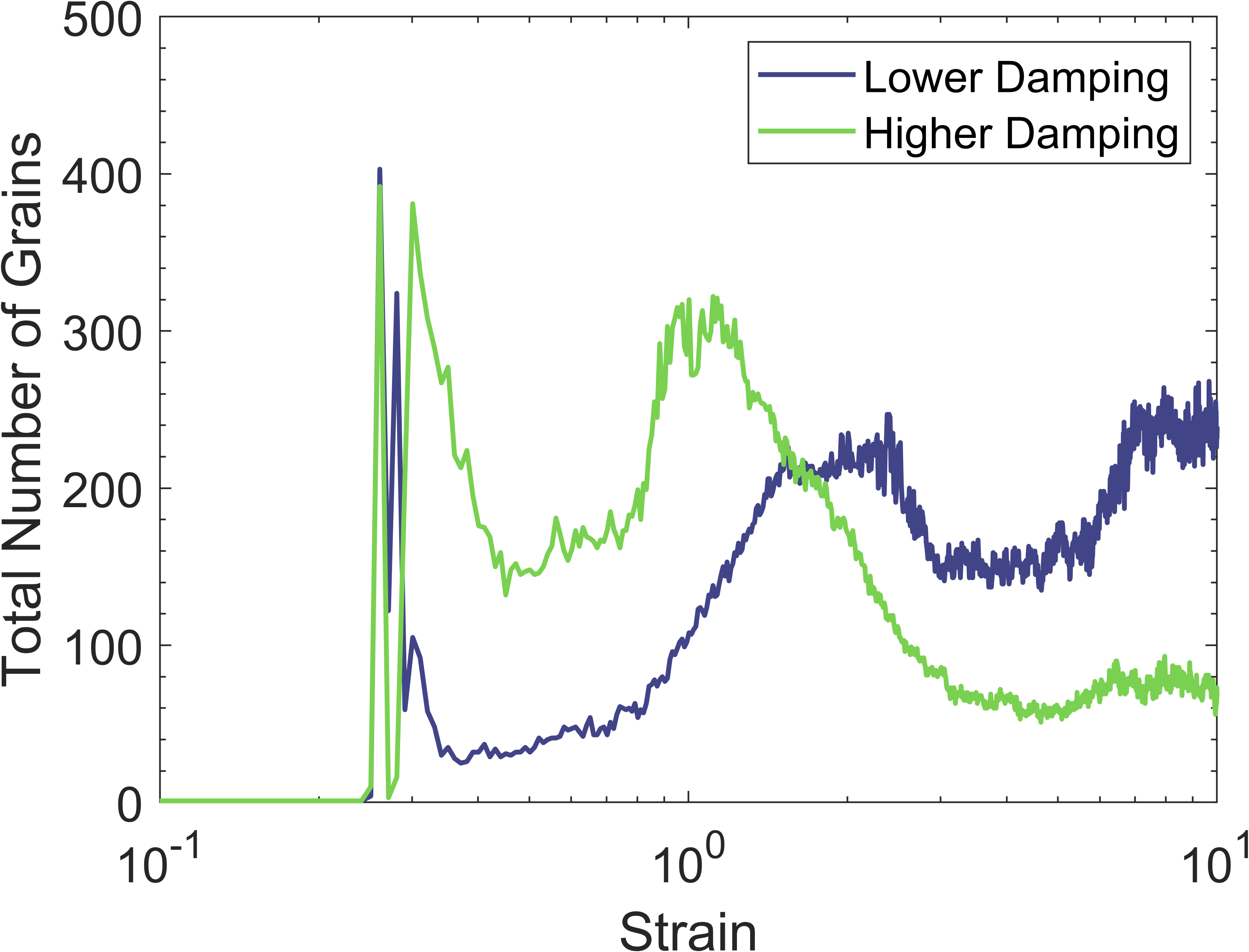}
\label{fig:Fig6c}
\end{subfigure}
\hspace{1em}
\begin{subfigure}{0.45\textwidth}
\caption{} 
\includegraphics[width=\linewidth]{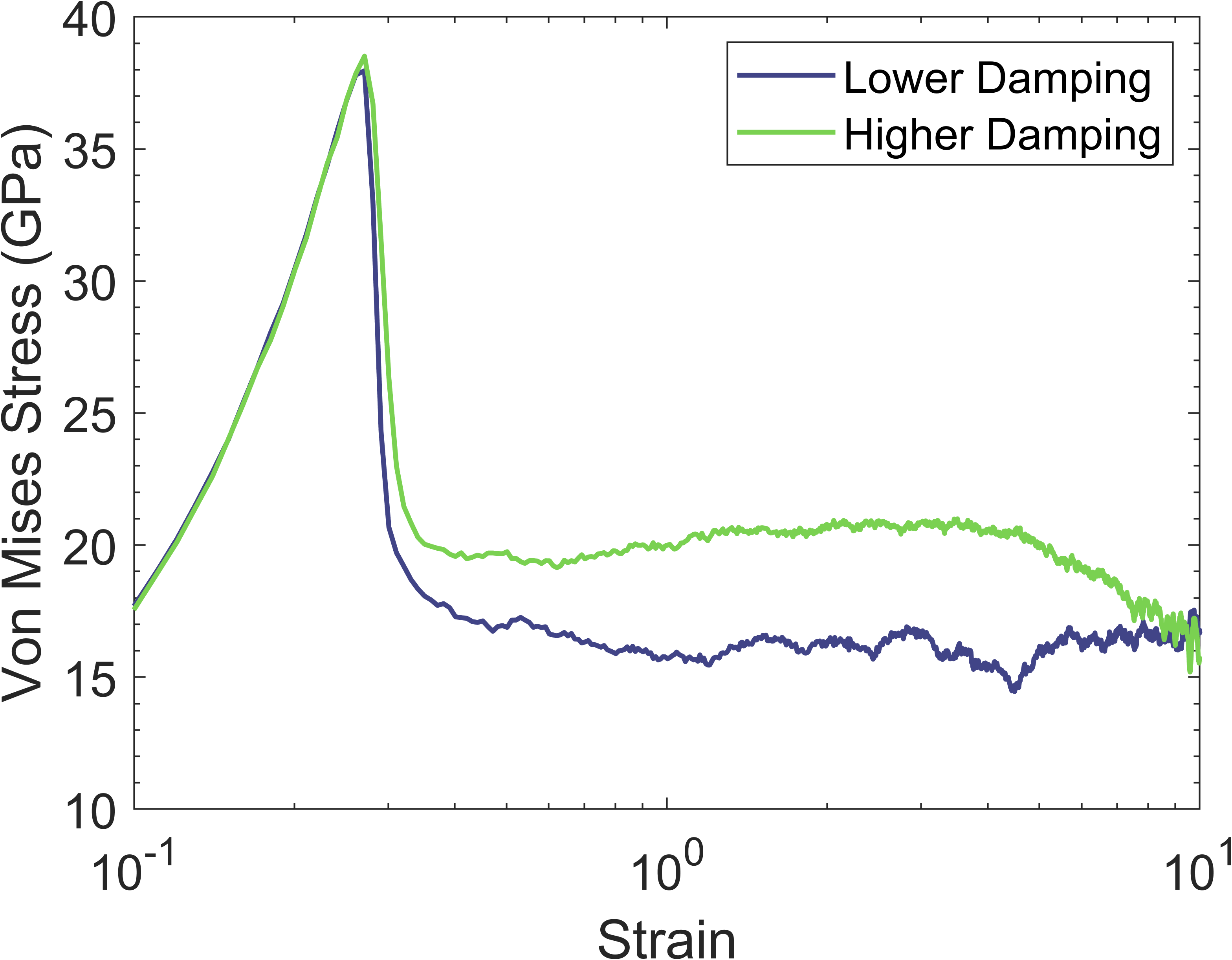}
\label{fig:Fig6d}
\end{subfigure}

\medskip
\begin{subfigure}{0.45\textwidth}
\caption{} 
\includegraphics[width=\linewidth]{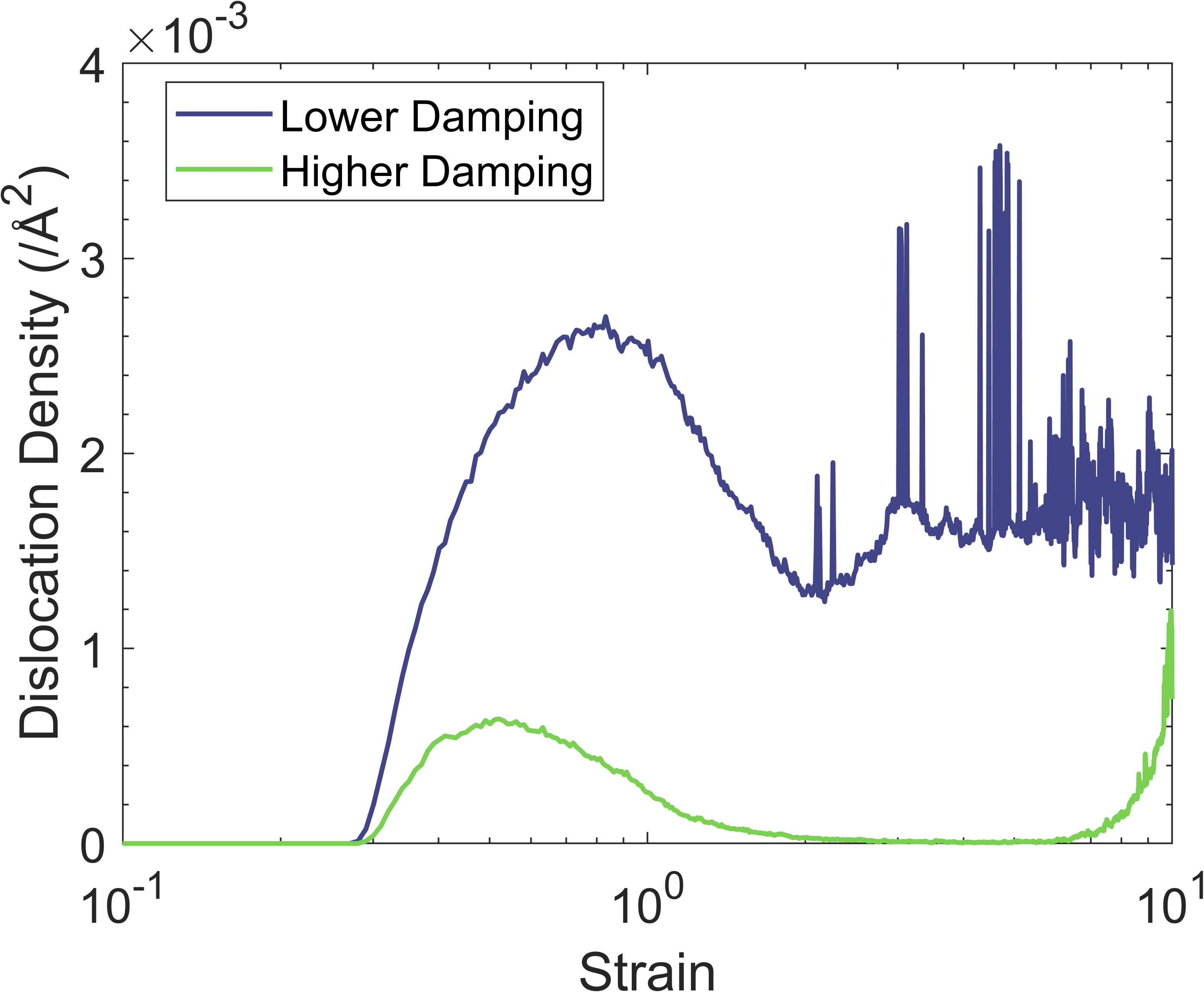}
\label{fig:Fig6e}
\end{subfigure}
\hspace{1em}
\begin{subfigure}{0.45\textwidth}
\caption{} 
\includegraphics[width=\linewidth]{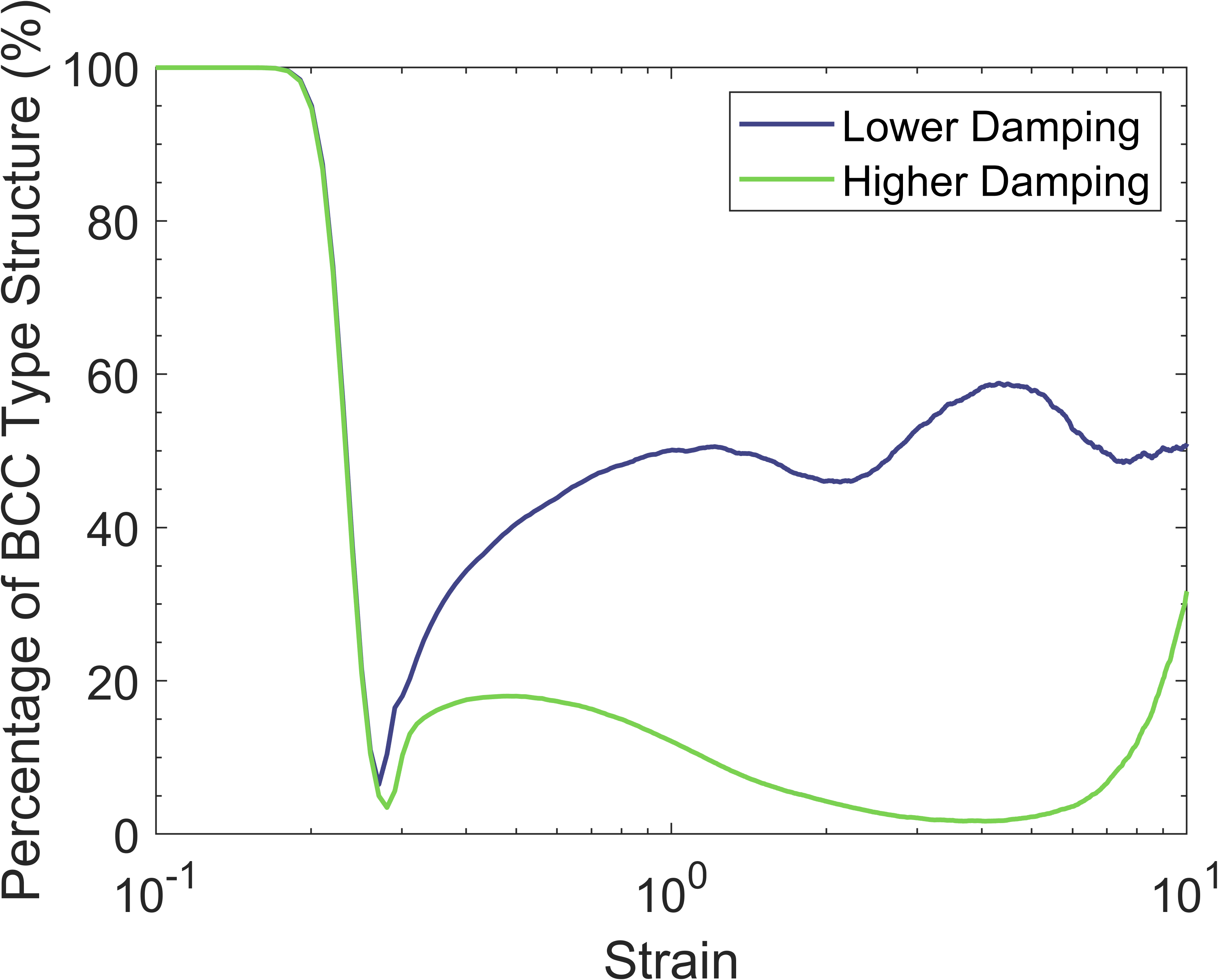}
\label{fig:Fig6f}
\end{subfigure}

\caption{Simulation data plotted as a function of strain using different damping parameters. Strain rate $d\gamma/dt= 1/33.5$ ps$^{-1}\approx 2.985\times 10^{10}$ s$^{-1}$. Damping parameter $b=6.875$ eV fs {\AA}$^2$ for purple line and damping parameter $b=68.75$ eV fs {\AA}$^2$ for green line. (a) Kinetic Energy  (b) Excess Potential Energy (c) Number of Grains  (d) Von Mises Stress (e) Dislocation Density  (f) Percentage of BCC Structure.} \label{fig:6}
\end{figure*}

Figure \ref{fig:5} shows atoms coloured according to their local crystallographic orientation, and only atoms in the body-centred cubic phase are visualized. The simulations, from left to right are at 300 K, 500 K, 800 K, and 1000 K, respectively. All simulations begin with a perfect box, thermalised to the desired temperature (Figure \ref{fig:Fig5a}). With applied shear strain, the cells all enter the highly disordered state (Figure \ref{fig:Fig5b}). The level of disorder appears to be larger for the lower temperature simulations. This is visible in that the number of body-centred cubic atoms is less for the 300 K simulation compared to, for example, the 1000 K simulation. This can again be explained by the stretching of atomic bonds. The higher temperature simulations inherently have a higher potential energy due to the equipartition of energy. Therefore, the bonds are not as stretched as for the higher temperature simulations at the yielding point, denoted by the lower excess potential energies in Figure \ref{fig:Fig4b} and lower yield stress in Figure \ref{fig:Fig4d}. As such, the PTM modifier can recognise more atoms as having a body-centred cubic structure at the onset of yielding.

Each simulation experiences a subsequent reorientation of atoms, followed by a grain growth up to $\gamma = 1$ (Figure \ref{fig:Fig5c}). At this point, by inspecting the colouring of atoms, the majority of atoms have similar local crystal orientation to the initially perfect crystal lattice, meaning that the misorientation between grains is small. As the shearing continues, it is evident that grains develop different orientations, as shown in Figure \ref{fig:Fig5d} and \ref{fig:Fig5e}. It is also evident that the grains are larger and longer for the higher temperature simulations compared to the lower temperature simulations. This is observed by comparing the 300 K and 1000 K simulations in Figure \ref{fig:Fig5e}, where the grain number and size are very different. This is also consistent with the lower grain numbers for the higher temperature simulations compared to the 300 K benchmark, seen at $\gamma = 10$ in Figure \ref{fig:Fig4c}.

As $\alpha$-iron is a BCC metal, it has a high stacking fault energy \cite{HUMPHREYS2004415} meaning that dislocation motion and cross-slip can readily occur. This means that dynamic recovery is the governing dynamic restoration mechanism in $\alpha$-iron and other ferritic metals \cite{WANG19941193, HUMPHREYS2004415} as the dislocations can easily move. Typically, dynamic recovery readily occurs in ferritic steels and iron at temperatures $>0.4T_{m}$ \cite{MCQUEEN20012375}. The data presented in Figure \ref{fig:Fig4e} suggests that this can also occur at room temperature, as this figure shows a marked annihilation of dislocations. The fundamental mechanisms of dynamic recovery are dislocation glide, climb, and cross-slip \cite{HUMPHREYS2004415}. We speculate that the accumulation of dislocations also contributes to nanocrystal formation, as dislocations organise into low-angle grain boundaries. In general, the dislocation density stays fairly constant after $\sim\gamma = 4$ for all simulations, whilst the grains continue to elongate, which is especially prominent for higher temperature simulations. This behaviour is expected for a metallic material under constant strain deformation \cite{HUMPHREYS2004415, Sellars1986167}. 

\subsection{Heat dissipation}

The heat dissipation from the lattice subsystem to the environment is represented by the Langevin thermostat. In metals, the dominant heat transfer mechanism is through electrons \cite{ZimanElectronPhonon}. Therefore, we use the damping parameter $b$, which governs the speed of heat dissipation, according to the phonon-electron coupling, as discussed in the methods section. The damping parameter has, so far, been kept at $b=6.875$ eV fs {\AA}$^{-2} = b_{1}$, which is derived from experimental values for iron \cite{Mason_JPCM_2015, AllenEph}. To investigate the effect of the rate of heat dissipation on nanocrystal formation, a further simulation was carried out with a damping parameter of $b=68.75$ eV fs {\AA}$^{-2} = b_{2}$. All other conditions were kept the same. The Langevin thermostat was set to 300 K. 

\begin{figure*}[t]
\begin{subfigure}{0.45\textwidth}
\caption{0 Strain} 
\includegraphics[width=\linewidth]{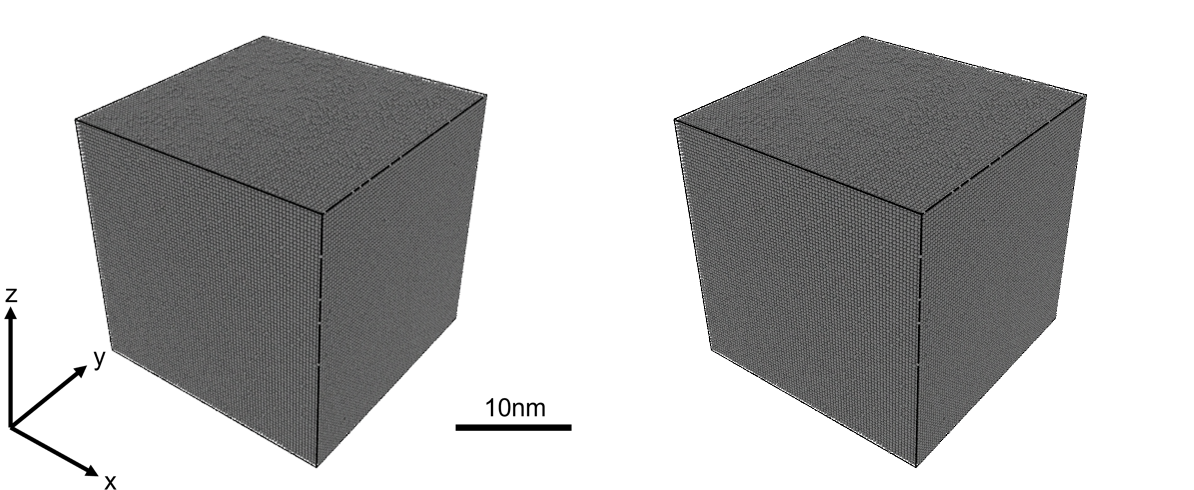}
\label{fig:Fig7a}
\end{subfigure}
\begin{subfigure}{0.45\textwidth}
\caption{Disordered Phase - 0.27 Strain} 
\includegraphics[width=\linewidth]{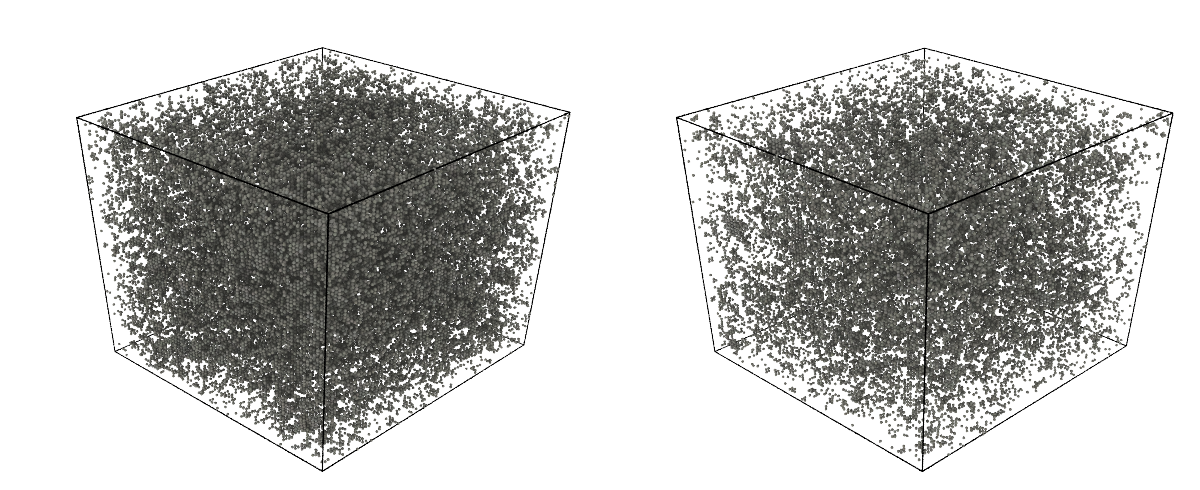}
\label{fig:Fig7b}
\end{subfigure}\

\medskip
\begin{subfigure}{0.45\textwidth}
\caption{Reordering - 0.32 Strain} 
\includegraphics[width=\linewidth]{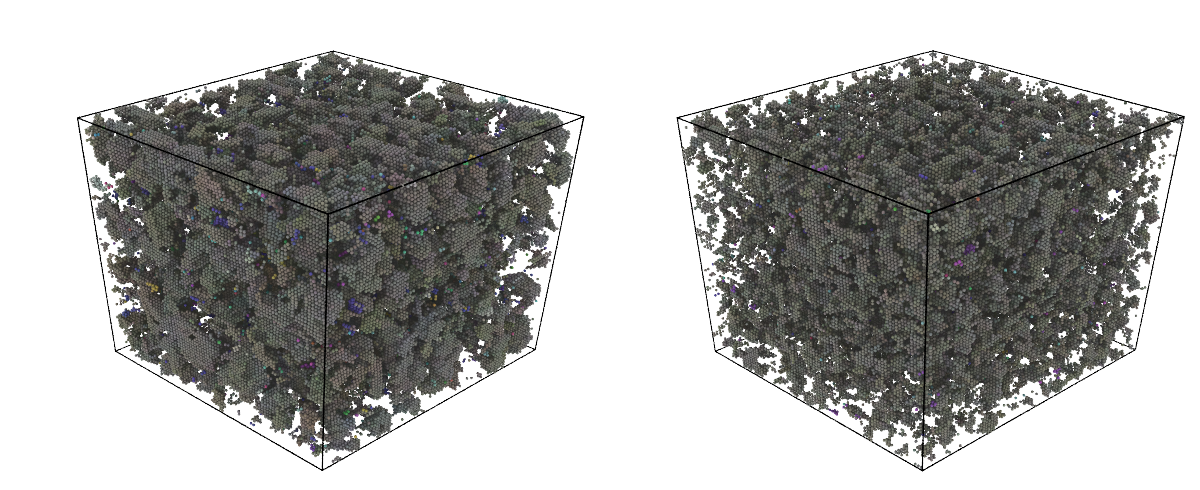}
\label{fig:Fig7c}
\end{subfigure}
\begin{subfigure}{0.45\textwidth}
\caption{1 Strain} 
\includegraphics[width=\linewidth]{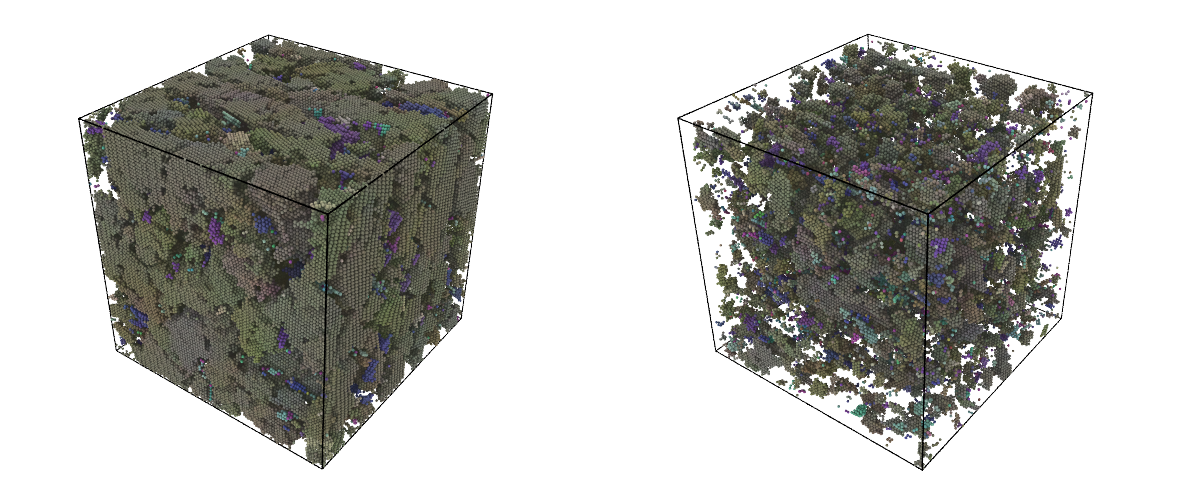}
\label{fig:Fig7d}
\end{subfigure}\

\medskip
\begin{subfigure}{0.45\textwidth}
\caption{4 Strain} 
\includegraphics[width=\linewidth]{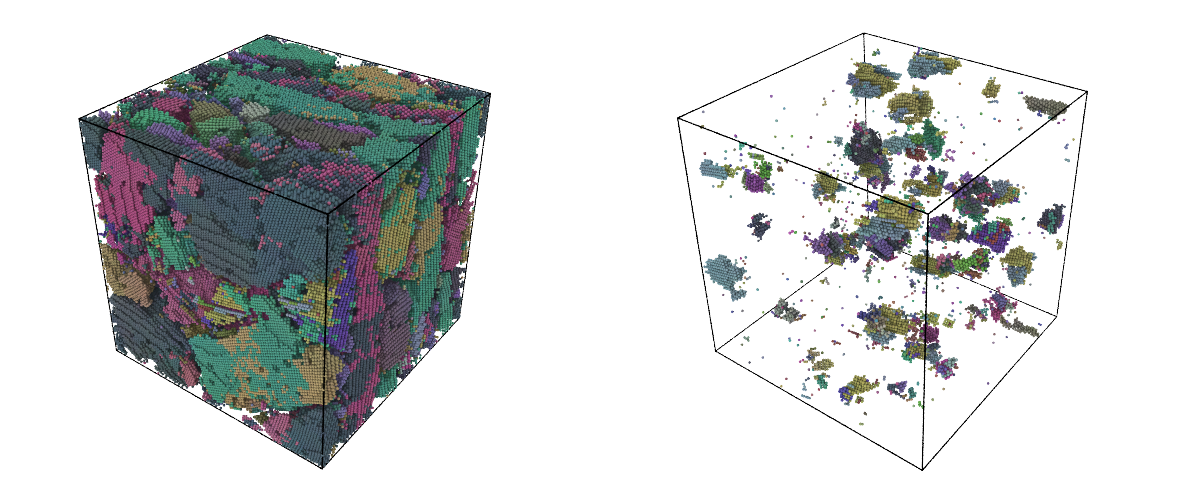}
\label{fig:Fig7e}
\end{subfigure}
\begin{subfigure}{0.45\textwidth}
\caption{7 Strain} 
\includegraphics[width=\linewidth]{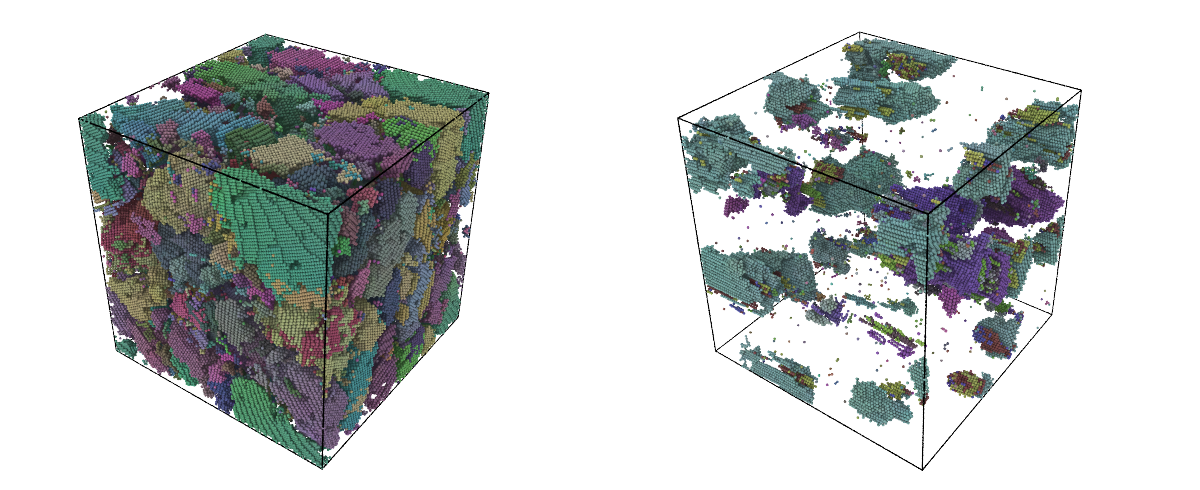}
\label{fig:Fig7f}
\end{subfigure}\

\medskip
\begin{subfigure}{0.45\textwidth}
\caption{10 Strain} 
\includegraphics[width=\linewidth]{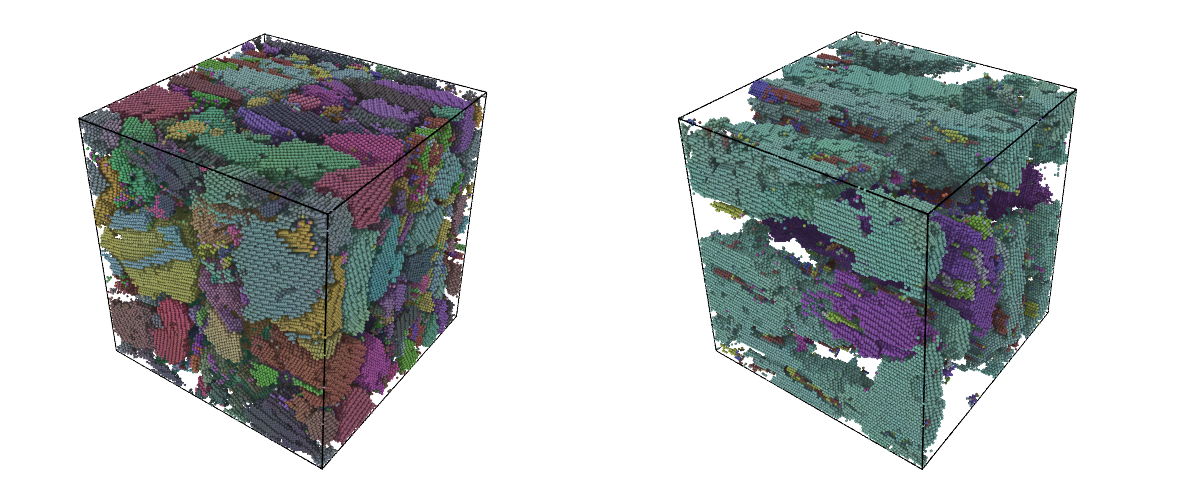}
\label{fig:Fig7g}
\end{subfigure}

\caption{Grain refinement process for different damping parameter simulations. Atoms are coloured according to local crystallographic orientations. Atoms not identified as BCC by polyhedral template matching are not shown. Strain rate $d\gamma/dt= 1/33.5$ ps$^{-1}\approx 2.985\times 10^{10}$ s$^{-1}$. Left - Benchmark simulation with damping parameter $b=6.875$ eV fs {\AA}$^2$; Right - Higher damping simulation with damping parameter $b=68.75$ eV fs {\AA}$^2$.} \label{fig:7}
\end{figure*}

Figure \ref{fig:6} shows various quantities plotted as a function of shear strain when the two different damping parameters are used. We also compare the percentage of body-centred cubic structure found in the simulation cell as a function of strain. This will become important when considering Figure \ref{fig:7}, which compares the evolution of the local crystal orientations in simulations using different damping parameters.

Figure \ref{fig:Fig6a} shows the change in kinetic energy between the lower (or benchmark) and higher damping simulations. The average kinetic energy in the lower damping simulation increases from $\sim 0.038$ eV/atom to 0.072 eV/atom, a difference of 0.034 eV/atom, before reducing to a dynamic quasi-steady state. Similarly, the higher damping simulation experiences an increase from 0.034 eV/atom to $\sim 0.045$ eV/atom, a difference of 0.011 eV/atom, before it drops. The increase in kinetic energy is much lower for the higher damping simulation because a higher damping parameter leads to a higher quenching rate. The kinetic energy is also noticeably lower for the higher damping simulation once both simulations have reached the dynamic quasi-steady state. The lower damping (benchmark) simulation plateaus at 0.042 eV/atom whilst the higher damping simulation sits at 0.039 eV/atom.

Figure \ref{fig:Fig6b} shows that both simulations experience an identical increase in excess potential energy with increasing shear strain up to a value of $\gamma = 0.27$, corresponding to the highly disorder state. The benchmark simulation peaks at 0.24 eV/atom, whilst the higher damping simulation peaks at 0.25 eV/atom; a small difference. However, whilst the benchmark simulation reduces to and subsequently stabilises at $\sim 0.09$ eV/atom, the higher damping simulation only decreases to $\sim 0.15$ eV/atom, which suggests a greater state of disorder. Furthermore, the excess potential energy shows a gradual decrease after $\sim\gamma = 4$, until it approximately reaches the same value as the benchmark simulation at $\gamma = 10$. As before, the average atomic von Mises stress, as shown in Figure \ref{fig:Fig6d}, and the excess potential energy per atom, are well correlated for both simulations.

Next, we consider the grain count in the simulation cells (Figure \ref{fig:Fig6c}). Both the benchmark and higher damping simulations exhibit significant disordered states at $\gamma = 0.27$. The benchmark cell contains about 400 grains, while the higher damping simulation shows 390. Both simulations show a rapid decrease in grain count. The benchmark simulation subsequently drops to $\sim 30$ grains, whereas the higher damping simulation only drops to $\sim 130$ grains. The benchmark simulation steadily increases in grain count until about $\gamma = 2$, reaching $\sim$220 grains. In contrast, the higher damping simulation rapidly rises to 320 grains at $\sim\gamma = 1.1$, then sharply declines. By $\gamma = 3$, the higher damping simulation maintains $\sim 70$ grains, plateauing until the end of simulation, a lower count than the benchmark simulation's plateau at 240 grains.

Figure \ref{fig:Fig6e} shows the dislocation density for the simulations. The initial spike in dislocation density is much lower for the higher damping simulation. Whilst the benchmark simulation increases to $\sim2.6\times10^{-3}/$\AA$^2$ at $\gamma = 0.82$, the higher damping simulation's dislocation density only increases to $6.4\times10^{-4}/$\AA$^2$ at $\gamma = 0.51$. With increased shearing, the dislocation density for the benchmark simulation decreases and then saturates at $\sim1.6\times10^{-3}/$\AA$^2$. This is in contrast to the higher damping simulation where the dislocation density decreases to $2.5\times10^{-5}/$\AA$^2$ at $\sim\gamma = 2$ and begins to increase at $\sim\gamma = 6.5$, with the dislocation density for the higher damping simulation peaking at $\sim1.2\times10^{-3}/$\AA$^2$ at $\gamma = 10$. 

Figure \ref{fig:Fig6f} shows the percentage of the overall structure in the body-centred cubic phase as a function of strain. Both simulations experience a highly disordered state transition at $\gamma = 0.27$, where the percentage of BCC atoms within the cells is below 10\% for both simulations. However, the benchmark simulation rapidly recovers, and around 50\% of the atoms are considered to be BCC after $\gamma = 10$. Conversely, the higher damping simulation makes a slight recovery to 17\% at $\gamma = 0.44$, before reducing to 2\% BCC. After $\gamma = 5$, the higher damping cell begins to recover its body-centred cubic structure, increasing to $\sim$32\% BCC at $\gamma = 10$.

A visual representation of atomic crystal orientation is shown in Figure \ref{fig:7}. Both simulations begin with a perfect cell (Figure \ref{fig:Fig7a}) before experiencing a disordered state (Figure \ref{fig:Fig7b}) followed by a reordering of atoms (Figure \ref{fig:Fig7c}). The number of atoms in the disordered phase is much larger for the higher damping simulation. During the reordering, the grains appear to be much larger for the benchmark simulation, with many small pockets of body-centred cubic atoms being present in the higher damping simulation. 

In Figure \ref{fig:Fig7d}, the higher damping simulation does not experience a grain growth similar to the benchmark simulation, with many fragmented grains and disordered atoms present. By Figure \ref{fig:Fig7e}, most of the cell is in the disordered phase. There are only a few pockets of body-centred cubic atoms still present within the box, as per the PTM modifier. This is consistent with Figure \ref{fig:Fig6f} which shows that the percentage of atoms in the body-centred cubic phase at $\gamma = 4$ was 2\%. Between $\gamma = 4$ (\ref{fig:Fig7e}) and $\gamma = 7$ (\ref{fig:Fig7f}), there is a noticeable grain growth, which is even more noticeable at $\gamma = 10$ (\ref{fig:Fig7g}), corresponding to a larger number of atoms transitioning into the body-centred cubic phase as also seen in Figure \ref{fig:Fig6f}.

Previous studies \cite{HUMPHREYS2004415, WANG19941193} consider dynamic recovery to be the dominant recovery mechanism in ferritic steels. However, the data presented in Figures \ref{fig:6} and \ref{fig:7} appears to show that dynamic recrystallisation mechanisms are at play, which agrees with the observations made by Tsuji \textit{et al.} \cite{TSUJI}, which confirmed the occurrence of dynamic recrytsallisation in body-centred cubic iron. 

Considering Figure \ref{fig:Fig7e} - \ref{fig:Fig7g}, the few remaining grains at $\gamma = 4$ grow much larger and recrystallisation occurs. This process continues to produce a small number of large grains at $\gamma = 10$ which coincides with an increase in the percentage of BCC phase, as shown in \ref{fig:Fig6f}. Interestingly, previous analysis of Figure \ref{fig:Fig6c} showed that the grain number plateaued at $\sim\gamma = 3$ for the higher damping simulation. This provides further evidence for dynamic recrystallisation, as the grain number stays constant whilst the body-centred cubic structure continues to recover. 

A larger damping parameter fundamentally means that the quenching rate is faster. This means that atoms lose their kinetic energy more quickly, and get trapped in a disordered state. As shown in Figure \ref{fig:Fig7c} and \ref{fig:Fig7d}, the higher damping simulation experiences reordering, similar to the benchmark simulation, but not to the same extent. Increased shear strain causes the atoms revert into a disordered state, as shown in Figure \ref{fig:Fig7e}. This is confirmed by considering Figure \ref{fig:Fig6f}, which shows a recovery of the BCC phase from the minimum at $\gamma = 0.27$, to 17\% BCC at $\gamma = 0.44$. This is followed by a return to the disordered phase. It is not immediately evident why the atoms return to the disordered state post yielding. In the following section, we will attempt to outline a possible explanation for this. 

As previously mentioned, the system is quenched at a higher rate for the higher damping simulation which causes many atoms to remain in the disordered phase. This essentially means that after reordering occurs at $\gamma = 0.32$, the grains are very small and the grain boundary volume is large. As such, the dislocations formed as a result of shearing are heavily constrained within small grains \cite{NIELSEN20217} and more readily meet and annihilate at the grain boundaries. This is evident from Figure \ref{fig:Fig6e} which shows an increase in dislocation density up to $6.0\times10^{-4}/$\AA$^2$ post yielding, followed by a quick drop in dislocation density to $2.6\times10^{-6}/$\AA$^2$ at $\gamma = 4$. 

As dislocations are annihilated at grain boundaries and an increasing number of atoms enter a disordered state, the percentage of body-centred cubic structure decreases. The growth in grain size for the higher damped simulation provides the necessary grain volume for the propagation of dislocations, and an increase in dislocation density can be observed with the increase in body-centred cubic structure at $\gamma = 7$. Since there are many highly disordered atoms that manifest as grain boundaries, the free energy in the simulation is high, which drives the recrystallisation.

\subsection{Strain Rate}

\begin{figure*}[t]
\begin{subfigure}{0.45\textwidth}
\caption{} 
\includegraphics[width=\linewidth]{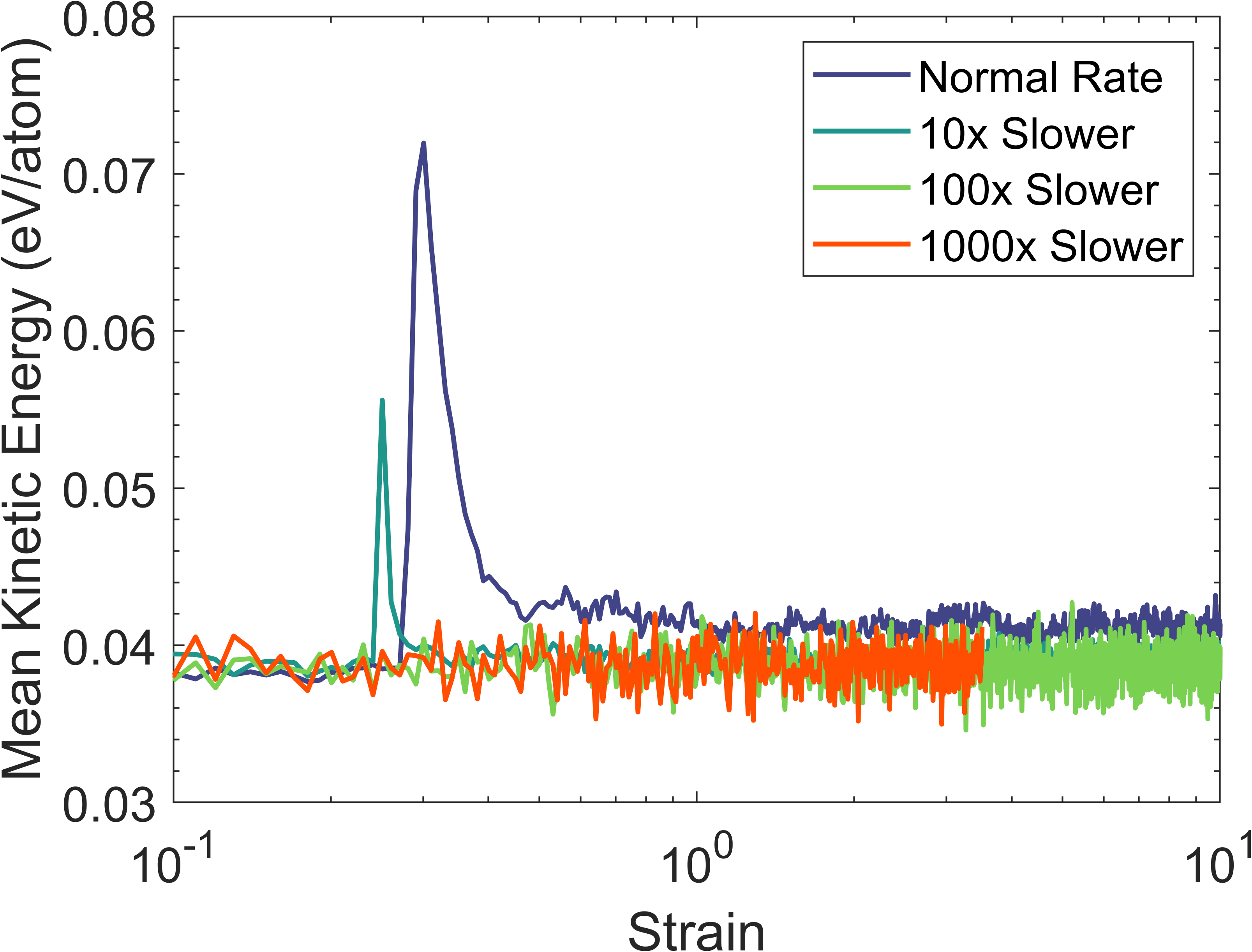}
\label{fig:Fig8a}
\end{subfigure}
\hspace{1em}
\begin{subfigure}{0.45\textwidth}
\caption{} 
\includegraphics[width=\linewidth]{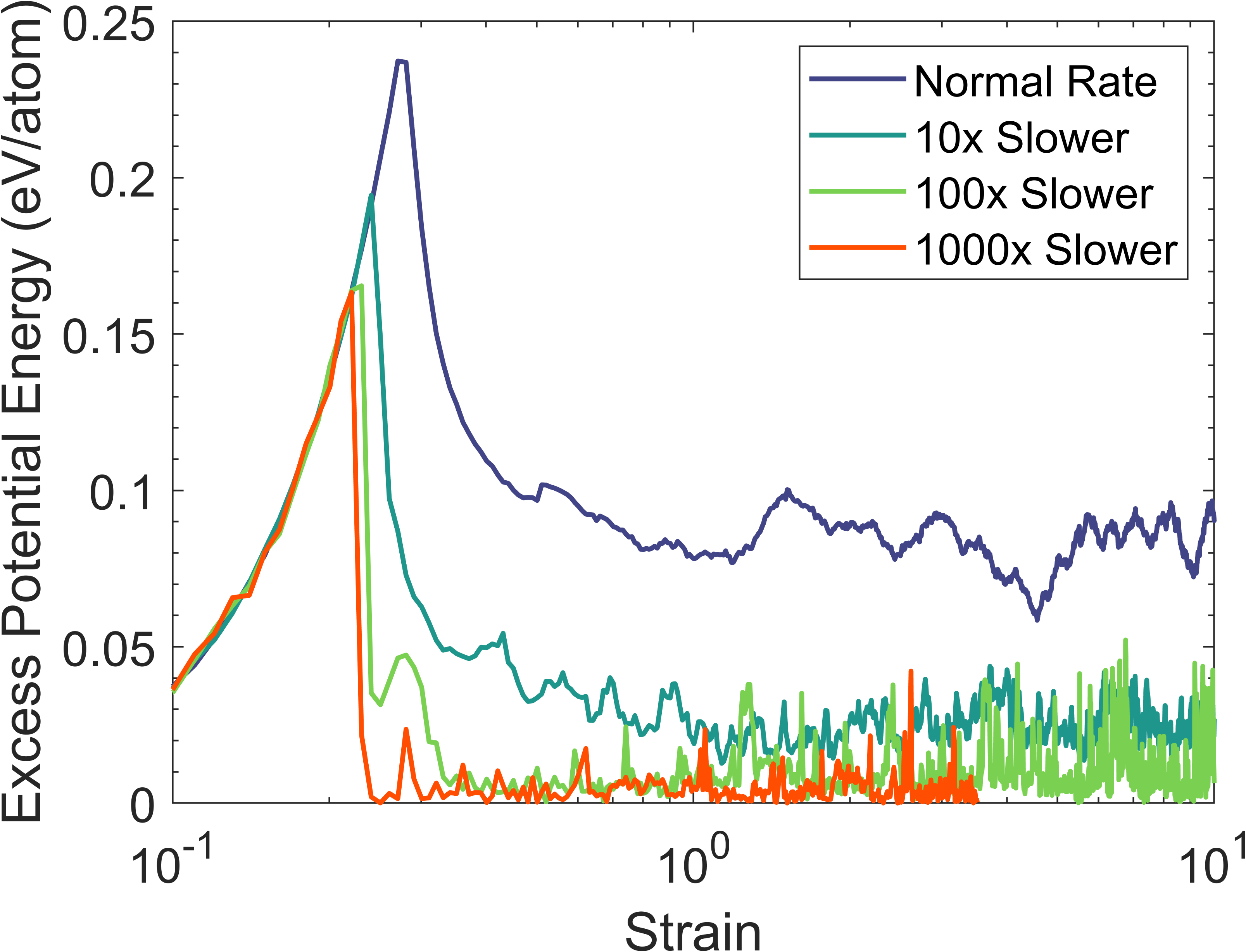}
\label{fig:Fig8b}
\end{subfigure}

\medskip
\begin{subfigure}{0.45\textwidth}
\caption{} 
\includegraphics[width=\linewidth]{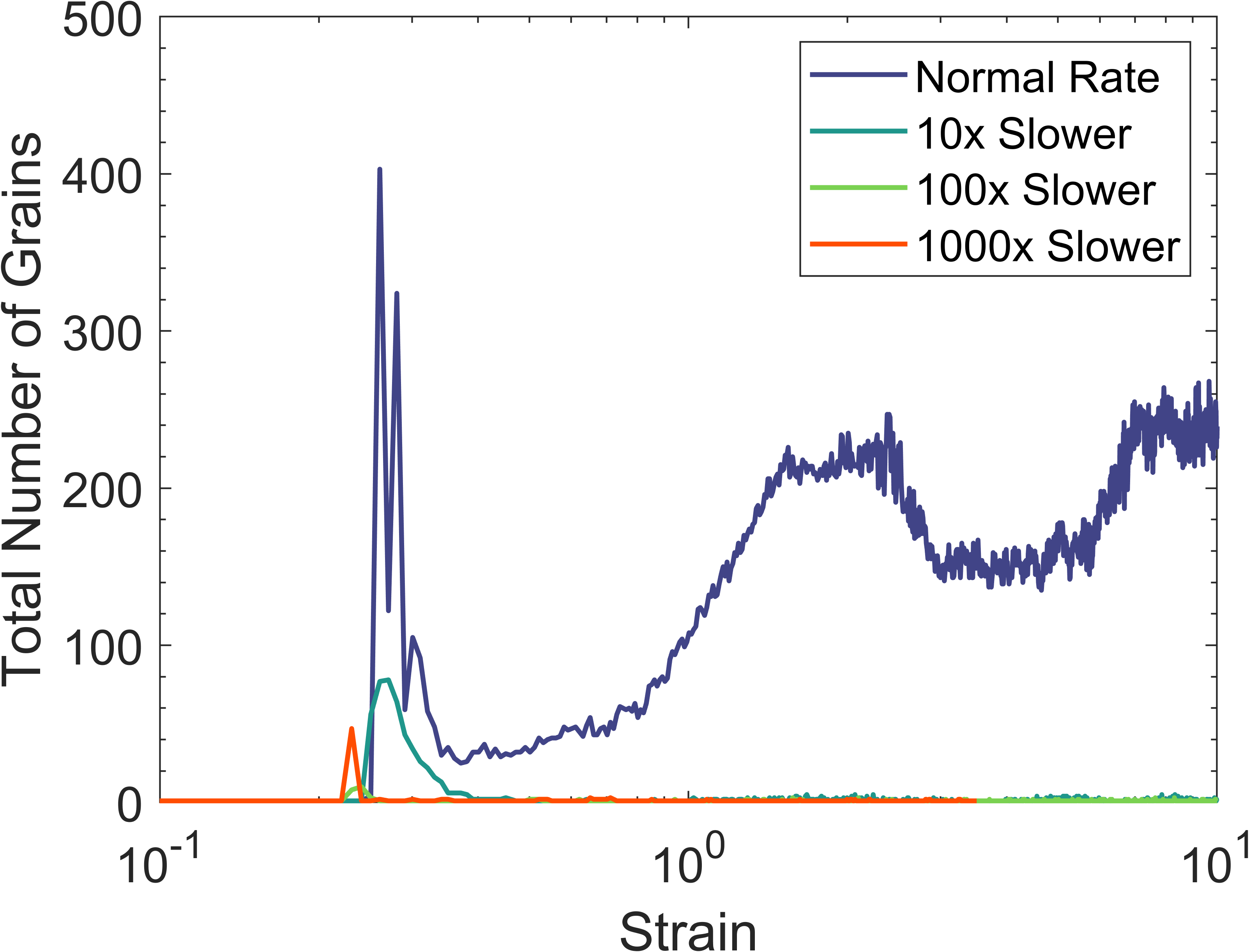}
\label{fig:Fig8c}
\end{subfigure}
\hspace{1em}
\begin{subfigure}{0.45\textwidth}
\caption{} 
\includegraphics[width=\linewidth]{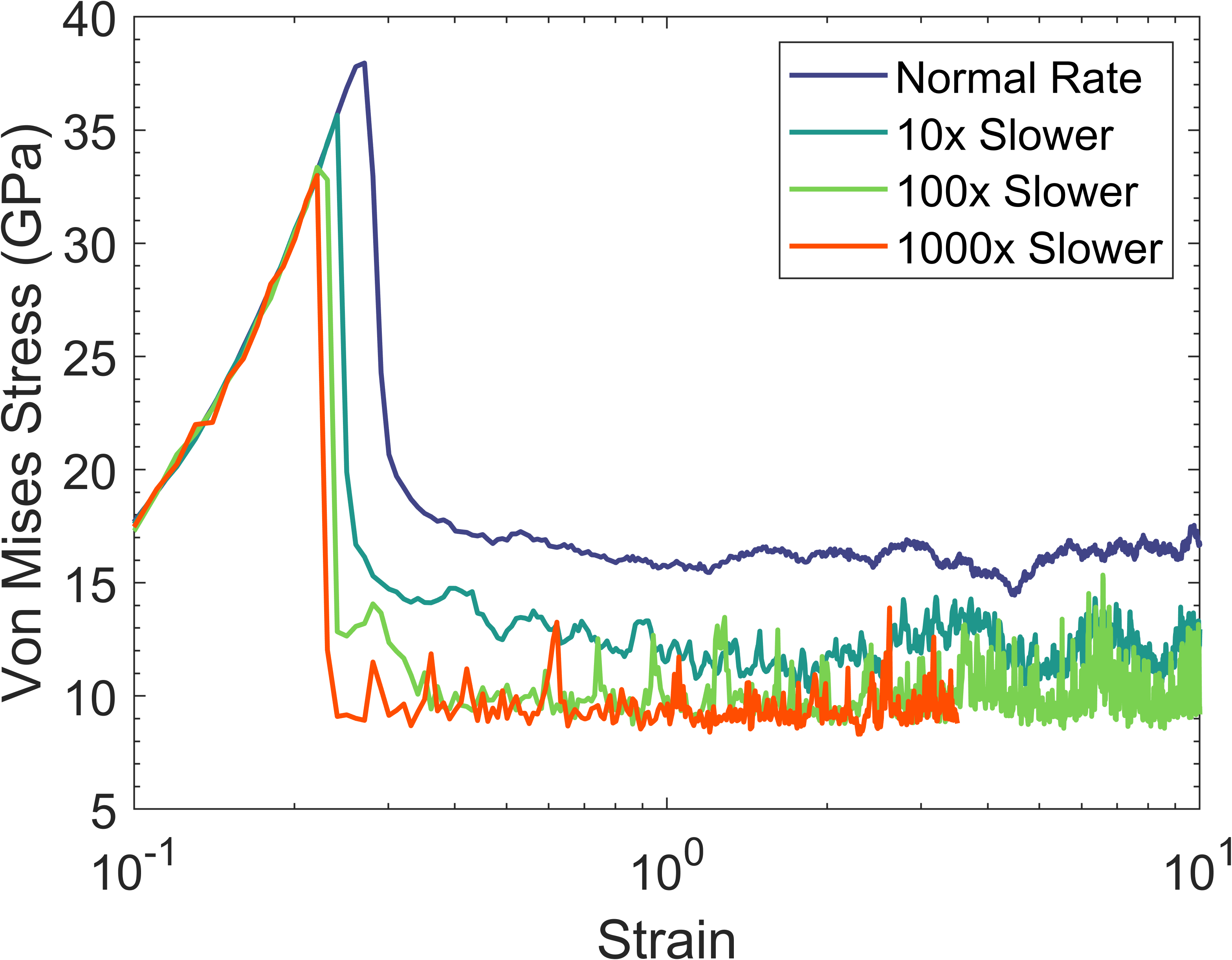}
\label{fig:Fig8d}
\end{subfigure}

\medskip
\begin{subfigure}{0.45\textwidth}
\caption{} 
\includegraphics[width=\linewidth]{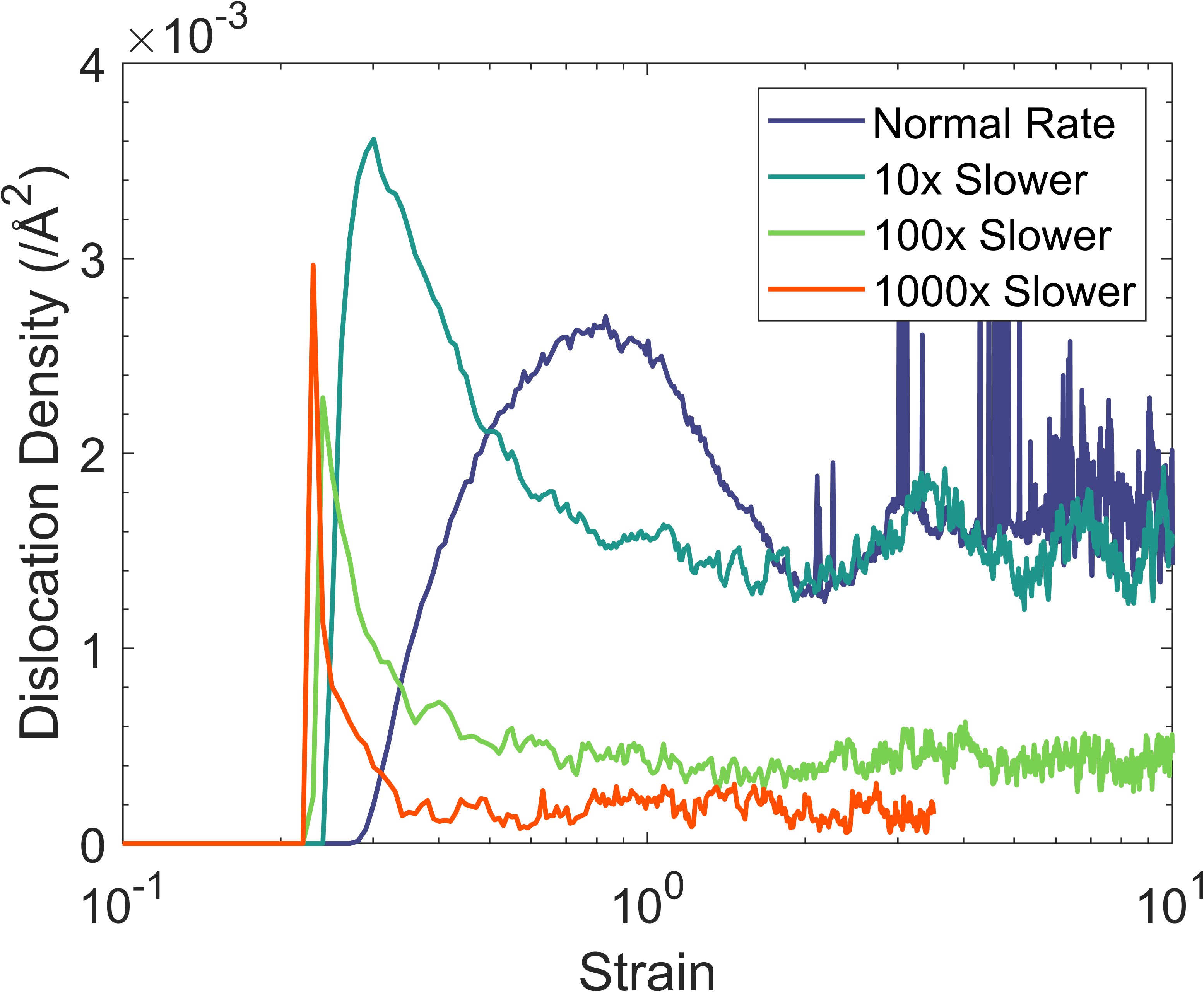}
\label{fig:Fig8e}
\end{subfigure}

\caption{Simulation data plotted as a function of strain for different strain rate simulations. Damping parameter $b=6.875$ eV fs {\AA}$^2$. Strain rate $d\gamma/dt= 1/33.5$ ps$^{-1}\approx 2.985\times 10^{10}$ s$^{-1}$ for purple line, $d\gamma/dt= 1/335$ ps$^{-1}\approx 2.985\times 10^{9}$ s$^{-1}$ for blue line, $d\gamma/dt= 1/3350$ ps$^{-1}\approx 2.985\times 10^{8}$ s$^{-1}$ for green line, and $d\gamma/dt= 1/33500$ ps$^{-1}\approx 2.985\times 10^{7}$ s$^{-1}$ for orange line. (a) Kinetic Energy (b) Excess Potential Energy (c) Number of Grains (d) Von Mises Stress (e) Dislocation Density.} \label{fig:8}
\end{figure*}

In this section we consider the effect of strain rate. We performed simulations at a shear strain rate 10 times and 100 times slower than the benchmark simulation, whilst keeping other parameters unchanged, up to $\gamma = 10$. We also performed a simulation 1000 times slower than the benchmark up to $\gamma = 3.5$ due to computational limitations. Figure \ref{fig:8} shows different quantities plotted as a function of strain for the normal and slower rate simulations.

\begin{figure*}[t]
\begin{subfigure}{1.0\textwidth}
\caption{Reordering} 
\includegraphics[width=\linewidth]{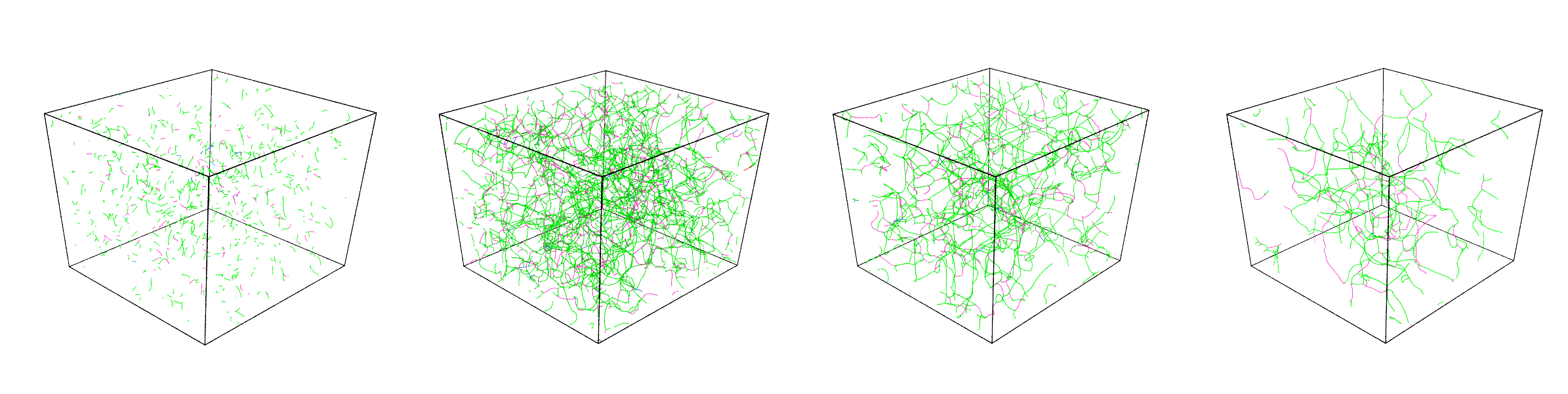}
\label{fig:Fig9a}
\end{subfigure}

\medskip
\begin{subfigure}{1.0\textwidth}
\caption{1 Strain}
\includegraphics[width=\linewidth]{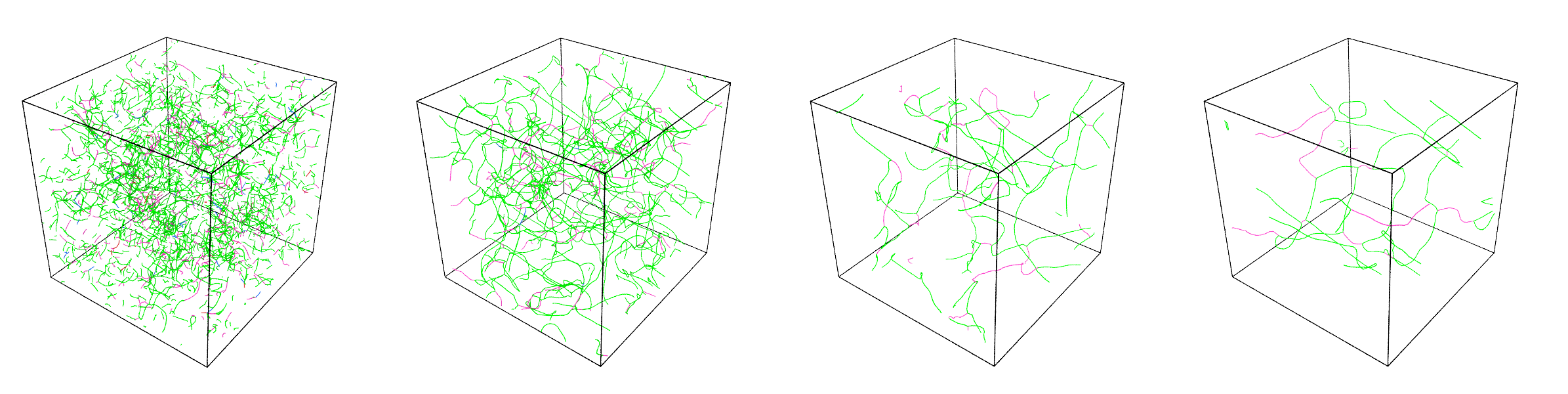}
\label{fig:Fig9b}
\end{subfigure}

\medskip
\begin{subfigure}{1.0\textwidth}
\caption{3 Strain} 
\includegraphics[width=\linewidth]{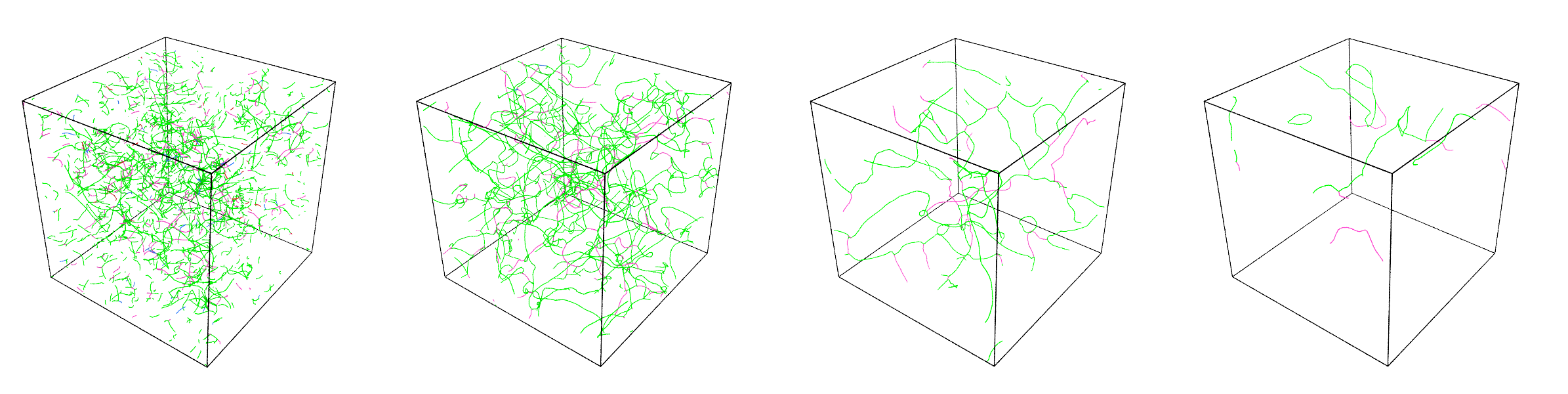}
\label{fig:Fig9c}
\end{subfigure}

\caption{Dislocation network comparison for different strain rate simulations. Damping parameter $b=6.875$ eV fs {\AA}$^2$. Green dislocation lines $\mathbf{b}=\frac{1}{2}\langle 111\rangle$ and pink dislocation lines $\mathbf{b}=\langle100\rangle$. Left - Benchmark simulation with strain rate $d\gamma/dt= 1/33.5$; Centre-left - 10x slower rate simulation with strain rate $d\gamma/dt= 1/335$; Centre-right - 100x slower rate simulation with strain rate $d\gamma/dt= 1/3350$; Right - 1000x slower rate simulation with strain rate $d\gamma/dt= 1/33500$.} \label{fig:9}
\end{figure*}

In Figure \ref{fig:Fig8a}, the initial kinetic energy spike is not as pronounced for the 10x slower strain rate simulation. Whilst the benchmark simulation shows an increase from 0.039 eV/atom to 0.072 eV/atom, the slow rate simulation only shows an increase from 0.039 eV/atom to 0.056 eV/atom. Furthermore, the spike for the benchmark simulation begins at $\gamma = 0.27$ and peaks at $\gamma = 0.3$, whilst the slow rate simulation experiences a rise starting at $\gamma = 0.24$, and peaking at $\gamma = 0.25$. After the initial increase and subsequent decrease in kinetic energy, the slow rate simulation plateaus at a lower kinetic energy of $\sim 0.039$ eV/atom, compared with the benchmark simulation at 0.042 eV/atom. We note that there is no spike in kinetic energy for the 100x and 1000x slower simulations, and the kinetic energy remains around the same value throughout the simulations. We attribute this to coarse sampling, as we sample our simulations based on strain values. The longer timescales associated with the slower rate simulations allow the reduced excess potential energy to be converted to kinetic energy during reordering between the sampled strain values. 

The difference in the excess potential energy is shown in Figure \ref{fig:Fig8b}. All of the simulations have identical trajectories in the initial stage as they experience the same elastic loading. However, the slower rate simulations reach their peak in potential energy at an earlier stage than the benchmark, with slower simulations reaching a lower peak excess potential energy. The von Mises stress correlates with the potential energy as shown in Figure \ref{fig:Fig8d}. As before, the benchmark simulation peaks at 0.24 eV/atom at $\gamma = 0.27$, while the 10x slower rate simulation peaks at 0.19 eV/atom at $\gamma = 0.24$. The 100x slower simulation peaks at 0.17 eV/atom at $\gamma = 0.23$, and the 1000x slower simulation peaks at 0.16 eV/atom at $\gamma = 0.22$. This is followed by a rapid decrease for all simulations. The slower rate simulations decrease and plateau at a lower excess potential energy, i.e. 0.02 eV/atom (for the 10x slower rate), than the benchmark, 0.09 eV/atom. This behaviour is consistent with previous observations \cite{BulatovPaper}, which also showed a decrease in yield strength for simulations with slower strain rates, as is shown in Figure \ref{fig:Fig8d}.

Though we can observe grain refinement in the benchmark simulation, this does not occur in the slower rate simulations, as shown in Figure \ref{fig:Fig8c}. Whilst we do observe a smaller initial peak of 80 grains at $\gamma = 0.27$ for the 10x slower rate simulation, this does not result in the formation of a largely polycrystalline structure, and after $\gamma = 0.38$, a single grain is present. Furthermore, we observe even smaller peaks for the 100x and 1000x slower rate simulations which, after $\gamma = 0.26$, return to being single crystalline. This is contrary to the data obtained for the benchmark, and other previously discussed simulations that experienced the generation of many nanocrystalline grains. We again propose that the observed peaks are a byproduct of PTM modifier failing to identify BCC structures when highly strained. Figure \ref{fig:Fig8d} shows that the yield point of the slower rate simulations is progressively lower than the benchmark with decreased strain rate. As such, the atomic bonds do not stretch to the same extent as in the benchmark simulation, allowing the PTM modifier to detect more of the body-centred cubic structure. Hence, the spike in grain number occurs at a lower strain and is lower in magnitude for the slower rate simulations.

Several differences between the simulations are observed when analysing the dislocation densities in Figure \ref{fig:Fig8e}. After a shear strain of $\gamma = 0.27$, the dislocation density for the benchmark simulation gradually increases up to $\sim2.6\times10^{-3}/$\AA$^2$ at $\gamma = 0.82$, after which it decreases to $1.2\times10^{-3}/$\AA$^2$ at $\sim\gamma = 2$ before plateauing at $1.6\times10^{-3}/$\AA$^2$. However, in the 10x slow rate simulation, dislocation density shows a rapid increase to $3.6\times10^{-3}/$\AA$^2$ at $\gamma = 0.3$ followed by a decrease to $\sim1.2\times10^{-3}/$\AA$^2$ at $\gamma = 1.6$. After this point, the dislocation densities for the benchmark and 10x slower simulations stay fairly similar. In contrast, the dislocation density in the 100x slower simulation increases sharply from $\gamma = 0.22$, up to $2.3\times10^{-3}/$\AA$^2$ at $\gamma = 0.24$, before decreasing rapidly to, and plateauing at $4.6\times10^{-4}/$\AA$^2$ at $\gamma = 0.5$. Similarly, the dislocation density in the 1000x slower simulation increases up to $2.9\times10^{-3}/$\AA$^2$ at $\gamma = 0.23$, before decreasing to $\sim4.6\times10^{-4}/$\AA$^2$ at $\gamma = 0.38$, and staying constant with subsequent shear.

A visual representation of dislocation distribution is shown in Figure \ref{fig:9}. During reordering, in Figure \ref{fig:Fig9a}, the benchmark simulation shows few dislocation lines, and they are sparse and disconnected. However, in the slower rate simulations, the dislocation density is much higher with a dislocation network being formed. This is in agreement with Figure \ref{fig:Fig8e} which depicts a large dislocation density after yielding for the slow rate simulations compared with the benchmark. 

At $\gamma = 1$ (Figure \ref{fig:Fig9b}), a significant increase in dislocation density is observed within the benchmark simulation. In contrast, the 10x slower rate simulation cell displays a reduction in dislocation density, consistent with Figure \ref{fig:Fig8e}. The 100x and 1000x simulations show a marked decrease in dislocation density, with few long dislocation lines being observed. From $\gamma = 1$ (Figure \ref{fig:Fig9b}) to $\gamma = 3$ (Figure \ref{fig:Fig9c}), the dislocation pattern in each simulation persists. The benchmark simulation shows short dislocation lines, with dislocation pile-ups attributed to grain boundary interactions. In contrast, the slower rate simulations display long dislocation lines, that form an extensive dislocation network. This behaviour persists due to the absence of grain boundaries constraining dislocation motion within the slow rate simulation cells. It is evident that the 100x and 1000x slower simulations have far fewer dislocation lines than the benchmark and the 10x slower simulation.

These observations can be rationalised as follows: the slower rate simulations do not show the formation of nanocrystals. During the initiation of plastic deformation, many dislocation lines are formed, which causes a spike in dislocation density, as seen in Figure \ref{fig:Fig9a}. These lines do not form in the benchmark simulation due to the presence of highly disordered atoms that manifest as grain boundaries. Nanocrystalline grains limit the volume in which the dislocations can form and propagate, resulting in the short dislocation lines observed in the benchmark simulation. With increased strain, the dislocations in the slower rate simulations are allowed to evolve and are not annihilated, as there are no grain boundaries. Instead, as shown in Figure \ref{fig:9}, they combine and form a large network of extended dislocations. The presence of grain boundaries does not permit this phenomenon to occur in the benchmark simulation, so short dislocations are observed.

\begin{figure}
\begin{subfigure}{0.45\textwidth}
\caption{} 
\includegraphics[width=\linewidth]{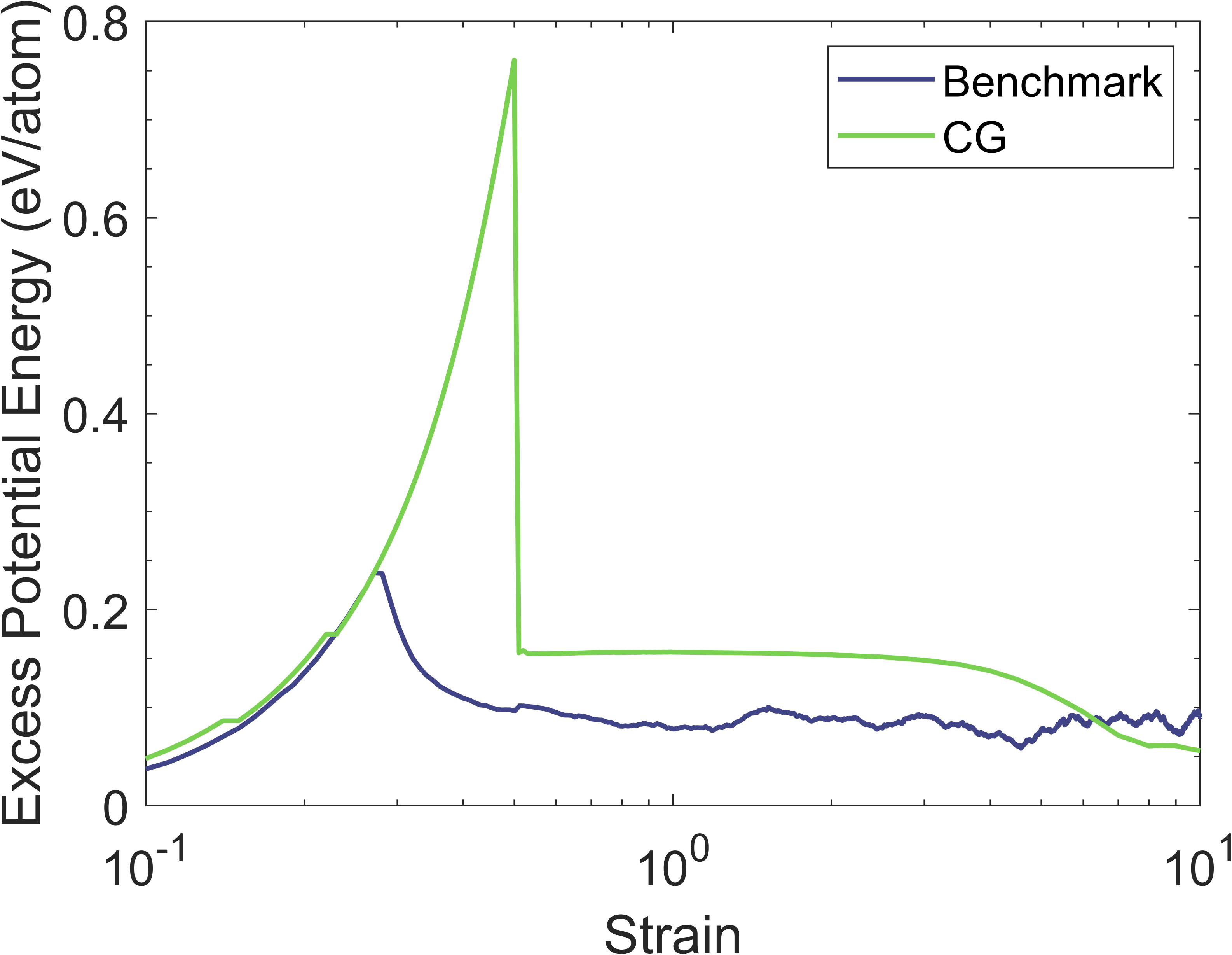}
\label{fig:Fig10a}
\end{subfigure}

\medskip
\begin{subfigure}{0.45\textwidth}
\caption{} 
\includegraphics[width=\linewidth]{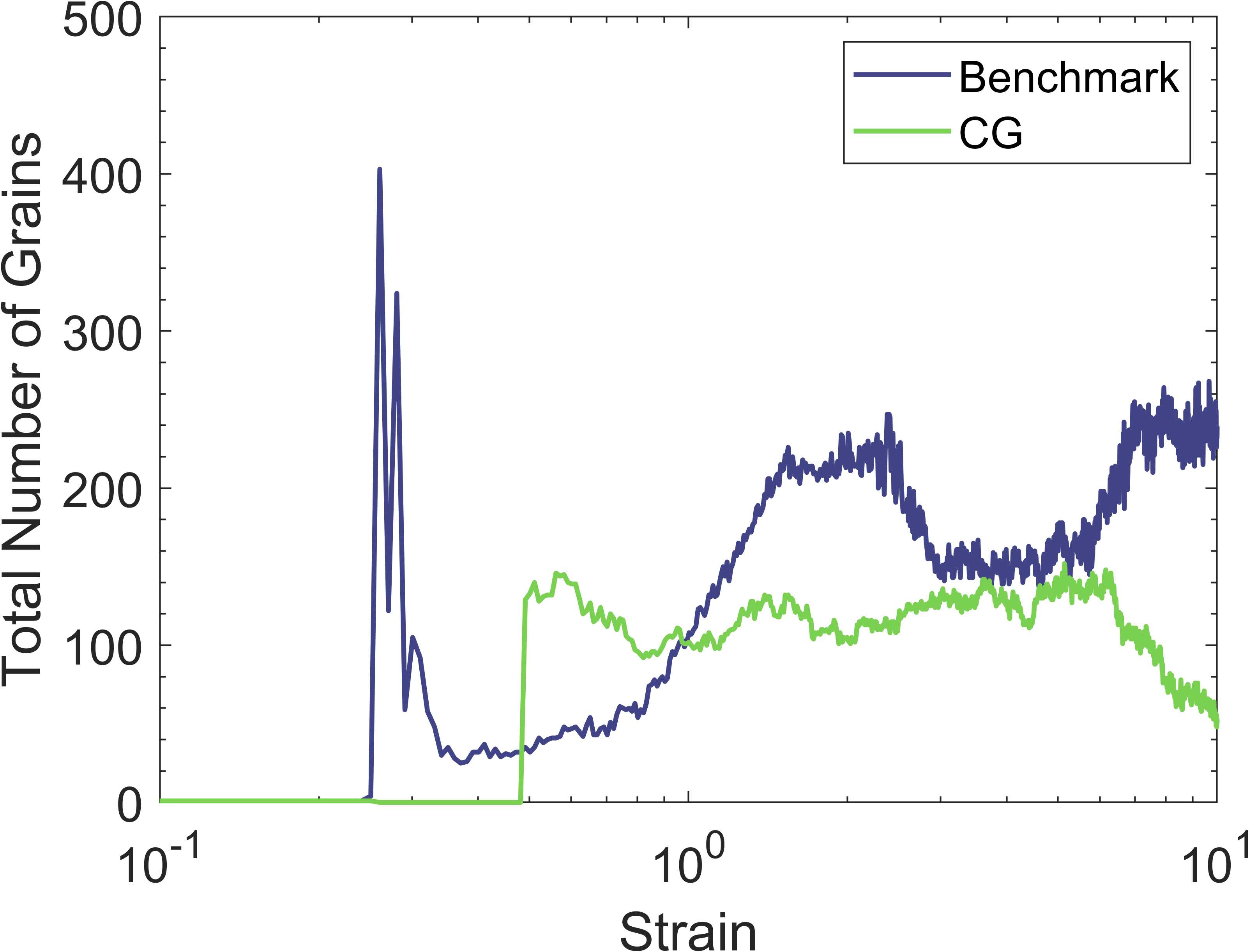}
\label{fig:Fig10b}
\end{subfigure}

\medskip
\begin{subfigure}{0.45\textwidth}
\caption{} 
\includegraphics[width=\linewidth]{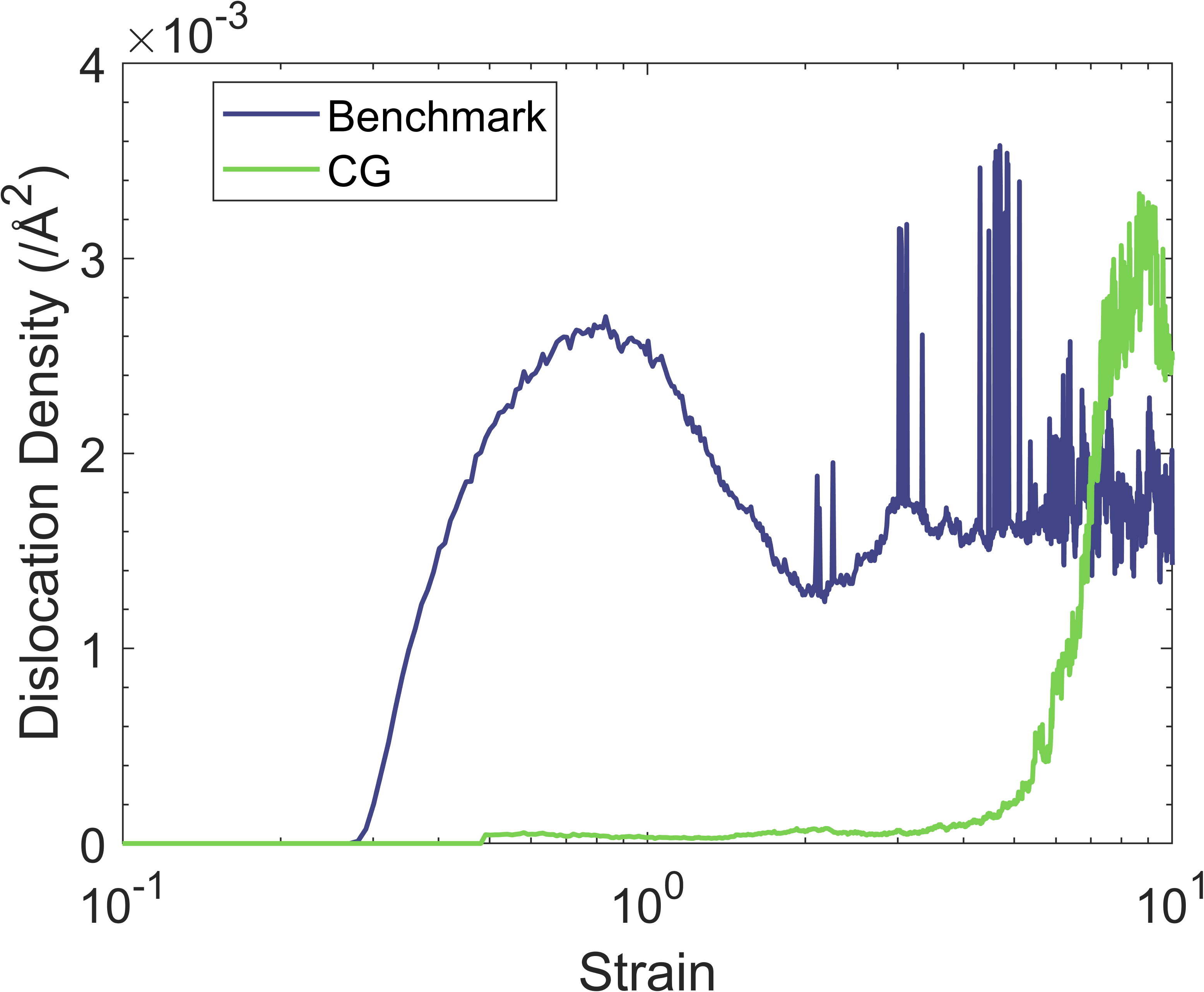}
\label{fig:Fig10c}
\end{subfigure}

\caption{Conjugate gradient simulation compared with benchmark simulation; (a) Excess Potential Energy (b) Number of Grains (c) Dislocation Density.} \label{fig:10}
\end{figure}

We may speculate about the reason why the slower strain rate simulation does not experience the same nanocrystal formation as the benchmark simulation. It has been shown experimentally that the yield strength of iron increases as the strain rate increases \cite{LUO2017142} which can be explained by strain rate sensitivity \cite{Borodin_2018}, and this is in agreement with Figure \ref{fig:Fig8d}. As the benchmark simulation is subjected to a higher strain rate, there is less time for the atoms to rearrange in response to the stress. Overall, there is less disorder in the slower rate simulations evidenced by the lower excess potential energy at yield, and stable nanocrystalline structures are not formed for the slowe rate simulations \ref{fig:Fig8c}. This allows larger dislocation lines to form in the slower rate simulations when compared to the benchmark simulation, as shown in Figure \ref{fig:Fig9a}. It appears that, with increasing strain, it is more energetically favourable for the atoms to retain their single crystal structure in the slower rate simulations. 

Dynamic recovery mechanisms subsequently appear to be present in the slower rate simulations, as the longer time scale gives dislocations more time to recover \cite{JIANG20221444}. As a result dislocations pile up less and thus do not form grain boundaries, limiting grain refinement. This is evident when considering Figures \ref{fig:Fig8e} and \ref{fig:9}, where the slower strain rate simulations show much lower dislocation densities. In fact, the lower the strain rate, the lower the dislocation density post yielding. It is also possible that the dislocations are simply unable to pile up, because the large dislocation network, on different slip planes, impedes extended dislocation glide \cite{WorkHardeningExplain}. As dislocation motion becomes constrained, the formation of low-angle grain boundaries and thus, further grain refinement, would then be inhibited. 

Experimental works suggest that severe plastic deformation at a lower strain rate correlates to a larger mean grain size \cite{ZHANG2013124, WAN201711} and it is possible that the present simulation cell size is the limiting factor. In the future, it may be beneficial to carry out shearing simulations with a much larger number of atoms to ascertain whether this is the case.

We also conducted a simulation using conjugate gradient shearing as discussed above. This simulation represents the athermal limit, and we show the results, in comparison to the benchmark simulation, in Figure \ref{fig:10}. 

\begin{figure}
\begin{subfigure}{0.4\textwidth}
\caption{} 
\includegraphics[width=\linewidth]{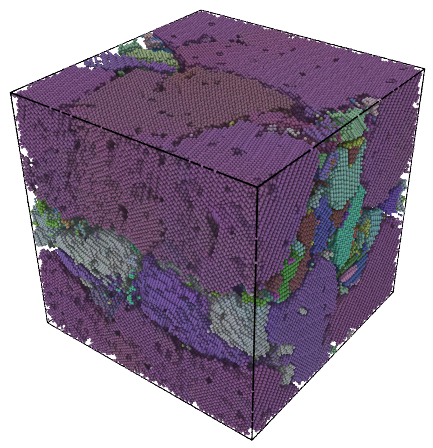}
\label{fig:Fig11a}
\end{subfigure}

\medskip
\begin{subfigure}{0.4\textwidth}
\caption{} 
\includegraphics[width=\linewidth]{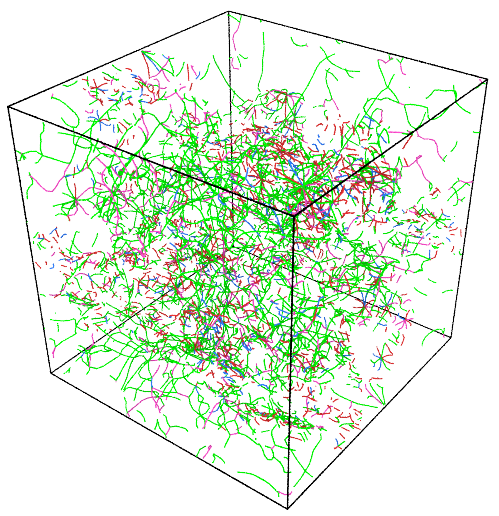}
\label{fig:Fig11b}
\end{subfigure}

\caption{Visualisation of the conjugate gradient simulation at $\gamma = 10$; (a) Local Crystal Orientation. Atoms not identified as BCC by polyhedral template matching are not shown; and (b) Dislocations.} \label{fig:11}
\end{figure}

Figure \ref{fig:Fig10a} shows a comparison of the excess potential energy. As before, the benchmark simulation peaks at 0.24 eV/atom at $\gamma = 0.27$ before reducing. However, rather dramatically, the CG simulation increases up to a peak of 0.76 eV/atom at $\gamma = 0.5$, before sharply decreasing to $\sim 0.16$ eV/atom and remaining there until $\sim\gamma = 8$, before decreasing and plateauing again at 0.06 eV/atom at $\gamma = 10$.

Figure \ref{fig:Fig10b} shows the presence of grains in the CG simulation, with $\sim 146$ grains at $\gamma = 0.56$. This number stays roughly the same until $\gamma = 6.2$ when the grain number begins to decrease until $\gamma = 10$, where $\sim 56$ grains remain. This is an interesting observation as the slower strain rate simulations above did not produce a nanocrystalline structure whilst the lowest limit CG simulation does. Figure \ref{fig:Fig11a} shows the atoms coloured by their local crystallographic orientation for the CG simulation at 10 strain and we can clearly see the presence of a nanocrystalline structure. At $\gamma = 10$, the excess potential energy for the CG simulation is lower than the benchmark. Figure \ref{fig:Fig10b} shows that this decrease in excess potential energy corresponds to a decrease in grain number. This suggests that the excess potential energy is predominantly associated with grain boundaries and as the total grain boundary area reduces, so too does the excess potential energy.

Considering Figure \ref{fig:Fig10c}, we note that the CG simulation initially has a low dislocation density up to $\gamma = 4$, after which there is a gradual increase. At $\sim$ $\gamma = 6$, we see a sharp increase in dislocation density until $\sim\gamma = 9$, where the dislocation density is  $\sim3.0\times10^{-3}/$\AA$^2$, much more than the benchmark or slower rate simulations. At $\gamma = 10$, the dislocation density is $\sim2.5\times10^{-3}/$\AA$^2$, which is still higher than the benchmark simulation. Figure \ref{fig:Fig11b} shows the identified dislocation lines in the CG simulation cell at $\gamma = 10$. We observe a dense dislocation network, with many piled up dislocations at the grain boundaries, similar to the benchmark simulation.

These results are very interesting in the context of the benchmark and slower rate simulations. We observe nanocrystal formation at very high strain rates with the benchmark simulation, and also in the athermal lower limit CG simulations, however, not in the intermediate rate simulations. Above, we proposed that dynamic recovery mechanisms are present in our molecular dynamics simulations of iron, and that the longer timescales associated with the slower rate simulations give dislocations more time to recover \cite{JIANG20221444} and thus limit grain refinement. In the CG simulation, on the other hand, thermal driving forces are absent, and thus nanograins are formed. As such, it appears that the benchmark simulation strain rate is sufficiently high to limit dislocation recovery mechanisms, which leads to the build up of dislocations and further grain refinement.

\subsection{Carbon Impurities}

\begin{figure*}[t]
\begin{subfigure}{0.45\textwidth}
\caption{} 
\includegraphics[width=\linewidth]{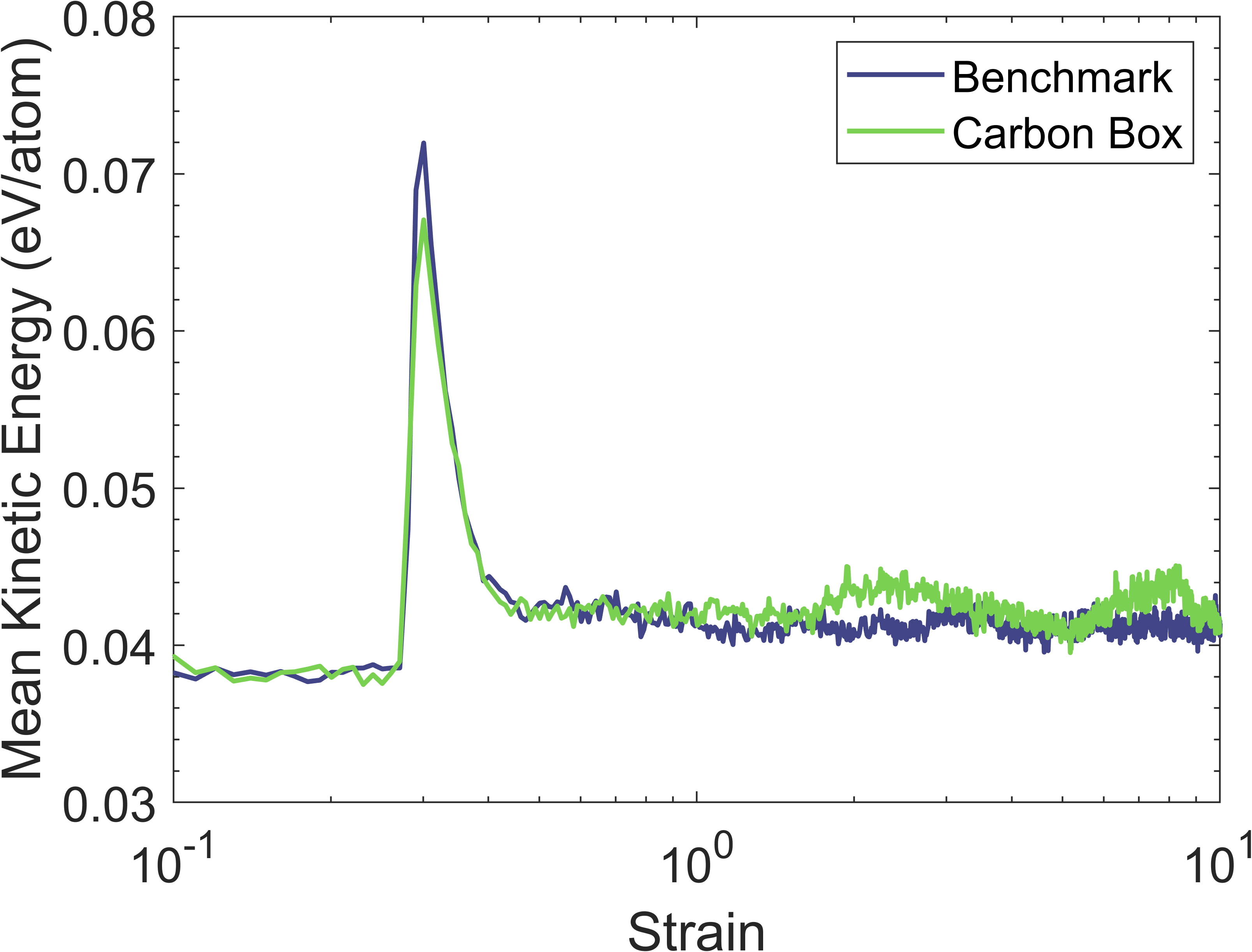}
\label{fig:Fig12a}
\end{subfigure}
\hspace{1em}
\begin{subfigure}{0.45\textwidth}
\caption{} 
\includegraphics[width=\linewidth]{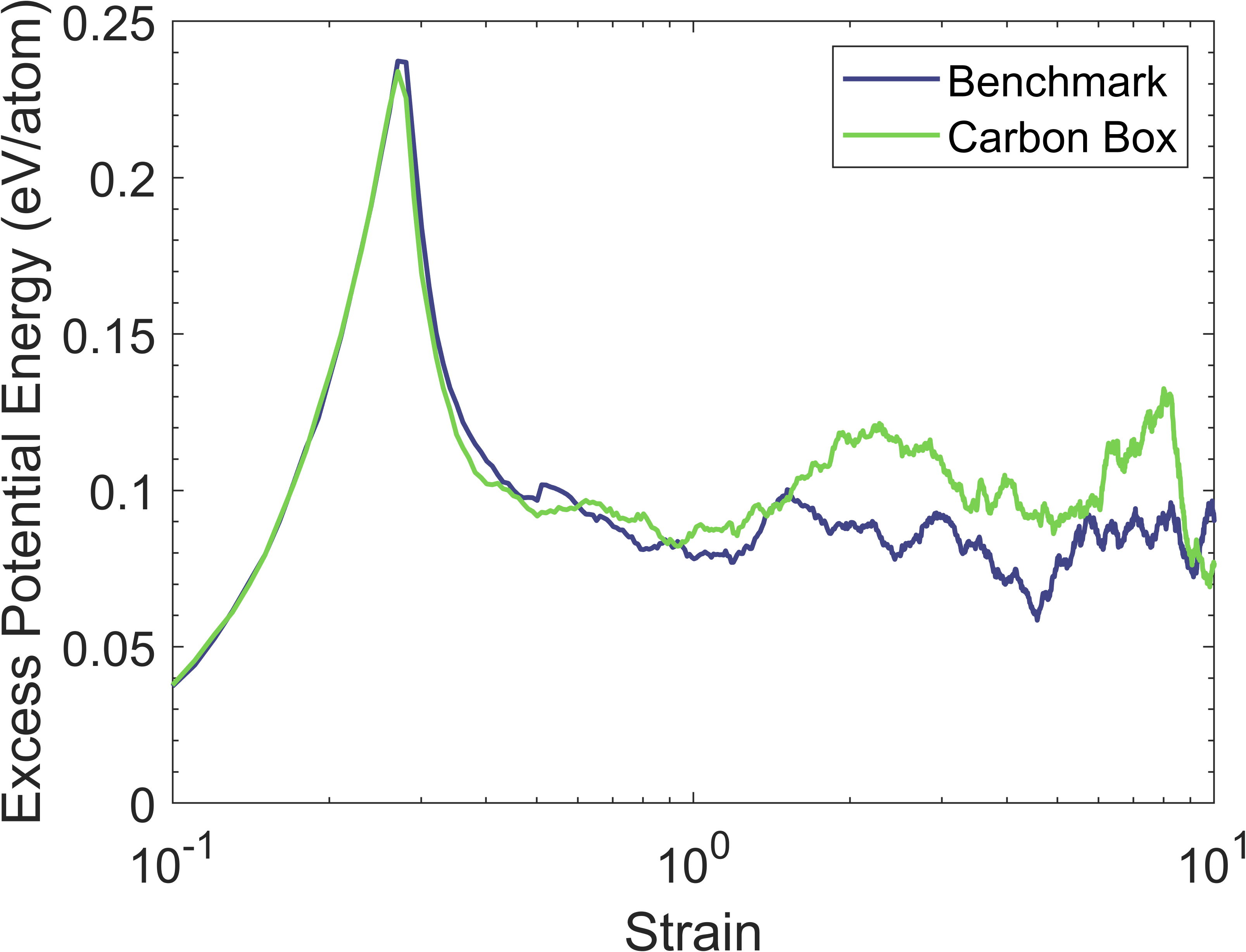}
\label{fig:Fig12b}
\end{subfigure}

\medskip
\begin{subfigure}{0.45\textwidth}
\caption{} 
\includegraphics[width=\linewidth]{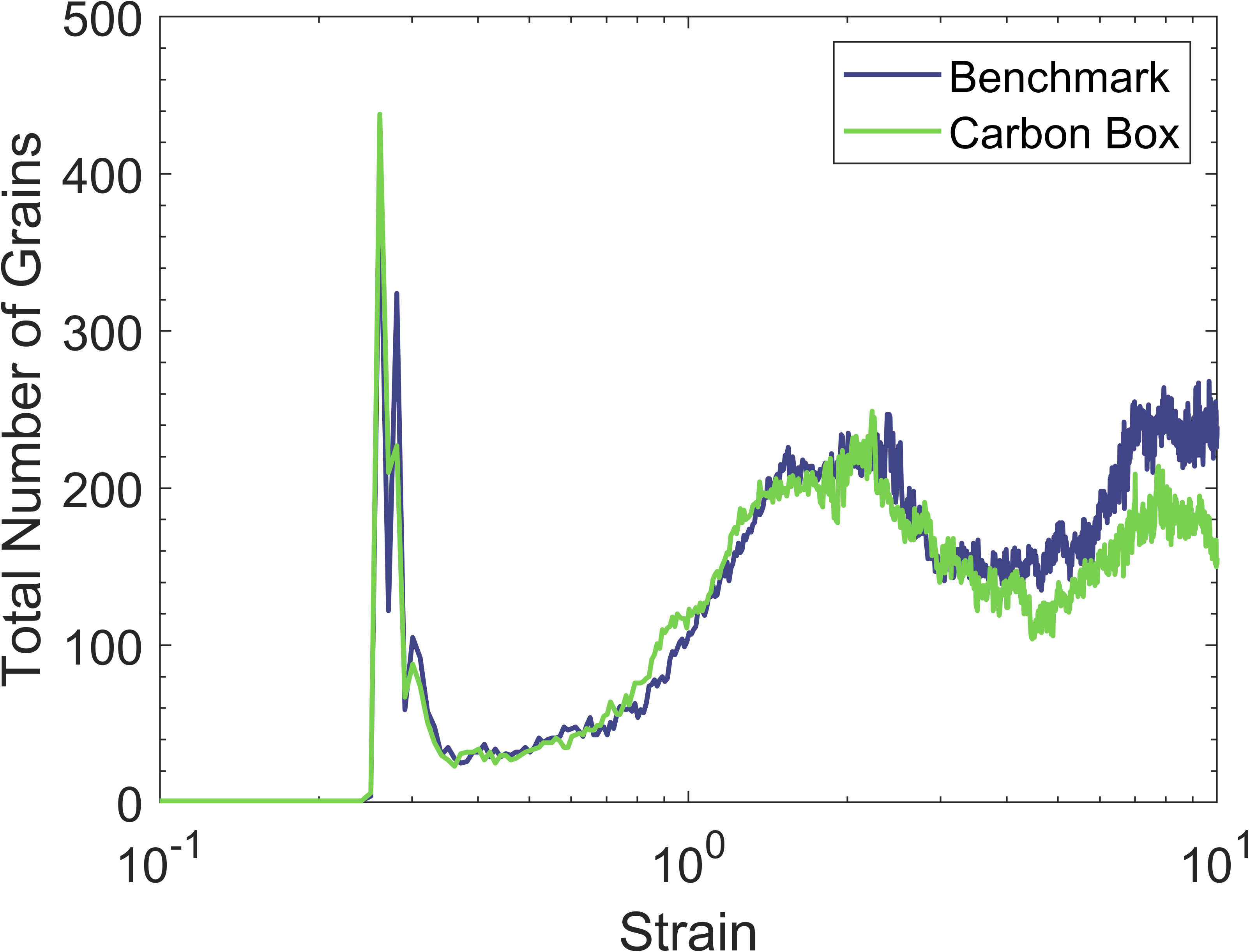}
\label{fig:Fig12c}
\end{subfigure}
\hspace{1em}
\begin{subfigure}{0.45\textwidth}
\caption{} 
\includegraphics[width=\linewidth]{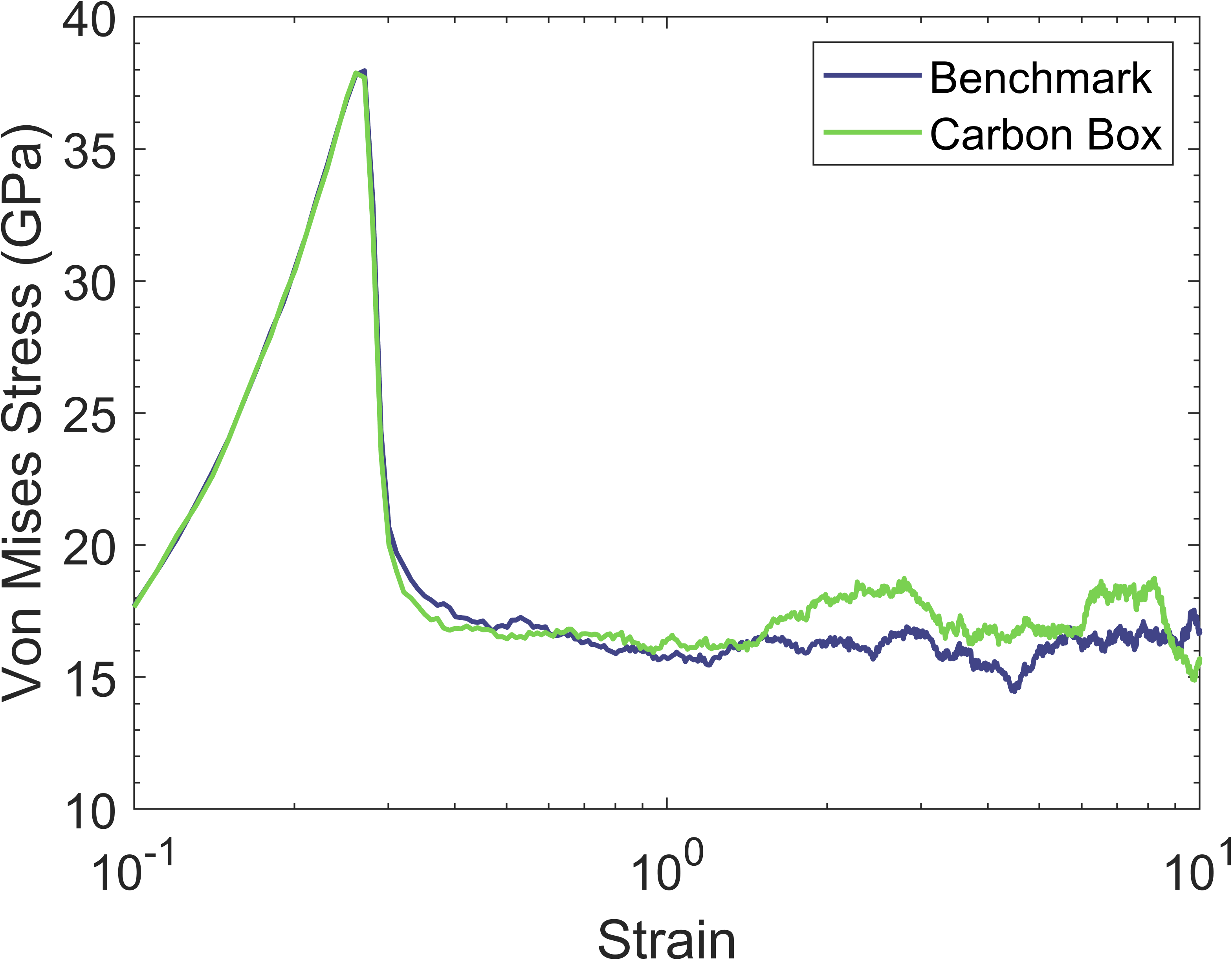}
\label{fig:Fig12d}
\end{subfigure}

\medskip
\begin{subfigure}{0.45\textwidth}
\caption{} 
\includegraphics[width=\linewidth]{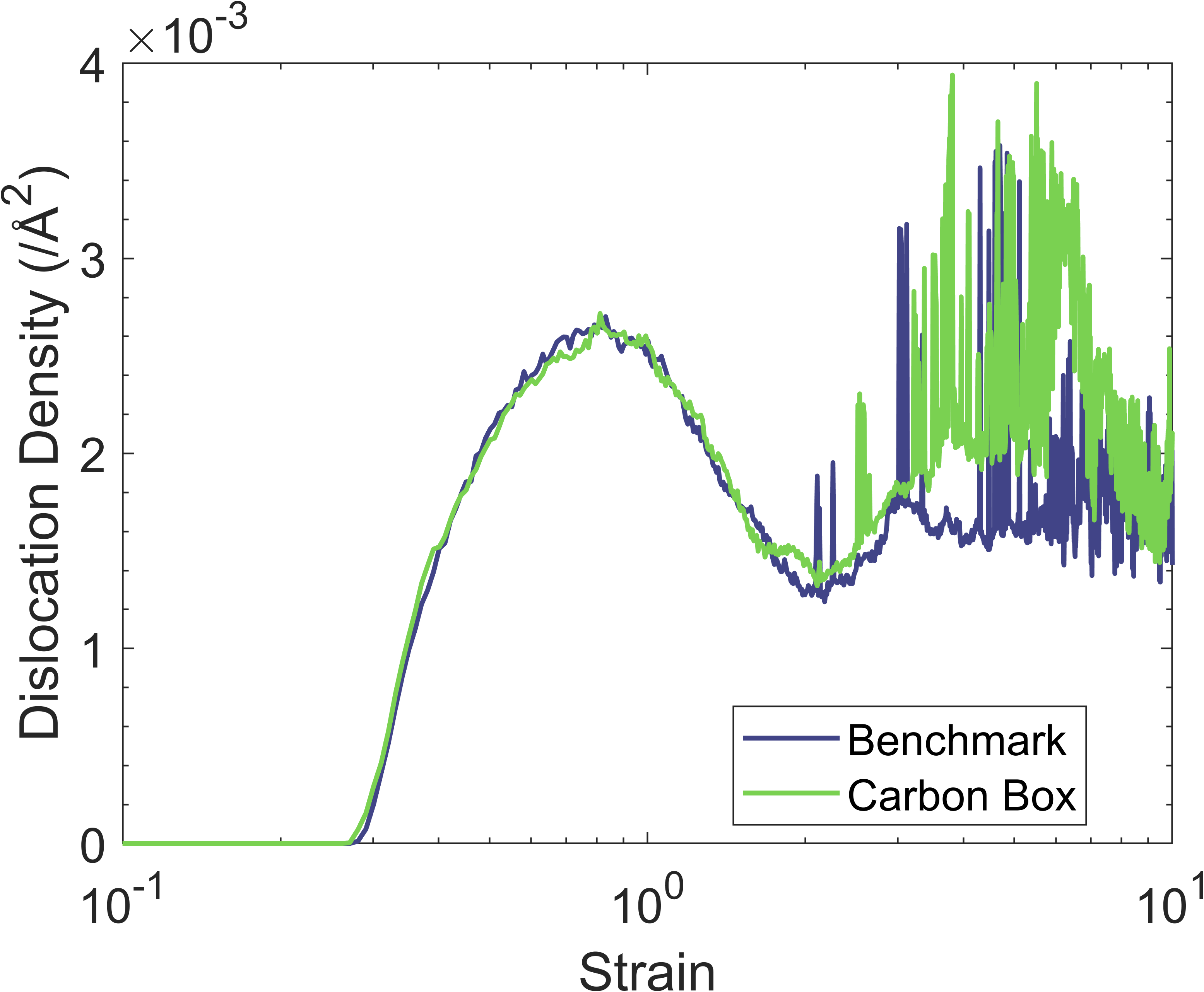}
\label{fig:Fig12e}
\end{subfigure}
\hspace{1em}
\begin{subfigure}{0.45\textwidth}
\caption{} 
\includegraphics[width=\linewidth]{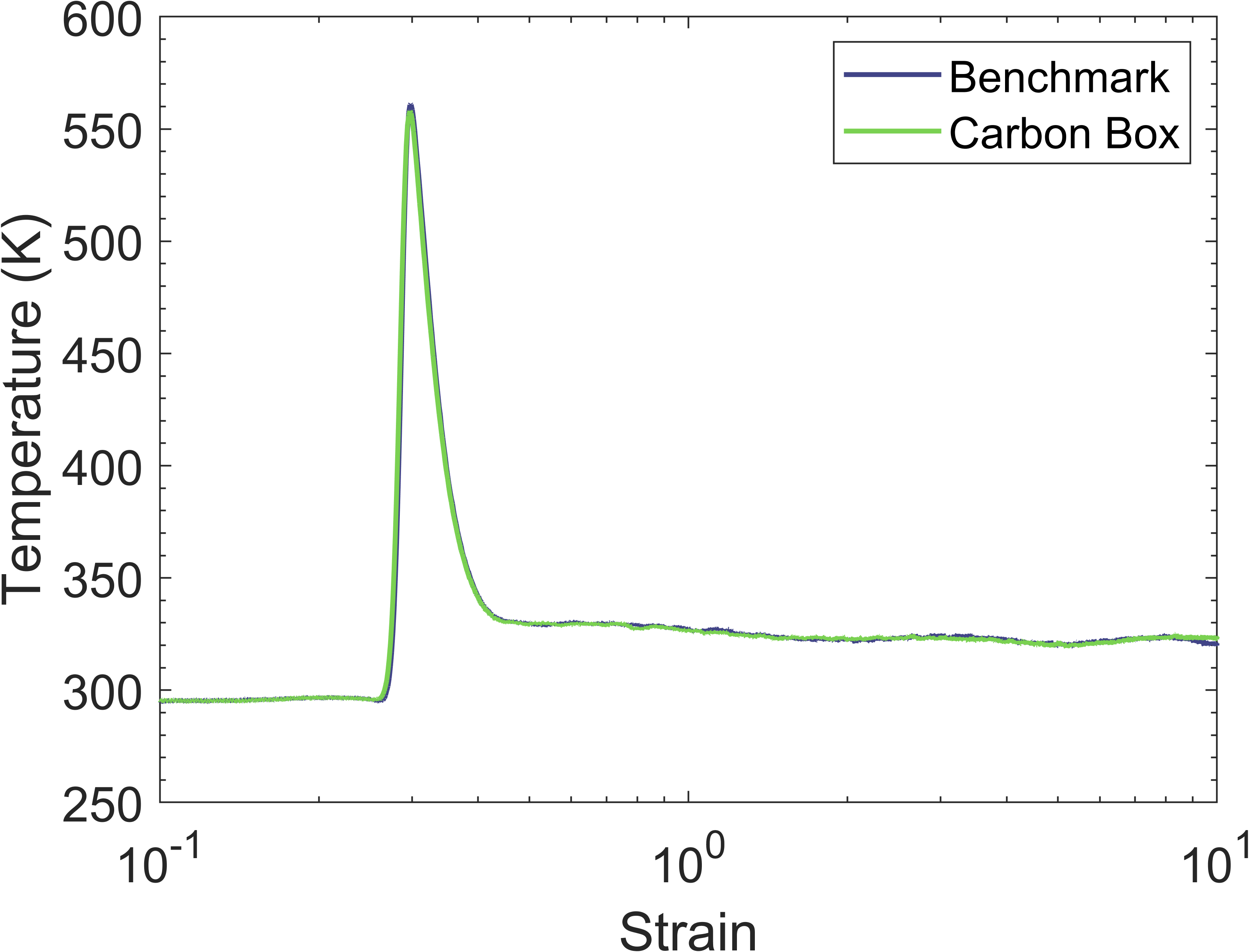}
\label{fig:Fig12f}
\end{subfigure}

\caption{Simulation data plotted as a function of strain for benchmark and carbon-containing simulations. Strain rate $d\gamma/dt= 1/33.5$ ps$^{-1}\approx 2.985\times 10^{10}$ s$^{-1}$ and damping parameter $b=6.875$ eV fs {\AA}$^-2 = b_{1}$. The purple line is for the benchmark and the green line is for 100 appm carbon in iron simulation. (a) Kinetic Energy (b) Potential Energy (c)  Number of Grains (d) Von Mises Stress (e) Dislocation Density (f) Temperature.} \label{fig:12}
\end{figure*}

In this section, we consider the effect of carbon impurities on nanocrystal formation during high shear. Some sources suggest carbon content of 0.006\% has some effect \cite{MILOSEV202165} whilst others argue that a carbon content as low as 0.001\% can already substantially affect the properties of iron \cite{CleavesIron}. Ferritic/martensitic steels that have been selected for fusion applications typically contain less than 0.15\% carbon \cite{CAUSEY2012511}. The atomic size of carbon is sufficiently small that the atom can enter the iron lattice as an interstitial solute atom \cite{Bhadeshia}. Experiments showed that carbon can greatly influence the microstructural behaviour of iron. For example, Stein \cite{STEIN196699} showed that the dislocation velocity exponent increases in iron with an increase in carbon content when held at room temperature. This alters the yield point and rate of crack propagation within the material. Molecular dynamics simulations have also explored the effect of carbon interstitials, for example they can block dislocation motion in an iron simulation cell \cite{shu2007}. Here we inserted 102 carbon atoms into the simulation cell, that is 100 appm or 0.01 atomic \%.

Figure \ref{fig:12} shows various simulation properties with and without carbon impurities. In this figure, we compared the benchmark simulation with a simulation cell containing 100 appm carbon. Figure \ref{fig:Fig12a} shows the differences in kinetic energy between the simulations. The benchmark simulation experiences a rapid increase in kinetic energy, from 0.039 eV/atom at $\gamma = 0.27$, to 0.072 eV/atom at $\gamma = 0.3$, before rapidly decreasing and plateauing at 0.042 eV/atom. Similarly, the kinetic energy for the carbon-containing simulation increases from 0.039 eV/atom at $\gamma = 0.27$, to 0.067 eV/atom at $\gamma = 0.3$, before rapidly decreasing. Hence, the difference in maximum kinetic energy between the simulations is $\sim 0.005$ eV/atom. Furthermore, both simulations experience kinetic energy reductions to 0.043 eV/atom at $\gamma = 0.32$. After this point, both kinetic energies remain very similar, with only minor differences between them. 

In Figure \ref{fig:Fig12b}, the behaviour of the excess potential energy is nearly identical for both simulations up to $\gamma = 1$. Both cells experience a rapid increase from $\gamma = 0$, followed by a rapid decrease at $\gamma = 0.27$, to a value of $\sim 0.09$ eV/atom at $\sim\gamma = 0.82$. However, the excess potential energy of the carbon-containing simulation increases again from this point, reaching 0.12 eV/atom at $\gamma = 2$, before decreasing slightly, and once again increasing to 0.13 eV/atom at $\gamma = 8$. This trend is also shown in Figure \ref{fig:Fig12d}, which shows the von Mises stress for both cells.

Figure \ref{fig:Fig12c} shows the number of grains found in each simulation cell. There are some minor differences between the two simulations. The initial grain number increases up to $\sim 400$ for the benchmark simulation, whilst the maximum number of grains found in the cell with carbon atoms is 440, a difference of 40 grains. Nevertheless, both simulations experience a nearly identical trajectory from $\gamma = 0.27$ to $\gamma = 3$. After this point, the number of grains present in the benchmark simulation is noticeably larger than the carbon-containing cell. For example, at $\gamma = 8$, the number of grains found in the benchmark simulation is $\sim 260$, whilst the carbon-containing cell only has $\sim 210$. 

A similar trend is observed when considering the dislocation densities of the simulations in Figure \ref{fig:Fig12e}. The trajectories of the simulations are nearly identical up to $\gamma = 3$. With continued shearing, the dislocation density of the carbon-containing cell is noticeably larger, until the values become similar for both simulations at $\sim\gamma = 8.5$. 

In Figure \ref{fig:Fig12f}, it is interesting to note very similar temperatures for the benchmark simulation and carbon-containing simulation, whilst the kinetic energy between them is not identical. It is not immediately apparent what could cause this.

Carbon atoms can move within the material by jumping between interstitial sites through diffusion \cite{SuttonInterfaces}. The work done by Wert \cite{Wert1950} characterised the diffusion coefficient of carbon in iron as a function of the temperature, and subsequent work has pointed to a diffusion barrier of $\sim 0.87$ eV for carbon in iron at room temperature \cite{KANDASKALOV2022126159}. This is consistent with the work of Tapasa \textit{et al.} \cite{TAPASA20071}, which found the activation energy of C migration in iron to be 0.82 - 0.86 eV. Fu \textit{et al.} \cite{ChuChun} performed density functional theory (DFT) calculations to explore the migration energy of C atoms in Fe. They found that C atoms will migrate from neighbouring octahedral sites through a tetrahedral site, with an energy barrier of 0.87 eV. This value agrees with the work of Wert \cite{Wert1950} and Tapasa \cite{TAPASA20071}. Other DFT calculations have also obtained similar values \cite{JiangDFT, DomainDFT}. In the carbon containing simulation, the highest kinetic energy exhibited by and individual carbon atom is 0.65 eV. This suggests that the carbon atoms in our simulation are inherently immobile, and will only move through the reordering of atoms through applied strain.

In Figure \ref{fig:Fig12e}, the carbon-containing cell initially shows a higher dislocation density compared to the benchmark simulation up to $\sim\gamma = 8.5$, after which they become similar. Carbon, behaving as a solute interstitial in the lattice, impedes dislocation motion \cite{ARGON19961877}. This prohibits dislocations from gliding effectively within the lattice, hampering their movement toward grain boundaries and limiting the formation of new low-angle grain boundaries. 


\section{Conclusion}

Nanocrystal formation in iron under high shear strain has been observed through molecular dynamic simulations. The process of nucleation and growth of nanograins during the shearing involves a disordered state, reordering of atoms, and grain coarsening. The disordered state is caused by the stretching of atomic bonds to very high levels due to the initially pristine simulation cell. This is followed by an increase in kinetic energy at yielding as dislocations are produced and atoms are reordered, leading to a sudden increase in kinetic energy gained from the drop in potential energy. Subsequently, this energy is dissipated into the environment, mimicked by the thermostat. Atoms rearrange locally to achieve energetically favourable configurations that lead to the formation of nanograins. 

We have examined the influence of various factors, such as thermostat temperature, heat dissipation rate, shear strain rate, and carbon content. Simulations at higher temperatures still form nanocrystalline microstructures, but with larger and longer grains. A faster rate of heat dissipation altered the grain refinement process, but still involved a disordered state, followed by a reordering of atoms and nanograin formation. 

Dynamic restoration mechanisms were observed to play a major role in nanocrystal formation. Simulations with a slower strain rate did not produce nanocrystalline material, with only a single crystal structure being observed. However, a conjugate gradient shear simulation, which captures the athermal, low deformation rate limit, did produce a nanocrystalline structure, which leads us to believe that the longer time scales of the slower rate simulations allow dislocation recovery processes to occur rather than allowing them to pile up and create nanograins. The inclusion of carbon interstitial atoms had little effect on nanocrystal formation, with the process of grain refinement being very similar to that of the pristine material.

The current simulations demonstrate a possible mechanism of nanograin formation using high-shear methods, and have explored large strains previously not computationally explored in literature. These results provide a clear indication of which factors are, and are not critical for nanograin formation in MD simulations

\section{Data Availability}

All input scripts and simulation data presented in the current work are available at \textit{A link will be provided after the review process and before publication.} 

\begin{acknowledgments}
The authors gratefully acknowledge the Department of Engineering Science at the University of Oxford for their contribution to the funding of the project. This work has been carried out within the framework of the EUROfusion Consortium, funded by the European Union via the Euratom Research and Training Programme (Grant Agreement No 101052200 — EUROfusion) and from the EPSRC [grant number EP/W006839/1]. To obtain further information on the data and models underlying this paper please contact PublicationsManager@ukaea.uk. Views and opinions expressed are however those of the author(s) only and do not necessarily reflect those of the European Union or the European Commission. Neither the European Union nor the European Commission can be held responsible for them. The authors acknowledge the use of the Cambridge Service for Data Driven Discovery (CSD3) and associated support services provided by the University of Cambridge Research Computing Services (www.csd3.cam.ac.uk) in the completion of this work. This work used the ARCHER2 UK National Supercomputing Service (https://www.archer2.ac.uk).
\end{acknowledgments}

\appendix
\section{Further analysis}
\subsection{Symmetry Breaking}

\begin{figure*}[t]
\begin{subfigure}{0.45\textwidth}
\caption{} 
\includegraphics[width=\linewidth]{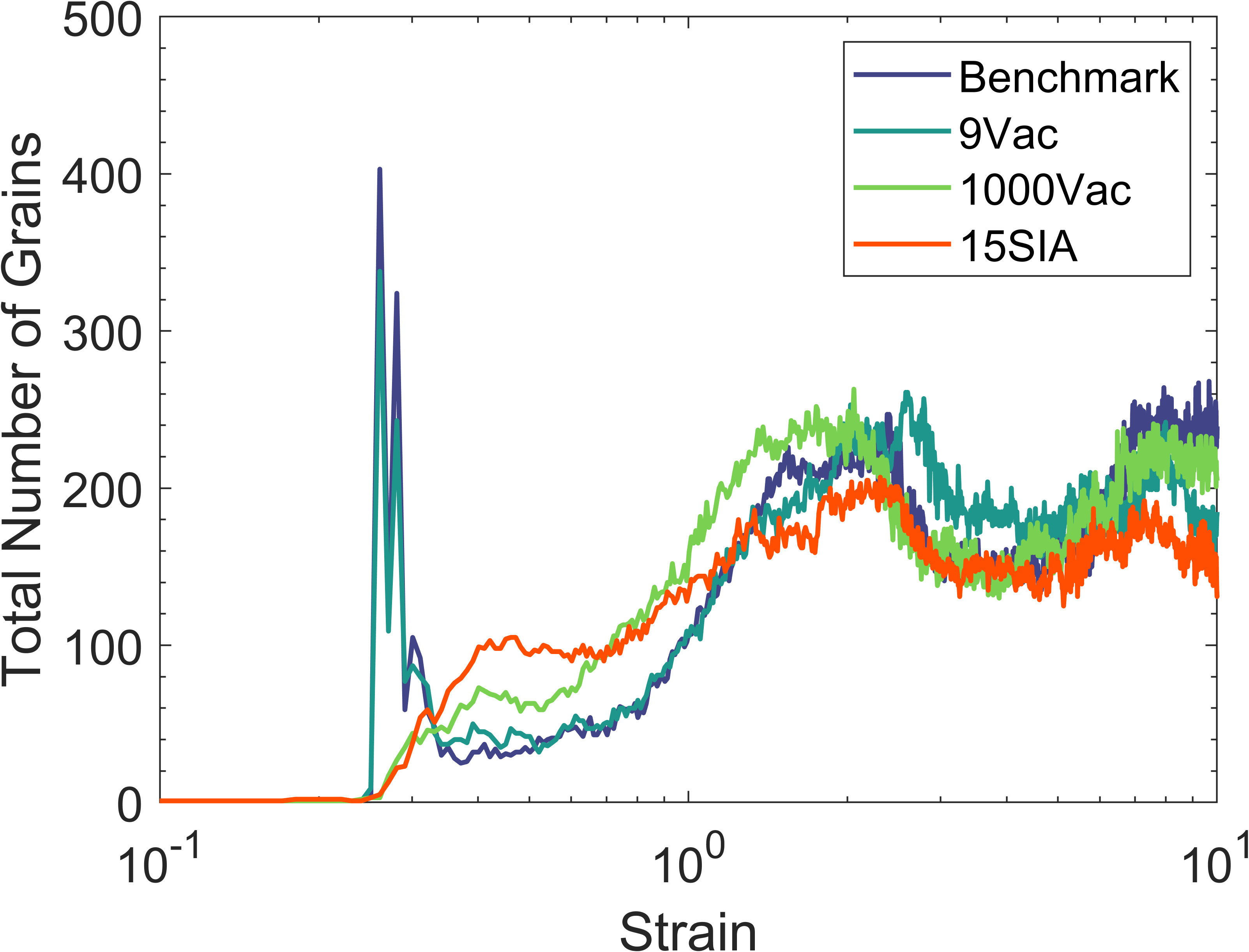}
\label{fig:Fig13a}
\end{subfigure}
\hspace{1em}
\begin{subfigure}{0.45\textwidth}
\caption{} 
\includegraphics[width=\linewidth]{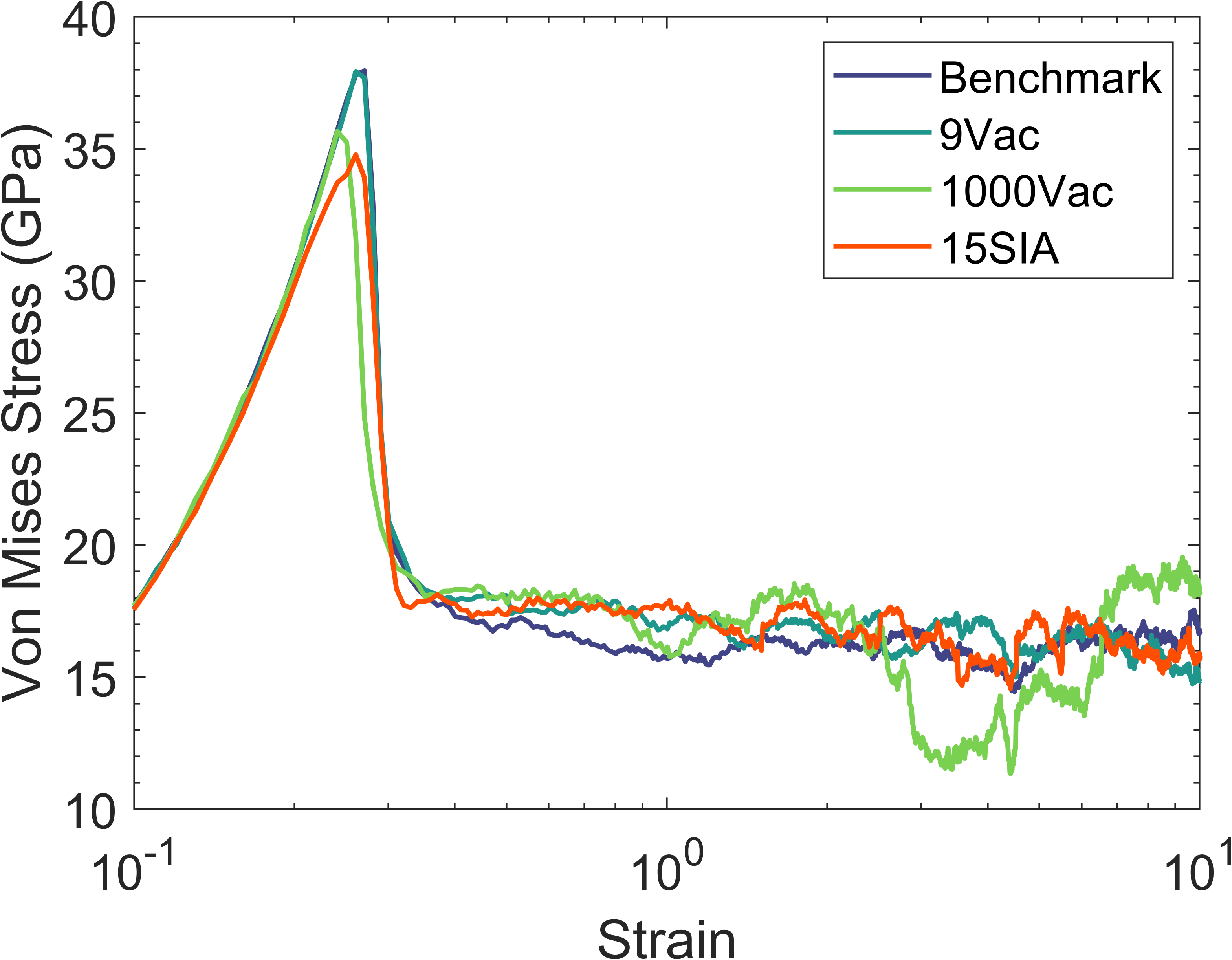}
\label{fig:Fig13b}
\end{subfigure}

\medskip
\begin{subfigure}{0.45\textwidth}
\caption{} 
\includegraphics[width=\linewidth]{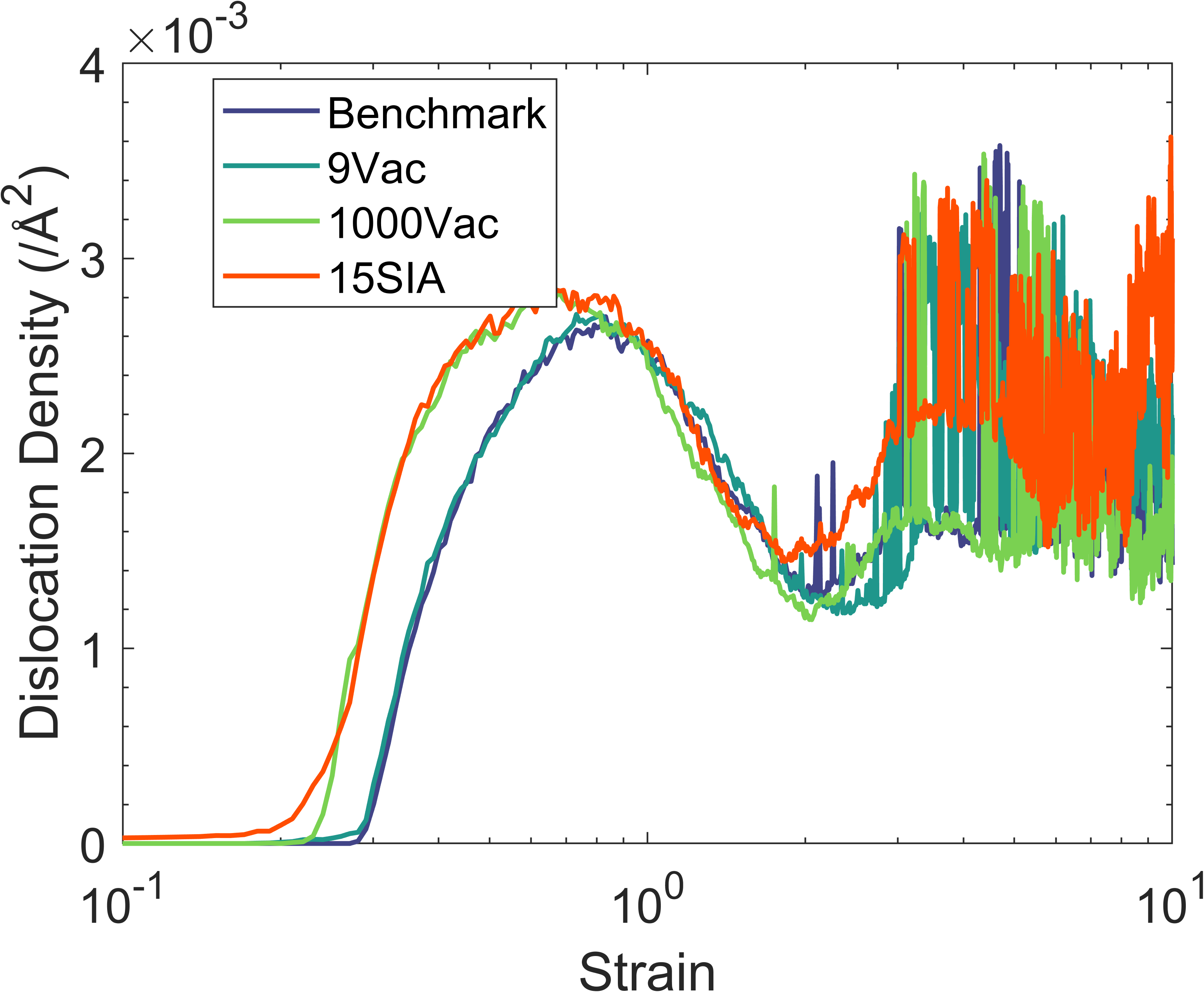}
\label{fig:Fig13c}
\end{subfigure}

\caption{Simulation data against strain for cells initially containing defects compared with the benchmark simulation; (a) Total Number of Grains (b) Von Mises Stress (c) Dislocation Density.} \label{fig:13}
\end{figure*}

Whilst we focus on initially pristine simulation boxes, in line with common practice for molecular dynamics simulations of metallic systems \cite{BulatovPaper}, we recognise that, apart from the previously mentioned defect free microwhiskers, most real metallic systems will have some initial concentration of defects. As such, we performed simulations with different levels of pre-existing defects for comparison. We carry out a simulation with 9 vacancies, located at the centre of the box, henceforth denoted by 9Vac. We also carried out a simulation with 1000 randomly distributed vacancies, denoted by 1000Vac. Lastly, we performed a simulation with a pre-existing self-interstitial prismatic $\frac{1}{2}\langle 111 \rangle$ dislocation loop with a diameter of 15 \AA, denoted as 15SIA. The results are shown in Figure \ref{fig:13}.

Figure \ref{fig:Fig13a} shows nanograin formation for all simulations, in spite of the fact that these simulations are seeded with defects. The behaviour of the Benchmark and 9Vac simulation are very similar in that at $\gamma = 0.27$, we see a large increase in the number of grains followed by a sharp decrease. This does not occur for the 1000Vac and 15SIA simulations, which show a gradual increase in grain number, starting at $\sim\gamma = 0.25$. 

Figure \ref{fig:Fig13b} shows that the yielding and post yielding stress-strain behaviour is again also very similar between the Benchmark and the 9Vac simulation. The 1000Vac simulation yields at a lower stress and at an earlier strain of $\gamma = 0.24$ when compared to the benchmark value of $\gamma = 0.27$. A decrease in yield stress is also observed for the 15SIA simulation however, this occurs at $\gamma = 0.27$, in line with the 9Vac and Benchmark simulation. This is expected as the simulations of \cite{BulatovPaper} previously showed that the yield stress decreases as defects are seeded into the system. This is not surprising as voids can act as dislocation sources \cite{TRAMONTINA20149, DENG2010234} whilst the presence of a dislocation loops reduces the stress, as generating new dislocations from perfect crystal requires greater stress than the multiplication of existing dislocations \cite{FrankRead}. This is further shown in Figure \ref{fig:Fig13c} where the production of dislocations occurs at a lower strain value for the 1000Vac and 15SIA simulations. Interestingly, we observe the creation of nancorystalline iron irrespective of the starting configuration. 

\subsection{Cell Size Influence}

\begin{figure}
\begin{subfigure}{0.45\textwidth}
\caption{} 
\includegraphics[width=\linewidth]{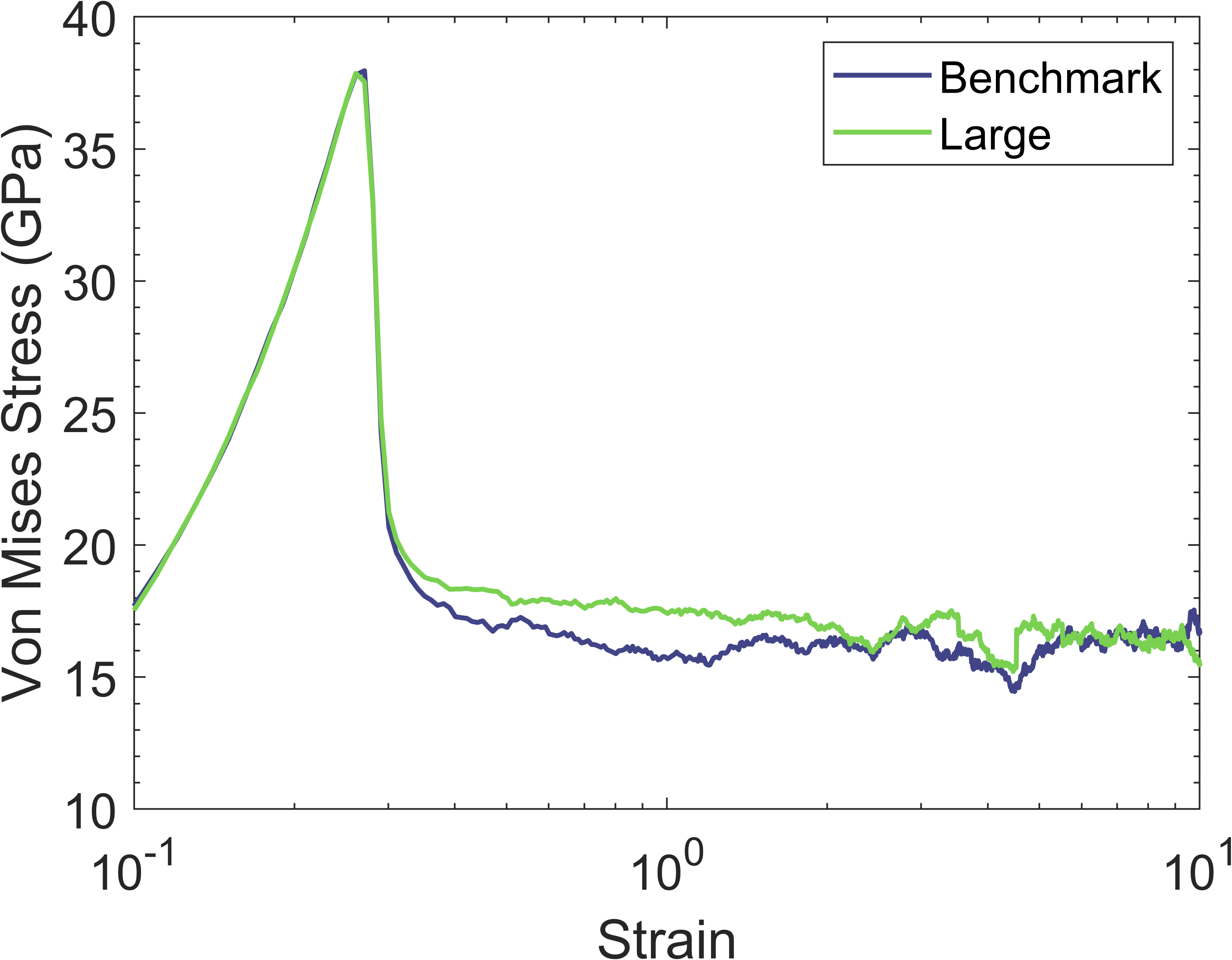}
\label{fig:Fig14a}
\end{subfigure}

\medskip
\begin{subfigure}{0.45\textwidth}
\caption{} 
\includegraphics[width=\linewidth]{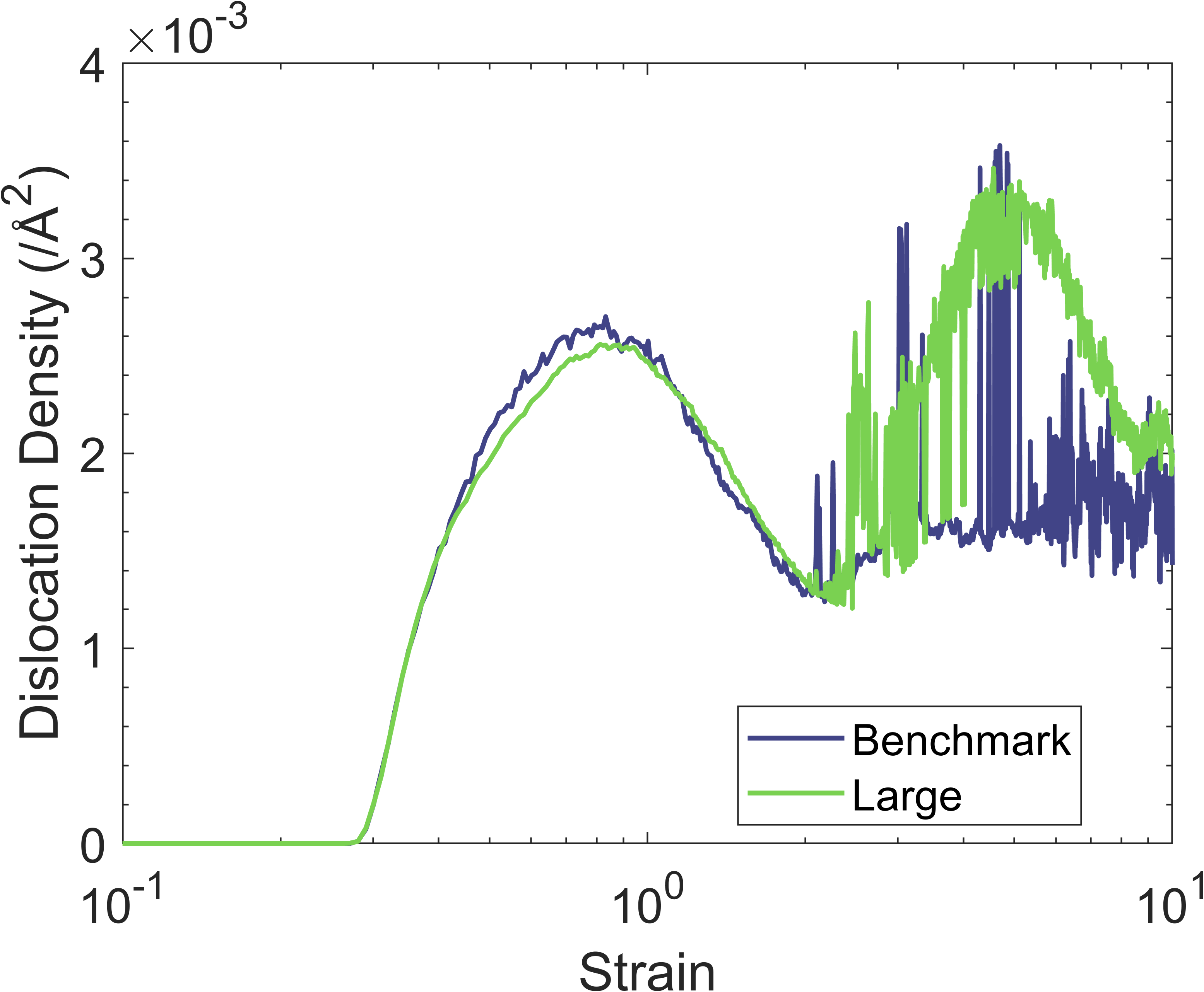}
\label{fig:Fig14b}
\end{subfigure}

\medskip
\begin{subfigure}{0.45\textwidth}
\caption{} 
\includegraphics[width=\linewidth]{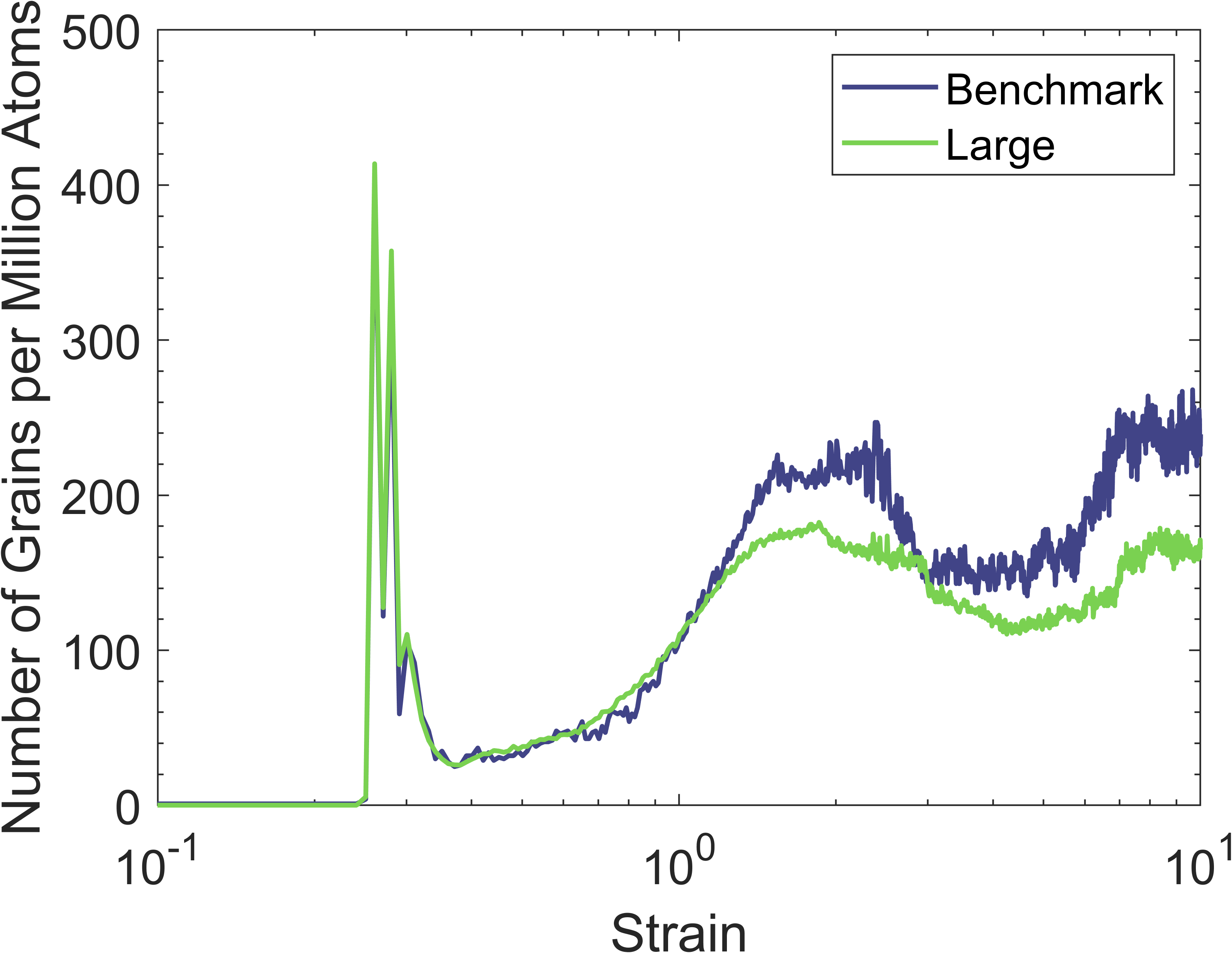}
\label{fig:Fig14c}
\end{subfigure}

\caption{Material behaviour dependence on simulation size; (a) Von Mises Stress (b) Dislocation Density (c) Total Number of Grains per 1 Million Atoms.} \label{fig:14}
\end{figure}

Selecting a simulation cell size that provides valid results, while remaining whithin the computational limitations is an important aspect of any molecular dynamics simulation. Streitz \textit{et al.} \cite{Streitz} recommended a simulation size of 8 million atoms for molecular dynamics simulations to produce size-independent results and as such, we performed a simulation with an a 8,192,000 atom simulation cell and subjected this to shear strain at the same strain rate as the benchmark simulation. The results are shown in Figure \ref{fig:14}. All other parameters were kept identical to the benchmark simulation.

Figure \ref{fig:Fig14a} shows the stress-strain response of the benchmark simulation compared to the large simulation box. Both simulations yield at 38 GPa at $\gamma = 0.27$, before the benchmark simulation reduces to and plateaus at $\sim 16$ GPa. The larger simulation box also reduces to $\sim 17$ GPa however, after $\gamma = 5.5$, both simulations saturate at $\sim 17$ GPa until the end of the simulation. 

Figure \ref{fig:Fig14b} also shows the dislocation density for each simulation. Overall, the dislocation density response in the benchmark and larger simulation are virtually identical up to $\gamma = 2.25$. After this point, the dislocation density for the larger simulation is higher than that of the benchmark simulation. It is not immediately evident why this difference may be occurring as there are no anomalies in the material response which can be attributed to this.  

Figure \ref{fig:Fig14c}, shows the total number of grains per 1 million atoms for the benchmark and large simulation. As expected, the large simulation has much more grains than the benchmark simulation so the data is normalised. Both simulations show a sharp increase in grain number to $\sim 410$ grains per million atoms at $\gamma = 0.27$, and exhibit very similar behaviour after this point. However, after $\sim\gamma = 1.25$, the benchmark simulation has a slightly greater number of overall grains per million atoms.

Overall, the material response of the smaller and larger box sizes are very similar and thus provide confidence that the 1 million atom cell exhibits representative behaviour. As previously mentioned, we are not ruling out the existence of size effects at larger simulation sizes, but we do not further explore these in this work. 

\subsection{Slip Systems}

To analyze the slip systems activated by shearing in the current orientation, we calculated the Schmid factor \cite{schmid1950plasticity}. We apply a constant shear strain in the \textit{xy} direction, resolving to principal strain at 45\degree. While the stress magnitude and direction changes as the simulation progresses, the initial applied stress direction is $\sigma = [1\overline{1}0]$. With shear strain confined to the \textit{xy} direction and periodic boundary conditions in place, plane stress is assumed. In body-centred cubic structures like $\alpha$-iron, slip direction predominantly align with the $\langle 111\rangle$ family, while slip planes encompass the \{110\}, \{112\}, and \{123\} families for iron \cite{barrett1937slip, theoryofdisloc}. Using these stress values alongside slip planes and directions, we calculated the Schmid factor using:
\begin{equation}
    m = cos(\phi)cos(\lambda),
\end{equation}
where \textit{m} is the Schmid factor, $\phi$ is the angle between the normal of the slip plane and the direction of applied stress, and $\lambda$ is the angle between the direction of applied stress and the slip direction. 

The Schmid factor can be calculated using:
\begin{eqnarray}
cos(\phi) = \frac{\vec{\sigma} \cdot \vec{n}}{|\vec{\sigma}| |\vec{n}|}, \\
cos(\lambda) = \frac{\vec{\sigma} \cdot \vec{d}}{|\vec{\sigma}| |\vec{d}|},
\end{eqnarray}
where $\vec{n}$ is the vector normal to the slip plane, and $\vec{d}$ is the vector in the direction of the slip. 

\begin{figure}
\includegraphics[width=0.45\textwidth]{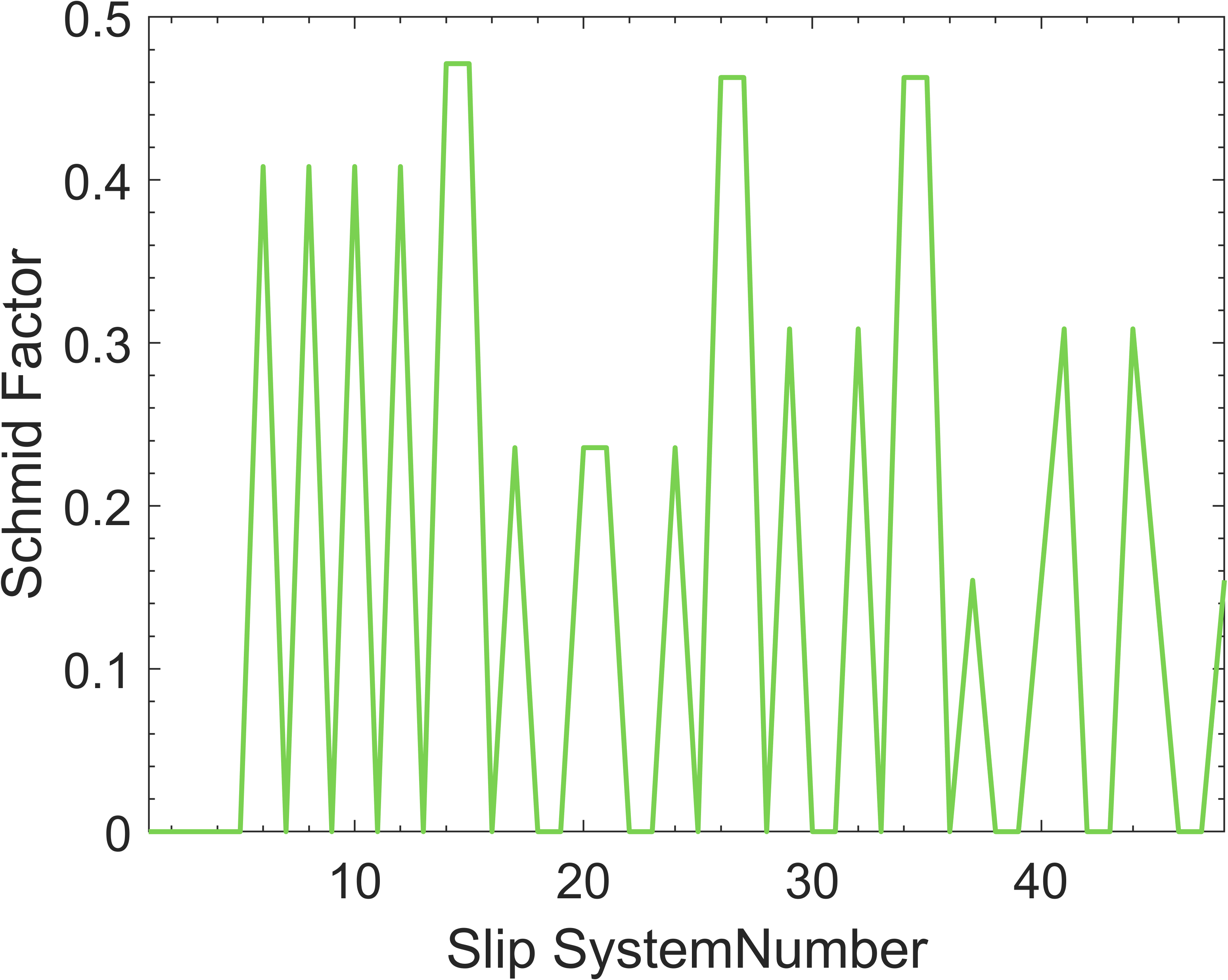}
\caption{Schmid Factor for all slip systems in BCC system under shear strain.} \label{fig:Fig15}
\end{figure}

There are a total of 48 slip systems in body-centred cubic metal. A full list of these can be found in \cite{SlipSystems}. The Schmid factor was calculated for each system and this is shown in Figure \ref{fig:Fig15}. The numbering of the slip systems matches that which is employed in Table 2 of \cite{SlipSystems}.

Figure \ref{fig:Fig15} shows that multiple slip systems can activate when the crystal is oriented along $x = [100]$ and the stress is applied in the $\sigma = [1\overline{1}0]$ direction. This observation may shed light on why the grain refinement process produces nanograins after a highly disordered state. During the shearing of the cell, a diverse set of slip systems is simultaneously engaged. Upon reaching the yield point, this leads to dislocations on multiple slip systems, resulting in stron forest hardening and subsequent dislocation organisation and grain boundary formation.
 
\subsection{Further Dislocation Analysis}

\begin{figure}
\includegraphics[width=0.45\textwidth]{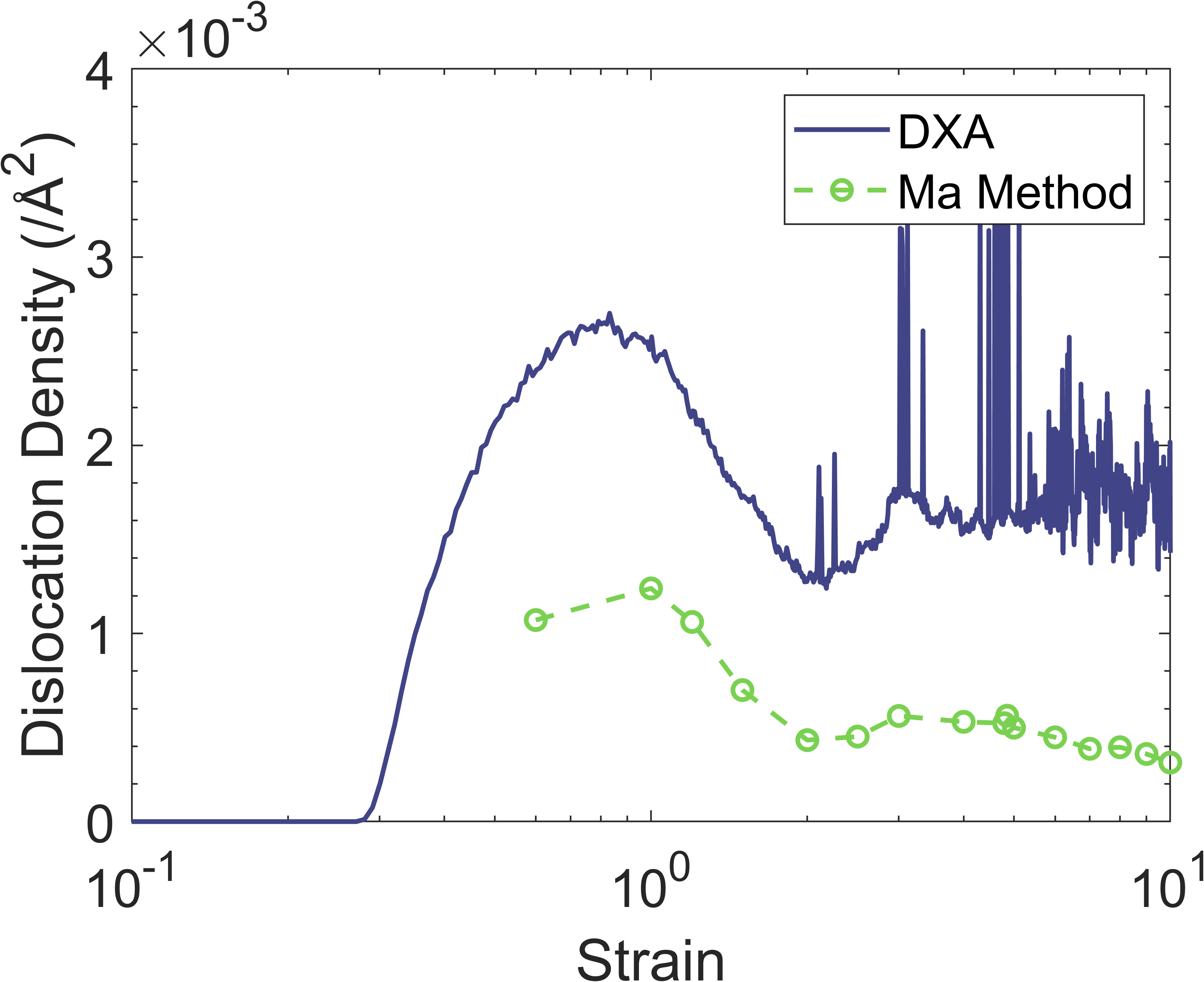}
\caption{Dislocation density comparison between standard dislocation analysis and dislocation analysis carried out when atoms near grain boundaries are removed.} \label{fig:16}
\end{figure}

We performed additional dislocation analysis calculations after eliminating grain boundaries. This is done by removing atoms close to grain boundaries using the code developed by Mason \cite{MasonCode,Ma_JNM_2023}. It provides the distance $d$ of each atom to the nearest grain boundary. Atoms with $d<1$ {\AA} were excluded, followed by a subsequent dislocation analysis calculation. Figure \ref{fig:16} displays the analysis outcomes. 

Figure \ref{fig:16} shows that the standard dislocation analysis tends to overestimate dislocation density, possibly due to grain boundaries. Dislocations can glide and accumulate at grain boundaries, with some leading to the formation of low-angle grain boundaries \cite{ITO200932, Borodachenkova17, XU20223506, XiuyanLi}. It is likely that dislocation analysis modifier interprets certain low-angle grain boundaries as dislocation lines, accounting for the discrepancy in dislocation count. Specifically, conspicuous spikes in dislocation density $\sim\gamma = 4.8$ are observed, absent when $d<1$ atoms are excluded. Thus, we conclude that the standard dislocation analysis encompasses dislocation pile-ups in its dislocation density computation. Nevertheless, the current analysis remains a robust foundation for comparing various simulations.

\subsection{What Constitutes a Grain?}

\begin{figure}
\includegraphics[width=0.45\textwidth]{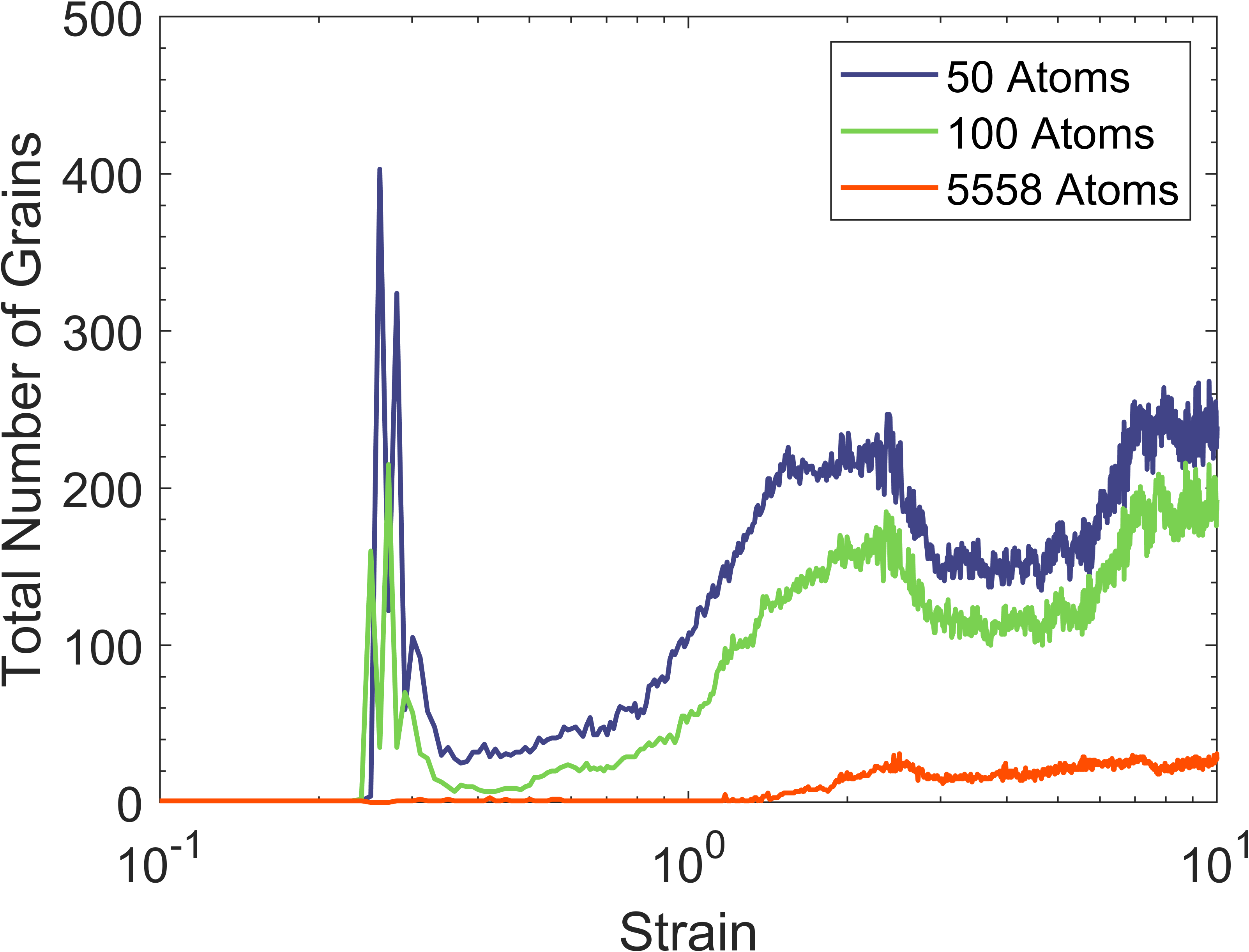}
\caption{Total number of grains based on minimum number of atoms per grain.} \label{fig:17}
\end{figure}

\begin{figure}
\begin{subfigure}{0.45\textwidth}
\caption{} 
\includegraphics[width=\linewidth]{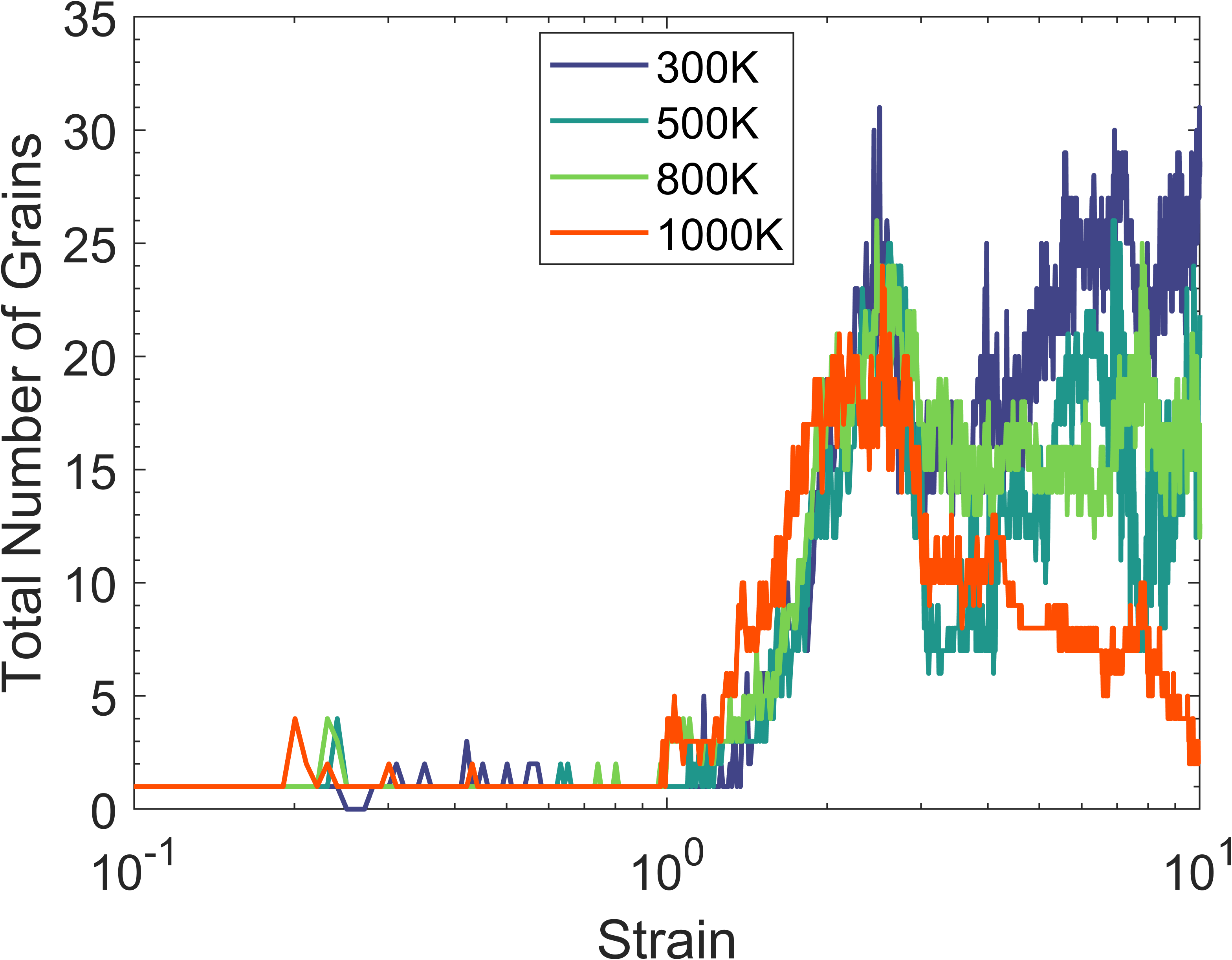}
\label{fig:Fig18a}
\end{subfigure}

\medskip
\begin{subfigure}{0.45\textwidth}
\caption{} 
\includegraphics[width=\linewidth]{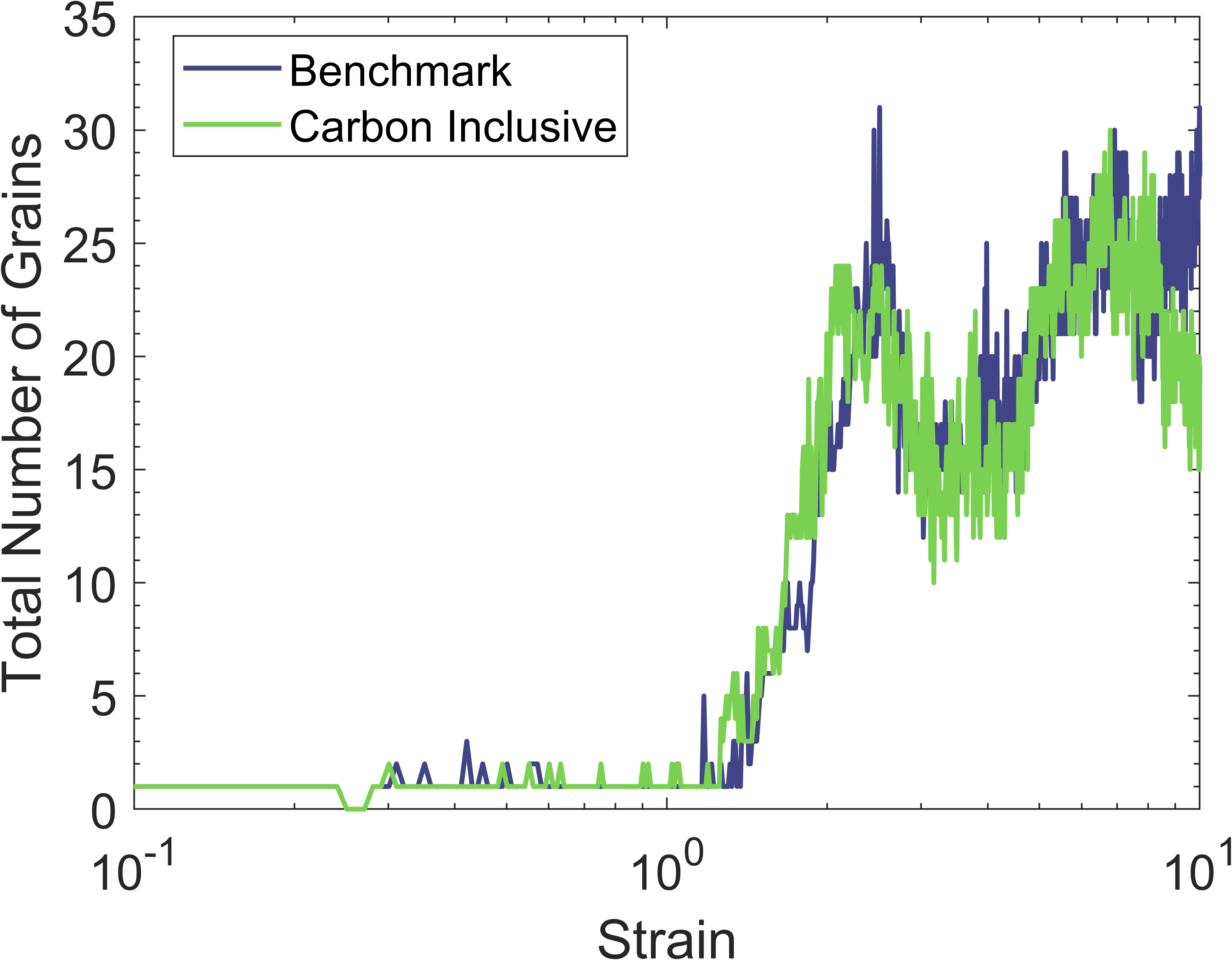}
\label{fig:Fig18b}
\end{subfigure}

\caption{Number of 5 nm spherical grains found in the simulations; (a) Dependence of temperature on 5 nm grains, and (b) Dependence of carbon inclusion on 5 nm grains.} \label{fig:18}
\end{figure}

In this work, we considered a minimum grain size of 50 atoms when determining the number of grains present within the system. Figure \ref{fig:17} shows the comparison of grain numbers for the benchmark simulation with different numbers of atoms selected to form the minimum grain size. The Ovito default of 100 atoms was selected for comparison, as was 5,558 atoms, which corresponds to a 5 nm diameter spherical grain. A 5 nm grain corresponds to a minimum grain size which can accurately be resolved using transmission Kikuchi diffraction (TKD, or t-EBSD) \cite{MORTAZAVI201542}, allowing comparisons with future experimental HPT data.

Figure \ref{fig:17} shows that the number of grains between the 50-atom and 100-atom analyses largely follow the same trajectory. At $\gamma = 0.27$, our-50 atom analysis showed 400 grains whilst the 100-atom analysis showed 215. This means that 185 grains were found to have less than 100 but more than 50 atoms at this point. This further suggests that the spike is caused by a highly disordered state of the atoms whereby only small pockets of body-centred cubic atoms remain, which is flagged as a grain. At this strain value, the 5 nm grain analysis cannot pick up any grains due to the highly disordered state, and the grain number is shown as 0. After $\gamma = 2$, the 50- and 100-atom trajectories follow the same trajectory, however, there is always a difference of $\sim 40$ grains between the analyses. Therefore, we conclude that there are always $\sim 40$ grains in the simulation cell which have less than 100 atoms but more than 50. Interestingly, the 5 nm grain analysis does not follow the same trajectory, and instead hovers around 1 grain up to $\sim\gamma = 1.37$. After this point, the grain number increases for the 5 nm analysis, as shown in Figure \ref{fig:17}. A local maximum of grain number is reached at $\sim\gamma = 2.5$, with 30 grains being present. This is followed by a minor decrease, after which the grain number sits steadily at $\sim 25$ after $\gamma = 4$. As such, we observe that the process of grain refinement is still observed, even if we define a grain with a minimum of 5,558 atoms.

By increasing the minimum number of atoms that constitute a grain, we can compare the effects of temperature and carbon on the formation of experimentally observable grains. Figure \ref{fig:18} shows the number of 5 nm spherical grains found in different temperature simulations and carbon-containing simulations.

In Figure \ref{fig:Fig18a}, there is no initial spike in 5 nm grains for any simulation, and the trajectories of the different simulations are largely similar until $\sim\gamma = 2.5$. This is contrary to the data shown in Figure \ref{fig:Fig4c} which shows that the 300 K simulation has the most grains at this point, suggesting that at 300 K, the grain numbers are increased due to the presence of many small grains. We also notice that the 1000 K simulation has $\sim 2$ grains larger than 5 nm at $\gamma = 10$, which is visible by considering Figure \ref{fig:Fig5e}. Ultimately, the 300 K simulation still shows the largest number of grains overall. However, this is not to the same extent as in Figure \ref{fig:Fig4c}, again suggesting that the shearing process stimulates the production of many small grains in the 300 K simulation. Nevertheless, the 500 K and 800 K simulations also have many small grains, and this is visible by comparing Figure \ref{fig:Fig18a} to Figure \ref{fig:Fig4c}. For example, at $\gamma = 6$, the total number of grains present for the 500 K simulation is 120 and that number drops to 20 when considering only 5 nm grains. 

Figure \ref{fig:Fig18b} also shows the comparison of 5 nm grains between the benchmark and carbon-containing simulation. We show here that the deviation in grain number observed in Figure \ref{fig:Fig12c} does not occur when considering 5 nm grain, up to $\gamma = 8.5$. This suggests that the benchmark simulation has many smaller grains than the carbon-containing simulation. After $\gamma = 8.5$, we notice that the carbon-containing simulation begins to dip in grain number whilst the benchmark simulation cell rises.

\subsection{Thermal Stability of Grains}

\begin{figure}
\vspace{1cm}
\includegraphics[width=0.45\textwidth]{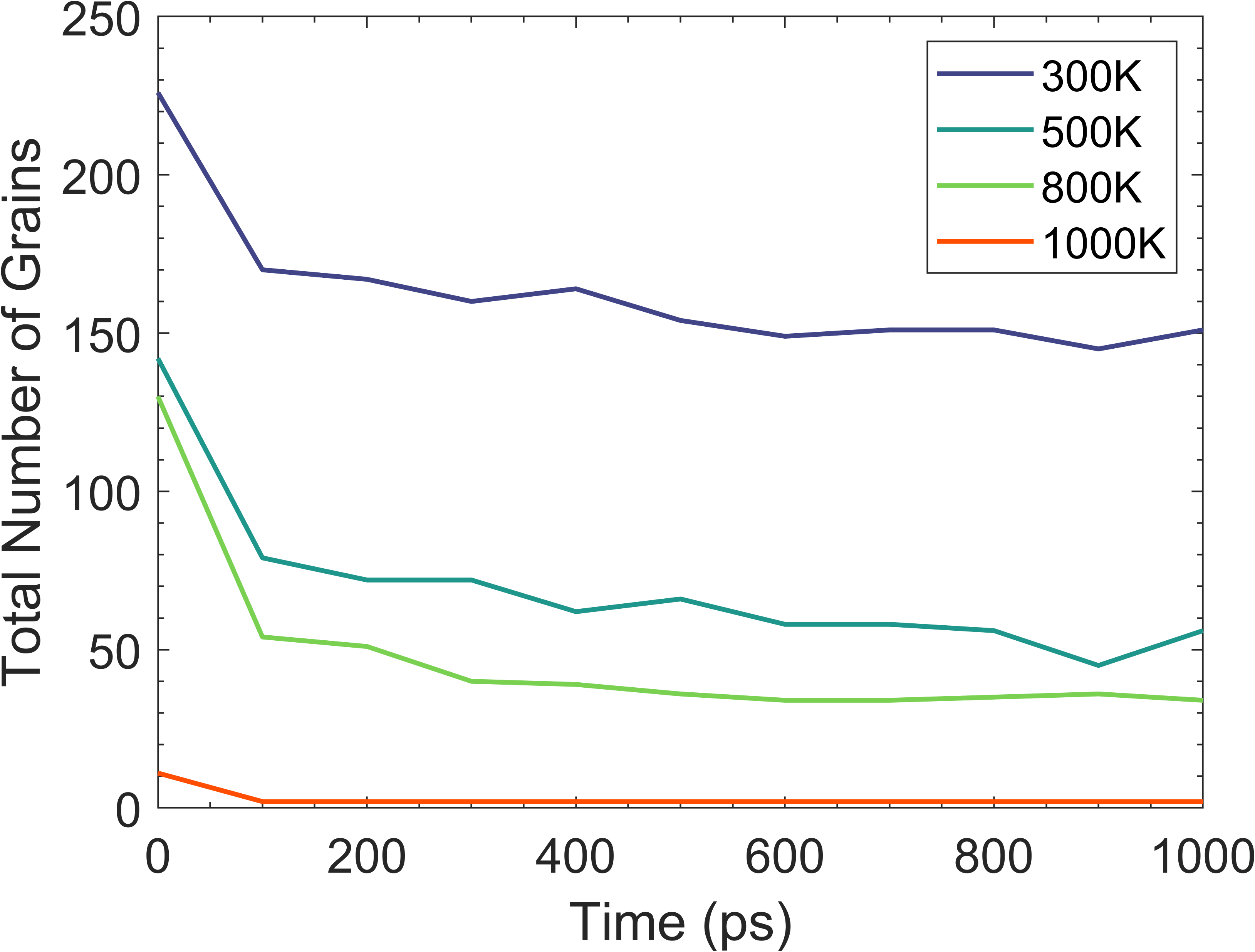}
\caption{Number of grains present in different temperature simulations after thermalising.} \label{fig:Fig19}
\end{figure}

It is important to consider the stability of the newly formed grains to assess their usefulness for future simulations. The benchmark simulation and the variable temperature simulations from Section 3 were chosen for assessment. These cells were thermalised for 1ns using the NPH ensemble, with the Langevin thermostat keeping the temperature at 300 K, 500 K, 800 K, and 1000 K, respectively. Figure \ref{fig:Fig19} shows the total number of grains present after the 1ns run time for each simulation cell. For the purposes of this comparison, the minimum grain size was taken as 50 atoms.

Figure \ref{fig:Fig19} shows that all cells follow similar trends. There is first a reduction in the number of grains found in the cell, after which the number of grains stabilises after $\sim 700$ ps of thermalisation. The 300 K, 500 K, and 800 K simulations stabilise at $\sim 150$, $\sim 56$, and $\sim 35$ grains, respectively. Interestingly, the 1000 K simulation stabilises much earlier, at a value of 2 grains. The results suggest that the nanocrystalline structure should remain stable at finite temperatures. 

\vfill

\bibliography{reference}

\end{document}